\documentclass{optica-article}

\journal{opticajournal}
\articletype{Research Article}

\usepackage{amsmath}
\usepackage{mathtools}

\usepackage{lineno}
\usepackage{graphicx}
\usepackage{amsthm}
\usepackage{placeins}
\usepackage{subcaption}
\usepackage{float}
\usepackage{array} 
\usepackage{cancel}
\usepackage{todonotes}
\usepackage{xcolor}

\usepackage{mdframed}

\newmdenv[
  backgroundcolor=green!5,
  linecolor=green!50!black,
  linewidth=2pt,
  roundcorner=5pt,
  skipabove=10pt,
  skipbelow=10pt,
  frametitle={Example},
  frametitlefont=\bfseries,
  splittopskip=\topskip,  
  splitbottomskip=0pt,
  skipbelowabove=10pt,
]{exampleblock}

\usepackage{pgfplots}
\pgfplotsset{compat = 1.5}
\usepackage{accents}

\usepackage[most]{tcolorbox}

\newtheorem{theorem}{Theorem}[section]

\tcolorboxenvironment{theorem}{
  enhanced,
  colback=blue!5, 
  colframe=blue!70, 
  boxrule=0.8pt,
  arc=4pt,
  left=6pt,
  right=6pt,
  top=6pt,
  bottom=6pt,
  fonttitle=\bfseries,
}

\newtheorem{criterion}{Criterion}

\makeatletter
\newenvironment{proofw}[1][\proofwname]{\par \pushQED{\qed}%
    \normalfont \topsep6\p@\@plus6\p@\relax
    \trivlist
    \item\ignorespaces
    }{%
    \popQED\endtrivlist\@endpefalse
}
\newcommand{\proofwname}{\mbox{}}
\makeatother

\definecolor{rebeccaGreen}{RGB}{38, 140, 36}

\usepackage{etoolbox}
	\patchcmd{\thebibliography}{\chapter*}{\section*}{}{}

\newcommand{\defeq}{\coloneqq}

\newcommand{\bs}[1]{\boldsymbol{#1}}
\newcommand{\bmat}{\begin{bmatrix}}
\newcommand{\ebmat}{\end{bmatrix}}
\newcommand{\calE}{\mathcal{E}}
\newcommand{\calN}{\mathcal{N}}
\newcommand{\bcalK}{\boldsymbol{\mathcal{K}}}
\newcommand{\p}{\partial}
\newcommand{\pp}[2]{\frac{\partial #1}{\partial #2}}
\newcommand{\ppn}[3]{\frac{\partial^{#1} #2}{\partial #3^{#1}}}

\newcommand{\dd}[2]{\frac{d #1}{d #2}}
\newcommand{\ddn}[3]{\frac{d^{#1} #2}{d#3^{#1}}}

\newcommand{\utilde}[1]{\underaccent{\tilde}{#1}}

\newcommand{\mbf}[1]{\mathbf{#1}}
\newcommand{\mbt}[1]{\mathbf{\tilde{#1}}}

\newcommand{\veps}{\varepsilon}
\newcommand{\pmp}{\text{p}}
\newcommand{\sig}{\text{s}}

\newcommand{\Is}{I_\sig}

\newcommand{\Ip}{I_\pmp}
\newcommand{\Iso}{I_{\sig 0}}
\newcommand{\lp}{\lambda_{\pmp}}
\newcommand{\ls}{\lambda_{\sig}}
\newcommand{\lt}{\lambda_{\text{t}}}
\newcommand{\FM}{\text{FM}}
\newcommand{\HOM}{\text{HOM}}
\newcommand{\nco}{n_{\text{core}}}
\newcommand{\ncl}{n_{\text{clad}}}

\newcommand{\rcl}{r_{\text{clad}}}

\newcommand{\dT}{\overline{{\Delta}T}}

\newcommand{\FT}[3]{\hspace{-1.5pt}\stackrel{\mathfrak{F}}{\mbox{}_{#1 \mapsto #2}}\hspace{-2.5pt}\big\{ #3 \big\}}

\newcommand{\fft}[3]{\hspace{-1.5pt}\stackrel{\text{fft}}{\mbox{}_{#1 \mapsto #2}}\hspace{-2.5pt}\big\{ #3 \big\}}

\definecolor{fgreen}{RGB}{63, 90, 54}

\newcommand{\tcfg}[1]{\textcolor{fgreen}{#1}}

\begin{document}

\title{Accelerated Coupled Mode Model for Fiber Laser Amplifiers as an Averaged Dynamical System}

\author{ Rebecca Bryant\authormark{1,*}, Jacob Grosek\authormark{2}, and Jay Gopalakrishnan\authormark{1} }

\address{
  \authormark{1}Portland State University, PO Box 751,  Portland, OR 97207, USA \\
  \authormark{2}Air Force Research Laboratory, 3550 Aberdeen Ave. SE, Kirtland Air Force Base, NM 87117, USA
}

\email{\authormark{*}rnb3@pdx.edu} %% email address is required; see note below about the corresponding author designation

\bigskip\bigskip\bigskip

\tableofcontents 
\newpage

%%%%%%%%%%%%%%%%%%%%%%%%%%%%%%%%%%%
% Abstract
%%%%%%%%%%%%%%%%%%%%%%%%%%%%%%%%%%%
\begin{abstract}
    We apply a known theorem for simplifying dynamical systems with bounded error to a specific optical fiber waveguide problem, supplementing the physical intuition and heuristics used in the optics community with proper mathematical justification.  
    Using techniques from averaging theory of dynamical systems, a reliable accelerated model based on the coupled mode theory (CMT) approach for a common fiber laser amplifier application is derived. 
    Computational testing reveals that this accelerated model achieves an ${\sim}4000$x increase in computational speed compared to the CMT model while preserving a high accuracy in key figures-of-merit such as output power and amplification efficiency. 
    Further, we argue that by adopting our recommended approximations within the reduced model framework enables the model to be applied a wider set of amplifier types and configurations than can the current (comparable) reduced models found in the literature.  
\end{abstract}

%%%%%%%%%%%%%%%%%%%%%%%%%%%%%%%%%%%
% Section: Introduction
%%%%%%%%%%%%%%%%%%%%%%%%%%%%%%%%%%%

\section{Introduction}\label{sec:Intro}

The vast majority of optical fiber models rely upon the \textit{slowly varying envelope approximations} (SVEAs) to reduce the computational burden of directly solving the Maxwell equations. 
\begin{subequations}\label{eq:SVEAs}
    For a highly coherent optical field with a wavenumber $\beta$ [rads/m] and a frequency $\omega$ [rads/s], there is a longitudinal SVEA (along the direction of propagation), 
    \begin{equation}\label{eq:LongitudinalSVEA}
        \Big| \ppn{2}{A}{z} \Big| \ll \Big| \beta \pp{A}{z} \Big| \quad \text{ or } \quad \Big| \ppn{2}{A}{z} \Big| \ll \Big| \beta \pp{A}{z} \Big| \ll \Big| \beta^{2} A \Big| \ \ \text{(\hspace*{-0.75mm}\cite[{\S}3.3]{shen1984principles}, \cite[Chpt. 1]{sutherland2003handbook}, \cite[p. 87]{banerjee2004nonlinear})} ,
    \end{equation}
    and a temporal SVEA, 
    \begin{equation}\label{eq:TemporalSVEA}
        \Big| \ppn{2}{A}{t} \Big| \ll \Big| \omega \pp{A}{t} \Big| \quad \text{ or } \quad \Big| \ppn{2}{A}{t} \Big| \ll \Big| \omega \pp{A}{t} \Big| \ll \Big| \omega^{2} A \Big| \ \ \text{(\hspace*{-0.75mm}\cite[Chpt. 1]{sutherland2003handbook}, \cite[p. 132]{banerjee2004nonlinear})} ,
    \end{equation}
    where the electric field $E$ is expressed as $E(x, y, z, t) = \text{real}\big( A(x, y, z, t) \exp[i \beta z - i \omega t] \big)$ and the complex-valued amplitude/envelope $A$ is considered to be slowly varying with respect to the frequencies $\beta$ and $\omega$.  
\end{subequations}
Laser light is by definition coherent, both longitudinally and temporally, which means that the beam ought to be highly collimated in its direction of travel ($z$-direction) and nearly monochromatic in its spectrum. 
Thus, a perfect laser field would be represented by Dirac delta functions in both the longitudinal and temporal frequency spaces centered about the wavenumber $\beta$ and frequency $\omega$, respectively; and the field's slowly varying envelope ($A$) would be unvarying in both $z$ and $t$. 
Note that the Fourier transform of a constant ($a$) is $\mathfrak{F}_{z \to k}\{ a \} = 2 \pi a \delta(k)$, where $\delta$ denotes the Dirac delta function. 
Thus, the electric field of this ideal laser could be represented by $E(x, y, z, t) = \text{real}\big( A(x, y) \exp[i \beta z - i \omega t] \big)$, where the slowly varying amplitude remains constant in $z$ and $t$. 
Perhaps in an ideal optical waveguide, a laser field can propagate through the fiber such that this amplitude can remain independent of any longitudinal and/or temporal variations. 
However, more realistically, optical fields do experience loss, gain (stimulated emission if the fiber is actively doped), thermal effects, and/or nonlinearities intrinsic to the fiber medium. 
Therefore, most optical models allow for the slowly varying amplitude to experience some dependence on its propagation direction and/or on time.  

The Fourier transform identity, $\mathfrak{F}_{t \to \Omega}\{ e^{i a t} \} = 2 \pi \delta(\Omega - a)$, indicates that by decomposing the electric field with a factor of $\exp[i \beta z - i \omega t]$, one is effectively centering the amplitude $A$ in frequency space about $k = 0$ rads/m and $\Omega = 0$ rads/s. 
The expectation is that the spectral content of $A$ is small compared to $\beta$ and $\omega$: 
\[
    \Big\{ k: \ \FT{z}{k}{A} \neq 0 \Big\} \ll \big| \beta \big| \quad \text{ and } \quad \Big\{ \Omega: \ \FT{t}{\Omega}{A} \neq 0 \Big\} \ll \big| \omega \big| \ \ .
\]
The spectral content of a given function is the set of frequencies needed to best represent this function as a linear combination of sinusoidal functions (i.e. as a Fourier series). 
Thus, if the given function has high frequencies within its spectral content, then that function is expected to also contain large derivative values (slopes) and second derivative values (concavities) since the sinusoidal functions from its Fourier series representation would also exhibit these behaviors: 
\begin{align*}
    \dd{\sin(\beta z)}{z} & = \beta \cos(\beta z) , \quad  \ddn{2}{\sin(\beta z)}{z} = -\beta^{2} \sin(\beta z) \quad \text{ and} \\
    \dd{\cos(\omega t)}{t} & = -\omega \sin(\omega t) , \quad  \ddn{2}{\cos(\omega t)}{t} = -\omega^{2} \cos(\omega t) \ \ .
\end{align*}
One can see that  
\begin{align*}
    \Big| \dd{\sin(\beta z)}{z} \Big| & \sim \mathcal{O}\big( \beta \big), \ \Big| \ddn{2}{\sin(\beta z)}{z} \Big| \sim \mathcal{O}\big( \beta^{2} \big) \quad \text{ and} \\
    \Big| \dd{\cos(\omega t)}{t} \Big| & \sim \mathcal{O}\big( \omega \big), \ \Big| \ddn{2}{\cos(\omega t)}{t} \Big| \sim \mathcal{O}\big( \omega^{2} \big) \ \ .
\end{align*}
Inversely, when the function, again think of a term in the Fourier expansion of the slowly varying amplitude ($A$), has only low frequencies within its spectral content, then one would determine that 
\begin{align*}
    \Big| \dd{\sin(k z)}{z} \Big| & \sim \mathcal{O}\big( k \big) \ll \beta, \ \Big| \ddn{2}{\sin(k z)}{z} \Big| \sim \mathcal{O}\big( k^{2} \big) \ll \beta^{2} \quad \text{ and} \\
    \Big| \dd{\cos(\Omega t)}{t} \Big| & \sim \mathcal{O}\big( \Omega \big) \ll \omega, \ \Big| \ddn{2}{\cos(\Omega t)}{t} \Big| \sim \mathcal{O}\big( \Omega^{2} \big) \ll \omega^{2} \ \ ,
\end{align*}
which is the impetus behind the SVEAs~\eqref{eq:SVEAs} since the amplitude is only expected to be comprised of low-frequency spectral content. 

This reasoning has given many physicists and optical engineers the correct intuition that convinced them to drop longitudinal and temporal derivatives from their governing equations to produce simplified optical fiber models. 
Indeed, by dropping the $\p^{2}A/{\p}z^{2}$ and $\p^{2}A/{\p}t^{2}$ terms, one achieves the classic \textit{paraxial approximation}. 
In cases where the amplitude varies over time scales much slower than the traverse time through the fiber ($t_{\text{traverse}} \sim \mathcal{O}(10 \text{ ns})$ for a $L = 10$ m fiber), e.g. for continuous-wave light, often the first temporal derivative is also omitted from the governing relations, citing the temporal SVEA~\eqref{eq:TemporalSVEA} as the justification. 
Indeed, a nondimensional analysis, where one sagaciously chooses characteristic magnitudes for all of the variables and parameters in the governing system, can support the validity of, and reasoning behind, the SVEAs. 
Thus, the optics practitioner tunes the model to the relevant spatial and temporal scales of the given problem (i.e. the physical phenomena) that are being considered. 

Another heuristic, related to the SVEAs, used within the optics community is the idea that \textit{phase-matched} terms in the governing relations capture the dominant physical processes, while \textit{non-phase-matched} terms generally have negligible contributions to the relevant phenomena, and so can be neglected. 
Recall that phase ($\phi$) is mathematically expressed in a factor: $\exp(i \phi)$. 
Similar to the SVEAs, if $\phi$ is changing ``rapidly'', then terms within the governing relations similar to $f(A) \exp(i \phi)$, where $f(A)$ is some function possibly of the amplitude $A$ that is slowly varying in comparison to the rate of changes in $\phi$, can be omitted without much loss to the validity of the model presuming that the ``rapidly'' changing phases do not strongly influence the relevant physics under investigation. 
Note that the high frequencies of $\beta$ and $\omega$ ensured that the phase within $\exp(i \beta z - i \omega t)$, namely $\phi = \beta z - \omega t$, would vary rapidly as the optical field propagates (in $z \in [0, L]$ and $t \in [0, t_{\text{final}}]$). 
One caveat to this phase-matching argument is that all of the phases of the model ought to be explicitly expressed as factors like $\exp(i \phi)$, which may require some manipulations of, and/or additional approximations to, the terms within the given system, and the governing relation must be simplified as much as possible before neglecting any terms. 
Sometimes, during the simplification process, terms with $\exp(i \phi)$ will be multiplied by $\exp(-i \phi)$, thus ``matching'' the phases ($\exp(i \phi) \cdot \exp(-i \phi) = 1$) and leaving behind a ``phase-matched'' term of $f(A) \cdot 1$. 
The non-phase-matched terms are those that have irreducible $\exp(i \phi)$ factors after the simplification process. 
Furthermore, the phase-matching process is also directly related to the conservation of momentum and energy~\cite{toulouse2005optical}. 
Phase-matching arguments are often found when deriving nonlinear phenomena such as self-phase modulation (SPM), cross-phase modulation (XPM), four-wave mixing (FWM), stimulated Brillouin scattering (SBS), and/or stimulated Raman scattering (SRS)~\cite{shen1965theory, headley1996unified, boyd2008nonlinear}. 
Moreover, these arguments have even been used for deriving a reduced transverse mode instability (TMI) model~\cite{menyuk2021accurate}, which helped inspire this effort. 

\begin{figure}% [H]
    \begin{center}
        \includegraphics[width = 0.85\linewidth]{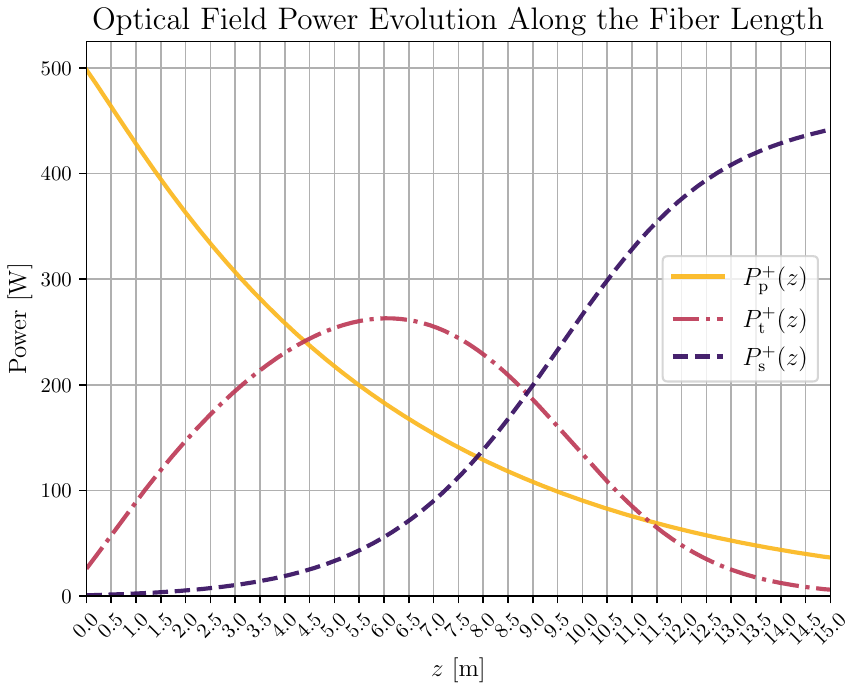}
        \caption{Power evolution along a two-tone configured fiber laser amplifier.}
        \label{fig:TwoTonePowerEvolution}
    \end{center}
\end{figure}
Unfortunately, neither the SVEAs~\eqref{eq:SVEAs} nor the phase-matching arguments are mathematically rigorous, and can be, at times, misleading. 
An example of this is for a well-functioning two-tone (laser gain competition) laser amplifier -- see Figure~\ref{fig:TwoTonePowerEvolution}. 
This fiber amplifier configuration consists of a cladding-pumped incoherent pump field, a core-pumped laser field, which can be called the ``tone field'', and the desired laser signal field such that $\lp < \lt < \ls$, where $\lambda_{\ell}$ is the wavelength of the pump field ($\ell = \pmp$), tone field ($\ell = \text{t}$), and signal field ($\ell = \sig$), respectively. 
The tone is typically seeded more strongly than the signal ($P_{\text{t}}(0) / P_{\sig}(0) > 1$) so that the tone field initially receives the majority of the energy from the incoherent pump field, but, eventually, transfers its energy to the signal such that, by the end of the fiber, most of the output power is in the signal: $P_{\pmp}(L), P_{\text{t}}(L) \ll P_{\sig}(L)$. 
In this scenario, the tone field will achieve a maximum power somewhere along the amplifier's length (near $z = 6$ m in Figure~\ref{fig:TwoTonePowerEvolution}), and will subsequently decrease in energy. 
Since $P \propto |A|^{2}$, at this maximal point, the derivative of the tone amplitude's magnitude would be zero-valued: ${\p}|A_{\text{t}}|/{\p}z = 0$. 
Near this point, the amplitude's phase will also occasionally be zero-valued (or, equivalently, a multiple of $2 \pi$), and so ${\p}A_{\text{t}}/{\p}z \simeq 0$. 
And yet, at the same point, the tone amplitude's concavity would be substantially negative: $\p^{2}A_{\text{t}}/{\p}z^{2} < 0$, which violates the longitudinal SVEA~\eqref{eq:LongitudinalSVEA} since 
\[
    \Big| \ppn{2}{A_{\text{t}}}{z} \Big| > 0 \ \cancel{\ll} \ \Big| \beta \pp{A_{\text{t}}}{z} \Big| \simeq 0 \ \ll \ \Big| \beta^{2} A_{\text{t}} \Big| > 0 \ \ .
\]
This violation has been numerically verified by the authors in an amplifier model that was derived by appealing to the SVEAs. 
So, clearly, one can produce viable fiber amplifier models of the two-tone configuration by implementing the SVEAs nonetheless. 
This article addresses this issue by augmenting the SVEAs and the phase-matching arguments with a known mathematical theorem as the proper citation for why certain terms can be neglected from the Maxwell system. 
 
There is a method for simplifying first-order dynamical systems that contain terms that are periodic with respect to an independent variable such that the average behavior of the solution is well-captured and the error, with respect to the original problem, remains bounded~\cite{sanders2007history}. 
This is accomplished by applying integral averaging to the oscillatory terms on the right-hand side of the given differential equations. 
The main benefit to this approach is that this averaged system can be numerically solved with a smaller number of discrete points, reducing the computational expense compared to the original problem, since Nyquist's sampling theorem~\cite{shannon2006communication} is more readily satisfied now that the frequencies from the periodic terms have been removed. 

We aim to apply this method of averaging to an active gain coupled mode theory (CMT) model for fiber laser amplifiers to provide rigorous mathematical justification to the ideas first presented in~\cite{menyuk2021accurate} and to give further explanation to the well-known SVEAs~\eqref{eq:SVEAs} and phase-matching arguments. 
We will demonstrate, by comparison to a higher-fidelity optical fiber model, that the integral averaging theorem approach produces a trustworthy reduced model that exhibits accurate performance metrics commonly used for fiber laser amplifiers. 
With this novel understanding of the averaging theorem applied to the amplifier problem, along with unique approximations on certain nonlinear quantities, this approach provides a path to producing trustworthy reduced models for a more diverse set of fiber configurations. 

In Section~\ref{sec:avg-thm}, we present the averaging theorem for dynamical systems with an outline of its proof, first detailed by~\cite{sanders2007history}, but also expanded to include differential equations over $\mathbb{C}^{n}$ and involving sums of periodic functions, which will be essential to the fiber amplifier application. 
A manufactured example of this model simplification by averaging, along with conversation on averaging as a frequency filtering technique, is provided. 
We also conduct a study for when averaging fails. 
Section~\ref{sec:CMTModel} outlines the CMT model which will be subject to the method of averaging in Section~\ref{sec:acm}, after some necessary approximations are made. 
Comparison results from the full CMT and accelerated CMT simulations, provided in Section~\ref{sec:comp-results}, support that averaging is an effective technique for model reduction, providing an increase in computational speed. 

\FloatBarrier
%%%%%%%%%%%%%%%%%%%%%%%%%%%%%%%%%%%
% Section: Generalized Averaging Theorem and Example
%%%%%%%%%%%%%%%%%%%%%%%%%%%%%%%%%%%

\section{Generalized Averaging Theorem and Example}\label{sec:avg-thm}

As a rigorous justification for applying the SVEAs and/or phase-matching arguments, we strongly recommend that the given dynamical system be simplified by means of integral averaging over terms with high-frequency periodicities that have negligible impacts on the relevant physics that are being simulated. 
This technique appeals to a theorem first presented by Sanders, Velhurst, and Murdock~\cite{sanders2007history} and concludes that the simplified dynamical system, after integral averaging, has a solution with a bounded error from the original system's solution. 
Presently, we reproduce this key theorem from the work of Sanders et al. with two modifications: (1) we use complex-valued, rather than real-valued, vector spaces and (2) we include sums of periodic functions on the right-hand side (r.h.s.) of our differential system, rather than just one periodic function.

\FloatBarrier
%%%%%%%%%%%%%%%%%%%%%%%%%%%%%%%%%%%
% Subsection: Generalized Periodic Averaging Theorem
%%%%%%%%%%%%%%%%%%%%%%%%%%%%%%%%%%%

\subsection{Generalized Periodic Averaging Theorem}\label{subsec:avg-thm}

Let $ D \subset \mathbb{C}^{n}$ be a connected, bounded, open set with compact closure $\overline{D}$, let $G \defeq  D \times \left[ 0, \infty\right)$, let $\mbf x_{0} \in D$ be constant, and let $S > 0$ and $\varepsilon_{0} > 0$ be constants. 
Consider the initial value problem (IVP) with $0 < \varepsilon \leq \varepsilon_{0}$, and 
\begin{equation}\label{eq:originalIVP}
    \frac{d \mbf x}{dt} = \varepsilon \sum^{N}_{i = 1} \mbf f_{i} \left( \mbf x\left( t \right), t \right), \text{ where } \mbf x \left( 0 \right) = \mbf x_{0} \ \ ,
\end{equation}
and where each $\mbf f_{i} : G \rightarrow \mathbb{C}^{n}$ is a periodic function in $t$ with a known period $\mathcal{T}_{i} > 0$ for $i \in \{1, ..., N \}$ and $N \in \mathbb{N}$, and $\mbf x_{0}$ is the known initial state of the dynamical system. 
The averaged initial value problem is  
\begin{align}\label{eq:averageIVP}
    \frac{d \mbf x}{dt} = \varepsilon \sum^{N}_{i = 1} \mbt f_{i} \left( \mbt x \left( t \right) \right), \text{ where } \mbt x \left( 0 \right) & = \mbf x_{0} \ \ ,
\end{align}
where $\mbt f_{i} \left( \mbt x \left( t \right) \right) \defeq \int^{\mathcal{T}_{i}}_{0} \mbf f_{i} \left( \mbt x \left( s \right), s \right) ds / \mathcal{T}_{i}$ is the integral average of each function $\mbf f_{i}$ over its own period. 
The following theorem gives a bound on the solutions to \eqref{eq:originalIVP} and \eqref{eq:averageIVP}. 
\begin{theorem}\label{thm:avg-Verl}
    Suppose, for $i \in \{1, ..., N \}$, that $\mbf f_{i}$ is Lipschitz continuous and
    that the following two criteria hold:
    \begin{criterion}\label{asmp:expl_period}
        Each $\mbf f_{i}$ has an explicit periodic dependence with associated period $\mathcal{T}_{i}$.
    \end{criterion}
    \begin{criterion}\label{asmp:small_param}
        There exists some small parameter $\varepsilon$ such that $\mbf x$, which solves~\eqref{eq:originalIVP}, remains in $D$ for all $0 \leq t \leq S / \varepsilon$.
    \end{criterion}
    Then there exists a constant $c > 0$, independent of $\varepsilon$, such that
    \begin{align*}
        \| \mbf x \left( t \right) - \mbt x \left( t \right) \| \leq c \varepsilon
    \end{align*}
    for $0 \leq \varepsilon \leq \varepsilon_{0}$ and $0 \leq t \leq \frac{S}{\varepsilon}$.
\end{theorem}
\noindent The value of the constant $c$ depends on the Lipschitz continuity properties of $\mbf f$, and on the boundedness of the parameters of the original ODE system. 
Specifically, 
\[
    c = \left[ \sum^{N}_{i = 1} \left(  2 S \lambda_{\mbf f_{i}} c_{\mbf x} + \mathcal{T}_{\mbf f_{i}} c_{\mbf f_{i} - \mbt f_{i}} \right) \right] e^{S \sum^{N}_{i = 1} \lambda_{\mbf f_{i}}} \ \ ,
\]
where $S$ represents the maximum value such that the solutions to the original ($\mbf x$) and averaged ($\mbt x$) IVPs remain in $D$ for $0 \leq t \leq S / \varepsilon$. Additionally, $c_{\mbf x}$ is the constant associated with slowly-varying $\mbf x$ such that $\|\mbf x \left( t_{1} \right) - \mbf x \left( t_{2} \right) \| \leq \varepsilon c_{\mbf x}$ for all $t_{1}, t_{2} \in \left[ 0, S / \varepsilon \right]$.
The constant $c_{\mbf f_{i} - \mbt f_{i}}$ is 
\[
    c_{\mbf f_{i} - \mbt f_{i}} \defeq \underset{\mbf x \in \mbf{D}, \ s \in \left[ 0, \frac{S}{\varepsilon} \right]}{\max} \left\Vert \left( \mbf f_{i} - \mbt f_{i} \right) \left( \mbf x \left(s \right) , s \right) \right\Vert \ \ .
\]
Finally, $\lambda_{\mbf f_{i}}$ is the Lipschitz constant associated with $\mbf f_{i}$. 
This theorem effectively says that the error between the solutions to the original dynamical system~\eqref{eq:originalIVP} and its integral averaged counterpart~\eqref{eq:averageIVP} is $\mathcal{O} (\varepsilon)$ over a certain span of time (span of the independent variable). 
If this error is acceptable for a given application, then the integral averaged dynamical system may be a viable reduced model surrogate to the original system. 

The following steps are a basic outline to the proof of this generalized averaging theorem: \vspace*{-2.5mm}
\begin{proofw}{\em{Synopsis of the Proof}:}
    \begin{enumerate}
        \item Define the error function: $\mbf E \left( t \right) \defeq \mbf x \left( t \right) - \mbt x \left( t \right)$. Rewrite $\mbf E$ as an integral equation and substitute the r.h.s. of the original and averaged ODEs to get $\mbf E \left( t \right) = \int^{t}_{0} \varepsilon \sum^{N}_{i = 1} \mbf f_{i} \left( \mbf x\left( s \right) , s\right) - \varepsilon \sum^{N}_{i = 1} \mbt f_{i} \left( \mbt x \left( s \right) \right)$. 
        \item Add zero in the form of $\varepsilon \sum^{N}_{i = 1} \mbf f_{i} \left( \mbt x \left( s \right), s\right) - \varepsilon \sum^{N}_{i = 1} \mbf f_{i} \left( \mbt x \left( s \right), s\right) ds$. 
        \item Take the norm of both sides, apply the \textit{triangle inequality}, and apply \textit{Fubini's theorem}: $\| \mbf E \left( t \right) \| \leq \varepsilon \sum^{N}_{i = 1} \left( A_{i} + B_{i} \right)$, where $A_{i} \defeq \left\Vert \int^{t}_{0} \mbf f_{i} \left( \mbf x \left( s \right), s \right) - \mbf f_{i} \left(\mbt x \left( s \right), s \right) ds \right\Vert$ and $B_{i} \defeq \left\Vert \int^{t}_{0} \mbf f_{i} \left(\mbt x \left( s \right), s \right) -  \mbt f_{i} \left( \mbt x \left( s \right) \right) ds  \right\Vert$. 
        \item Bound each $A_{i}$ using Lipschitz continuity: $A_{i} \leq \lambda_{\mbf f_{i}} \int^{t}_{0} \left\Vert \mbf E \left( s \right) \right\Vert ds$. 
        \item Bound each $B_{i}$ using the result that $\mbf f_{i} - \mbt f_{i}$ is bounded for each $i$: $B_{i} \leq  2 S \lambda_{\mbf f_{i}} c_{\mbf x} + \mathcal{T}_{\mbf f_{i}} c_{\mbf f_{i} - \mbt f_{i}}$. 
        \item Apply a specific case of the \textit{Gr{\"o}nwall's inequality}~\cite[p.~5]{sanders2007history}: $\left\Vert \mbf E \left( t \right) \right\Vert \leq c \varepsilon$, where 
        \[
            c = \left[ \sum^{N}_{i = 1} \left( 2 S \lambda_{\mbf f_{i}} c_{\mbf x} + \mathcal{T}_{\mbf f_{i}} c_{\mbf f_{i} - \mbt f_{i}} \right) \right] e^{S \sum^{N}_{i = 1} \lambda_{\mbf f_{i}}} \ \ .
        \]
    \end{enumerate}
\end{proofw}

\FloatBarrier
%%%%%%%%%%%%%%%%%%%%%%%%%%%%%%%%%%%
% Subsection: A Complex-Valued Vectorial Averaging Example
%%%%%%%%%%%%%%%%%%%%%%%%%%%%%%%%%%%

\subsection{A Complex-Valued Vectorial Averaging Example}\label{subsec:avg-ex}
 
To exemplify this integral averaging technique, consider a complex-valued vector ODE system with multiple periodic terms such that the dependent variable $\mathbf{z} = \big[ z_{1} , z_{2} \big]^{\text{T}}$ evolves over time ($t$) according to     
\begin{equation}\label{vecODE}
    \dd{\mathbf{z}}{t} = \varepsilon 
        \begin{bmatrix}
            - z_{2} z_{1} - 3 z_{1} + \frac{7}{3} \sin \left( \pi t \right) -  \frac{3i}{5} \cos \left( \frac{1}{8} t \right) + \frac{9}{5} e^{\frac{5i}{6}t} + 1 \\
            \frac{i}{2} z_{1} + \frac{1}{8z_{2}} + \frac{i}{7} \sin \left( \frac{\pi}{10} t \right) - \frac{1}{65} z^{2}_{2}e^{\frac{\pi i }{5}t}
        \end{bmatrix} \ , \ 
    \mathbf{z} \left( 0 \right) = 
        \begin{bmatrix}
            2 - 3i \\
            -2 - i
        \end{bmatrix} \ , 
\end{equation}
and $\varepsilon = 0.15$. 
This dynamical system~\eqref{vecODE} can be re-written as 
\begin{equation*}
    \frac{d \mathbf{x}}{dt} = \varepsilon \left( \mathbf{f}_{1} + \mathbf{f}_{2} + \mathbf{f}_{3} + \mathbf{f}_{4} + \mathbf{f}_{5} \right) \ , \  
    \mathbf{x} \left( 0 \right) = 
        \begin{bmatrix}
            2 - 3i \\
            -2 - i
        \end{bmatrix} \ , 
\end{equation*}
where $\mbf f_{1}$ through $\mbf f_{5}$ are given in terms of an $\mbf f_{0}$ by
% \vspace*{-3.5mm}
\begin{align*}
    \mathbf{f}_{0} & \defeq 
        \begin{bmatrix}
            - z_{2} z_{1} - 3 z_{1} + 1 \\
            \frac{i}{2} z_{1}  + \frac{1}{8 z_{2}}
        \end{bmatrix} \ , &   
        \mathbf{f}_{1} & \defeq \frac{1}{5} \mathbf{f}_{0} + 
            \begin{bmatrix}
                \frac{7}{3} \sin \left( \pi t \right) \\
                0
            \end{bmatrix} \ , &  
            \mathbf{f}_{2} & \defeq \frac{1}{5} \mathbf{f}_{0} + 
                \begin{bmatrix}
                    - \frac{3i}{5} \cos \left( \frac{1}{8} t \right) \\
                    0
                \end{bmatrix} \ , \\
    \mathbf{f}_{3} & \defeq \frac{1}{5} \mathbf{f}_{0} + 
        \begin{bmatrix}
            \frac{9}{5} e^{ \frac{5i}{6} t } \\
            0
        \end{bmatrix} \ , &  
        \mathbf{f}_{4} & \defeq \frac{1}{5} \mathbf{f}_{0} + 
            \begin{bmatrix}
                0 \\
                \frac{i}{7} \sin\left( \frac{\pi}{10} t \right)
            \end{bmatrix}, \text{ and} &   
            \mathbf{f}_{5} & \defeq \frac{1}{5} \mathbf{f}_{0} + 
                \begin{bmatrix}
                    0 \\
                    \frac{1}{65} z^{2}_{2} e^{\frac{\pi i }{5}t}
                \end{bmatrix} \ .
\end{align*}
Applying Theorem~\ref{thm:avg-Verl} yields the corresponding averaged dynamical system, where $\tilde{\mathbf{z}} = \big[ \tilde{z}_{1},\tilde{z}_{2} \big]^{\text{T}}$ evolves according to  
\begin{equation*}
    \dd{\tilde{\mathbf{z}}}{t} = \varepsilon \left( \tilde{\mathbf{f}}_{1} + \tilde{\mathbf{f}}_{2} + \tilde{\mathbf{f}}_{3} + \tilde{\mathbf{f}}_{4} + \tilde{\mathbf{f}}_{5} \right) = \varepsilon \hspace{1pt} \mathbf{f}_{0}(\tilde{\mathbf{z}}) \ , \
    \tilde{\mathbf{z}}\left( 0 \right) = 
        \begin{bmatrix}
            2 - 3i \\
            -2 - i
        \end{bmatrix} \ , \text{ where}
\end{equation*}
\vspace*{-3.5mm}
\begin{align*}
    \tilde{\mathbf{f}}_{1} & \defeq \frac{1}{\mathcal{T}_{1}} \int^{\mathcal{T}_{1}}_{0} \mathbf{f}_{1} ds = \frac{1}{2} \int^{2}_{0} \frac{1}{5} \mathbf{f}_{0} + 
        \begin{bmatrix}
            \frac{7}{3} \sin \left( \pi t \right) \\
            0
        \end{bmatrix} ds = \frac{1}{5} \mathbf{f}_{0}(\tilde{\mathbf{z}}) \ , \\
    \tilde{\mathbf{f}}_{2} & \defeq \frac{1}{\mathcal{T}_{2}} \int^{\mathcal{T}_{2}}_{0} \mathbf{f}_{2} ds = \frac{1}{16 \pi} \int^{16 \pi}_{0} \frac{1}{5} \mathbf{f}_{0} + 
        \begin{bmatrix}
            -\frac{3i}{5} \cos\left( \frac{1}{8} t \right) \\
            0
        \end{bmatrix} ds = \frac{1}{5} \mathbf{f}_{0}(\tilde{\mathbf{z}}) \ , \\
   \tilde{\mathbf{f}}_{3} & \defeq \frac{1}{\mathcal{T}_{3}} \int^{\mathcal{T}_{3}}_{0} \mathbf{f}_{3} ds = \frac{5}{12 \pi} \int^{ \frac{12 \pi}{5}}_{0} \frac{1}{5} \mathbf{f}_{0} + 
        \begin{bmatrix}
            \frac{9}{5} e^{\frac{5i}{6}t} \\
            0
        \end{bmatrix} ds = \frac{1}{5} \mathbf{f}_{0}(\tilde{\mathbf{z}}) \ , \\
    \tilde{\mathbf{f}}_{4} & \defeq \frac{1}{\mathcal{T}_{4}} \int^{\mathcal{T}_{4}}_{0} \mathbf{f}_{4} ds = \frac{1}{20} \int^{ 20}_{0} \frac{1}{5} \mathbf{f}_{0} + 
        \begin{bmatrix}
            0 \\
            \frac{i}{7} \sin \left( \frac{\pi}{10} t \right)
        \end{bmatrix} ds = \frac{1}{5} \mathbf{f}_{0}(\tilde{\mathbf{z}}) \ , \text{ and} \\
    \tilde{\mathbf{f}}_{5} & \defeq \frac{1}{\mathcal{T}_{5}} \int^{\mathcal{T}_{5}}_{0} \mathbf{f}_{5} ds = \frac{1}{10} \int^{ 10}_{0} \frac{1}{5} \mathbf{f}_{0} + 
        \begin{bmatrix}
            0 \\
            \frac{1}{65} z^{2}_{2} e^{\frac{\pi i}{5}t}
        \end{bmatrix} ds = \frac{1}{5} \mathbf{f}_{0}(\tilde{\mathbf{z}}) \ .
\end{align*}
Therefore, $\tilde{\mathbf{f}}(\tilde{\mathbf{z}}) = \tilde{\mathbf{f}}_{0}(\tilde{\mathbf{z}})$, and so 
\begin{equation}\label{AvgVecODE}
    \frac{d \tilde{\mathbf{z}}}{dt} = \varepsilon 
        \begin{bmatrix}
            -\tilde{z}_{2} \tilde{z}_{1} - 3 \tilde{z}_{1} + 1 \\
            \frac{i}{2} \tilde{z}_{1}  + \frac{1}{8 \tilde{z}_{2}}
        \end{bmatrix} \ , \text{ where } \ \
    \tilde{\mathbf{z}}(0) = 
        \begin{bmatrix}
            2 - 3i \\
            -2 -i
        \end{bmatrix} \ .
\end{equation}
\begin{figure}%[H]
    \centering
    \includegraphics[width=0.5\linewidth]{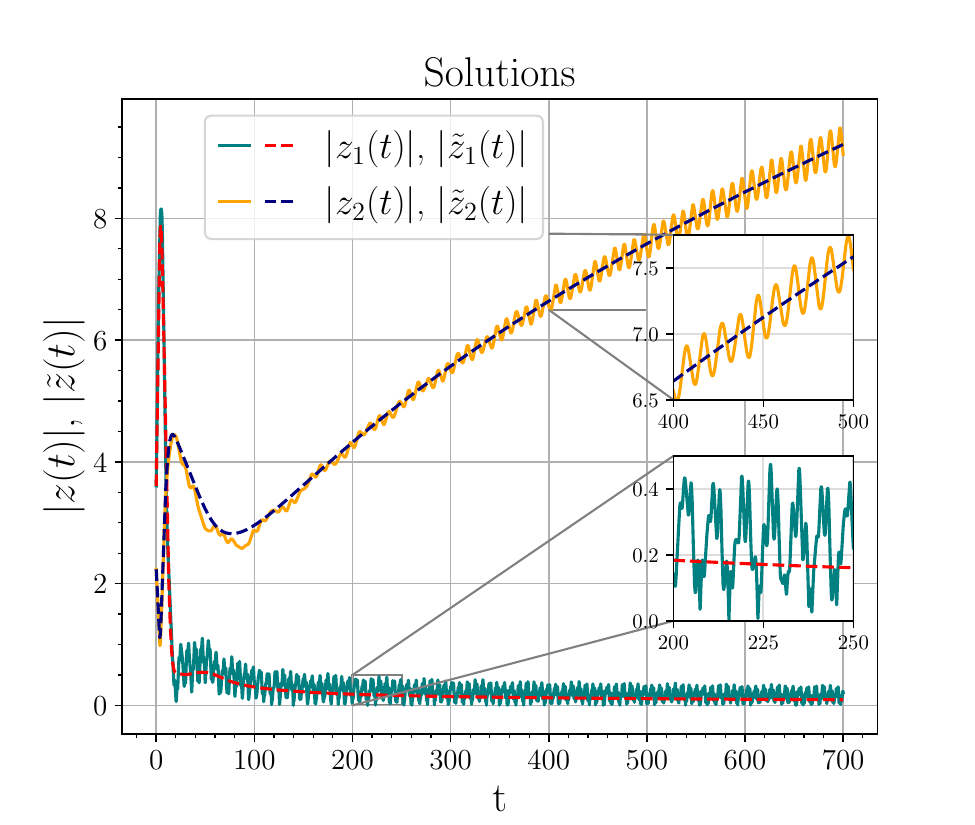}
    \caption{Magnitude of the solutions to the original~\eqref{vecODE} and averaged~\eqref{AvgVecODE} ODEs.}
    \label{fig:vecMag}
\end{figure}
\FloatBarrier

\begin{figure}%[H]
    \centering
    \begin{subfigure}[b!]{0.49\linewidth}
        \centering
        \includegraphics[width=\linewidth]{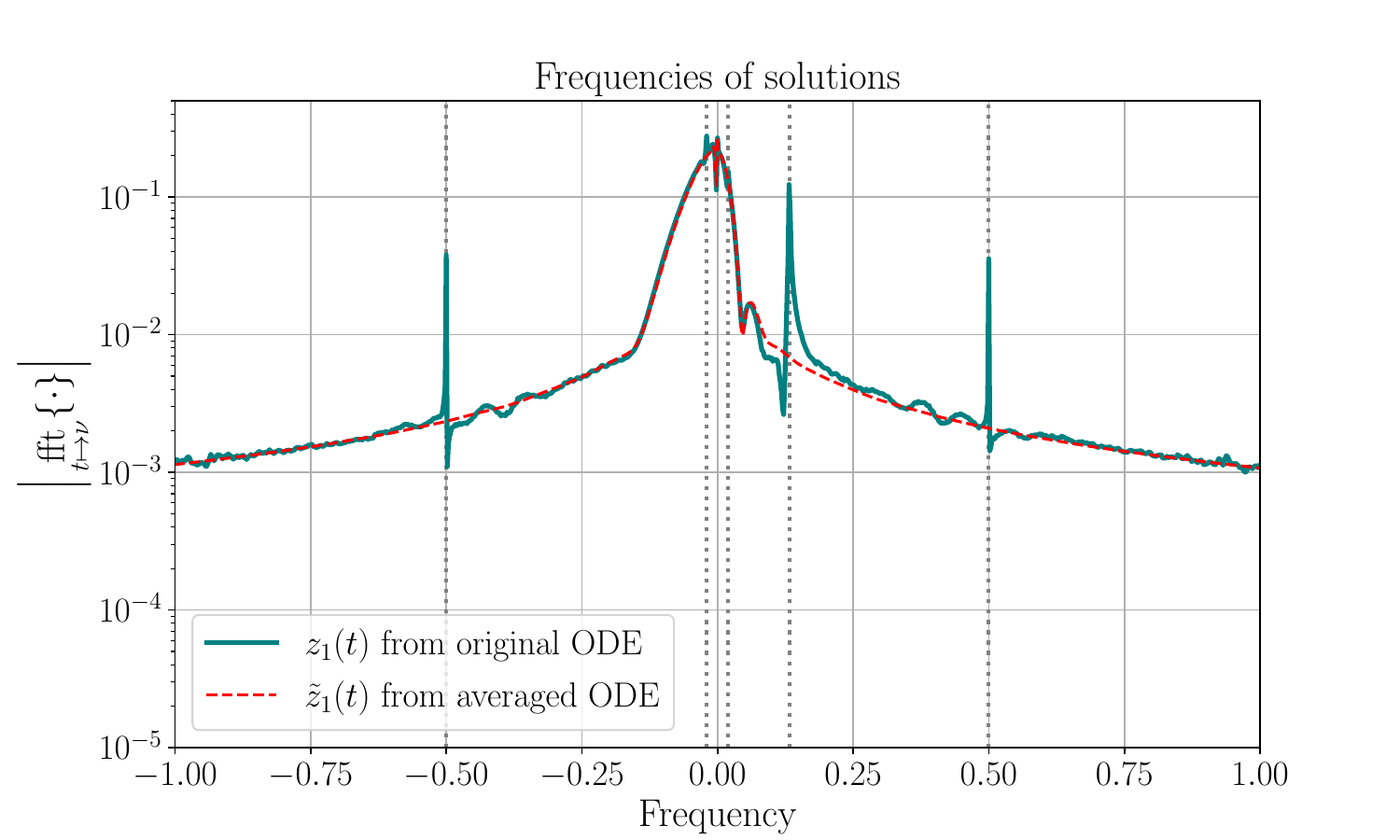}
        \caption{\hspace*{1pt}$\left| \hspace{3pt} \fft{t}{\nu}{z_{1}(t)} \right|$ and \hspace*{1pt}$\left| \hspace{3pt} \fft{t}{\nu}{\tilde{z}_{1}(t)} \right|$ \ .}
    \end{subfigure}
    \hfill
    \begin{subfigure}[b!]{0.49\linewidth}
        \centering
        \includegraphics[width=\linewidth]{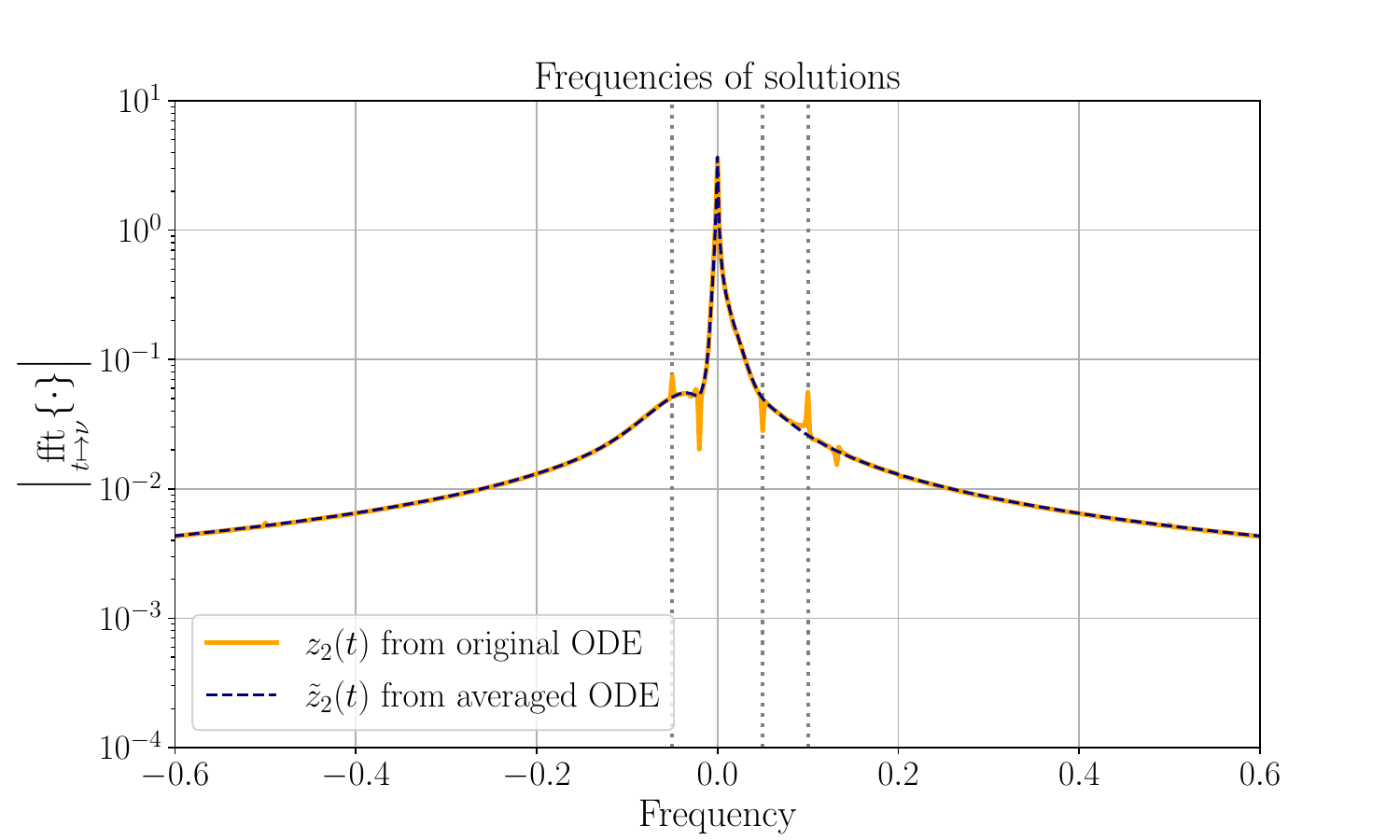}
        \caption{\hspace*{1pt}$\left| \hspace{3pt} \fft{t}{\nu}{z_{2}(t)} \right|$ and \hspace*{1pt}$\left| \hspace{3pt} \fft{t}{\nu}{\tilde{z}_{2}(t)} \right|$ \ .}
    \end{subfigure}
    \caption{Frequencies of the solutions to the original~\eqref{vecODE} and averaged~\eqref{AvgVecODE} ODEs.}
    \label{fig:AvgEx3freq}
\end{figure}

In Figure~\ref{fig:vecMag}, the solution to~\eqref{AvgVecODE} has the same qualitative behavior as the solution to~\eqref{vecODE} without any of the rapid oscillations. 
In the frequency space of the solutions, which are depicted in Fig.~\ref{fig:AvgEx3freq}, the original solution has peaks at higher frequencies that correspond to the periods present in the original ODE system. 
Any additional frequency peaks in the original solution are beat frequencies that are caused by interference between two periodic terms from the r.h.s. of the original ODE. 
All of these higher frequencies have been filtered for the averaged solution. 

\FloatBarrier

\begin{figure}% [H]
    \centering
    \begin{subfigure}[b!]{.45\textwidth}
        \centering
        \includegraphics[width=\linewidth]{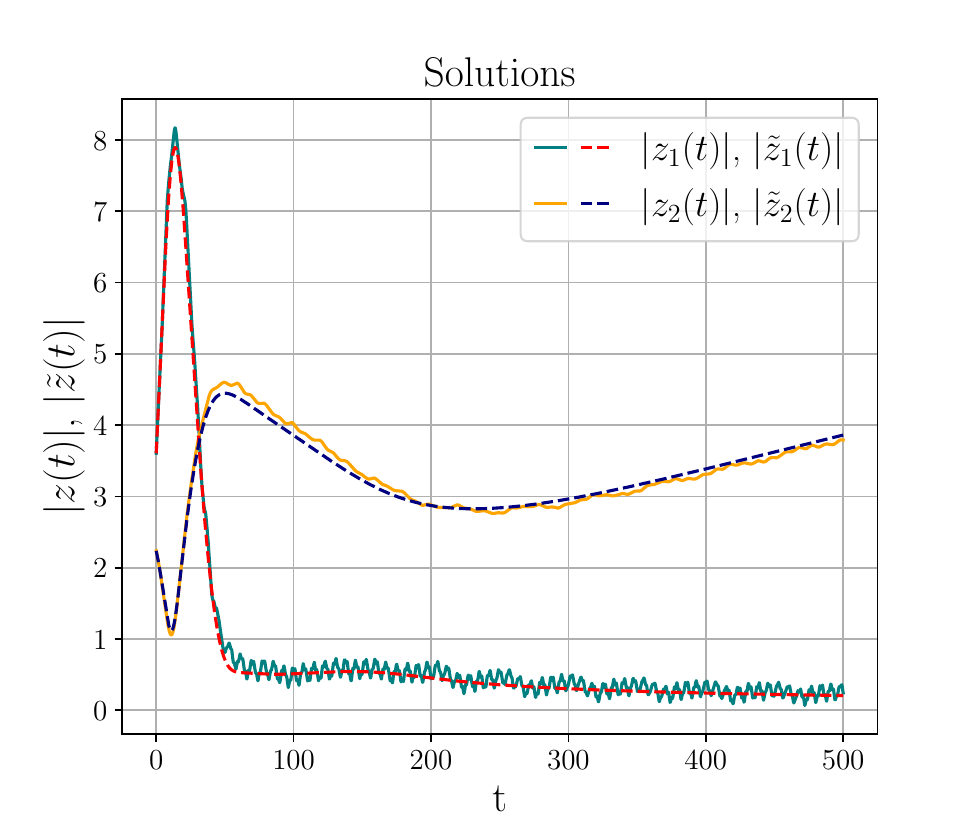}  
        \caption{$\varepsilon = 0.05$.}
        \label{fig:sub-second}
    \end{subfigure}
    \hfill
    \begin{subfigure}[b!]{.45\textwidth}
        \centering
        \includegraphics[width=\linewidth]{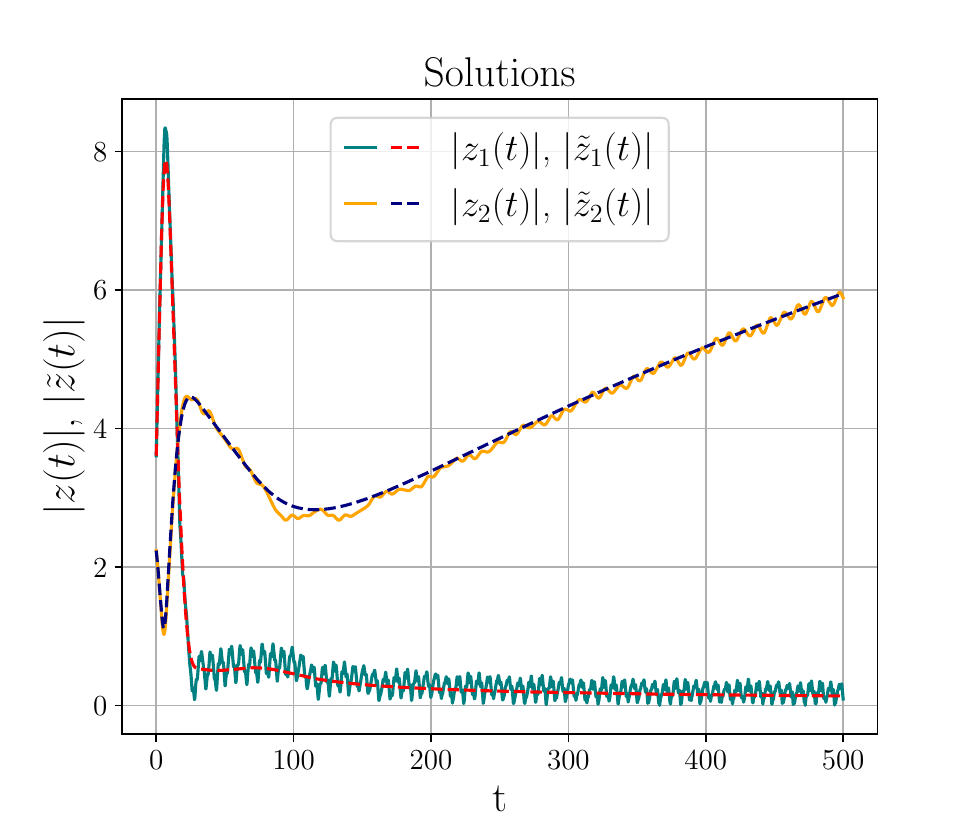}  
        \caption{$\varepsilon = 0.1$.}
        \label{fig:sub-fourth}
    \end{subfigure}
    \mbox{} \\
    \begin{subfigure}{.45\textwidth}
        \centering
        \includegraphics[width=\linewidth]{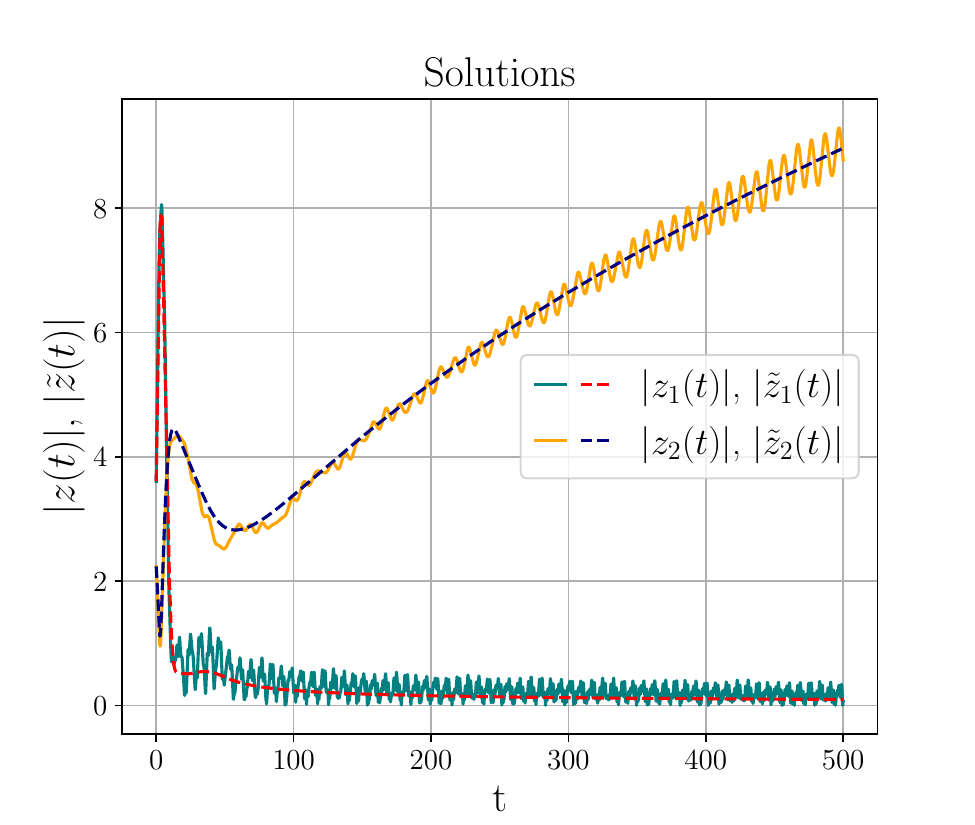}  
        \caption{$\varepsilon = 0.2$.}
        \label{fig:sub-fifth}
    \end{subfigure}
    \caption{Solutions to the original~\eqref{vecODE} and averaged~\eqref{AvgVecODE} ODEs for various values of $\varepsilon$.}
    \label{fig:changing_eps}
\end{figure}

The span over which averaging is an effective technique for approximating the original periodic ODE system is another point of interest since the averaging theorem only guarantees a bounded error between solutions over a problem-dependent subset of time. %: $0 \leq t \leq S/ \varepsilon$. 
Increasing the value of $\varepsilon$ increases the amplitude of the periodic terms in the original ODE, which then increases the error between the original and averaged solutions, as can be seen in Figure~\ref{fig:changing_eps}. 
To understand how the choice of $\varepsilon$ impacts the accuracy of the averaged system, we compute the maximum relative error between the solutions of the original~\eqref{vecODE} and averaged~\eqref{AvgVecODE} ODEs over the last maximum period in the original ODE system: 
\begin{align}\label{eq:z_error}
    \text{error}_{\mbf z, \mbt z} (\varepsilon)  = \underset{t \in \left[ t_{\text{final}} - \max_{i}(\mathcal{T}_{i}), t_{\text{final}} \right]}{\max}  \frac{\| \mbf z \left( t, \varepsilon \right) - \mbt {z} \left( t, \varepsilon \right) \|}{\|\mbf z \left( t, \varepsilon \right) \|} \ \ .
\end{align}
For our example dynamical system~\eqref{vecODE}, the longest period is $\max_{i}(\mathcal{T}_{i}) \equiv \mathcal{T}_{2} = 16 \pi$. 
\FloatBarrier

\begin{figure}%
    \centering
    \includegraphics[width=0.5\linewidth]{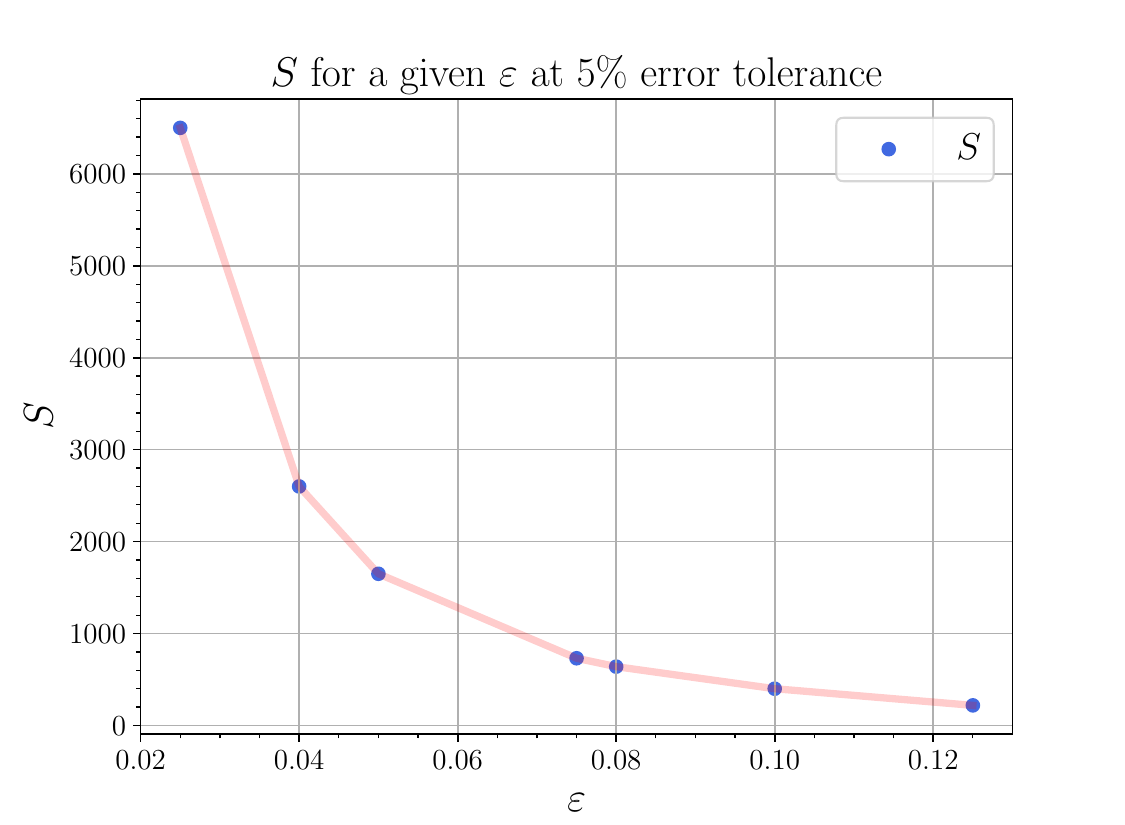}
    \caption{The maximum span of time ($S$) for which the relative error~\eqref{eq:z_error} remains below $5\%$ for various values of $\varepsilon$.}
    \label{fig:veps_vs_S}
\end{figure}

In order to determine a span of time ($S / \varepsilon$) over which the averaged solution remains an acceptably accurate representation of the original solution, the value of $\varepsilon$ is varied and the two solutions are propagated out to the time when the relative error~\eqref{eq:z_error} is calculated to exceed a user-chosen error tolerance. 
In this case, we will choose this tolerance to be $5\%$. 
Again, solving our governing systems~\eqref{vecODE} and~\eqref{AvgVecODE}, Figure~\ref{fig:veps_vs_S} conveys a strong inverse relationship between the value $\varepsilon$ and the maximum time over which the averaged solution maintains a suitably low error. 
\FloatBarrier

%%%%%%%%%%%%%%%%%%%%%%%%%%%%%%%%%%%
% Section: Coupled Mode Theory (CMT) Model
%%%%%%%%%%%%%%%%%%%%%%%%%%%%%%%%%%%

\section{Coupled Mode Theory (CMT) Model}
\label{sec:CMTModel}

Since this effort investigates propagating optical fields in a fiber waveguide, our model leverages the standard \textit{coupled mode theory} (CMT) modeling approach~\cite{snyder1972coupled, ward2012origin, NaderDajanMadde13, goswami2021simulations}, which is derived from Maxwell's equations. The principle idea behind CMT is that the electric field is projected onto a finite set of transverse guided orthonormal modes, presuming that this modal basis well-represents the original field. 
The interaction of these modes leads to periodic mode beatings over millimeter-length distances, which contribute very little to the physics of the stimulated emission (i.e. active gain) process, indicating that they would negligibly affect the power in the optical fields. 
We start from a scalar CMT model for a multi-moded fiber (i.e. large mode area) laser amplifier. 

Consider a straight, weakly-guiding ($(\nco - \ncl)/\nco \ll 1$), step-index, double-clad, ytterbium (Yb)-doped optical fiber laser amplifier that guides light down the length of the fiber in the +$z$-direction. 
Let $\ell \in \{\pmp, \sig\}$ denote the pump ($\pmp$) and the laser signal ($\sig$) fields, respectively. The derivation of the CMT model requires a complexification of the electric field in which the following time-harmonic convention is adopted:
\[
    \mathbf{E}_{\ell}(x, y, z, t)  = \text{Real}\left[ \bs{\calE}_{\ell}(x, y, z) e^{- i \omega_{\ell} t} \right] \ \ \left[ \frac{\text{V}}{\text{m}} \right] .
\]
Furthermore, the electric field is presumed to be linearly polarized and to be comprised of a finite set of orthonormal modes such that 
\[
    \bs{\calE}_{\ell} \left( x,y,z \right)  = \vec{e}_{x} \sum^{M_{\ell}}_{m=1} A^{m}_{\ell} \left( z \right) \varphi^{m}_{\ell} \left(x,y \right) e^{i \beta^{m}_{\ell} z} \ \ ,
\]
where $\vec{e}_{x}$ is the unit vector in the direction of the linear polarization state. 
The scalar CMT model for a lossless co-pumped fiber amplifier is expressed as an ODE system for the $z$-dependent mode amplitudes $A_{\ell}^{j}$ [V], where $j \in \{ 1, ..., M_{\ell} \}$~\cite{snyder1972coupled, ward2012origin, NaderDajanMadde13, goswami2021simulations}. 
These ODEs depend on the transverse mode profiles $\varphi^{j}_{\ell} = \varphi^{j}_{\ell}(x, y)$ [1/m] and their corresponding propagation constants $\beta^{j}_{\ell} \in \mathbb{R}$ [rads/m]: 
\begin{subequations}\label{eq:GeneralCMTModel}
    \begin{equation}\label{eq:CMTODEs}
        \frac{\partial A_{\ell}^{j}}{\partial z}(z) = \sum^{M_{\ell}}_{m = 1} \kappa_{\ell}^{j, m}(z) A_{\ell}^{m}(z) e^{i \left( \beta^{m}_{\ell} - \beta^{j}_{\ell} \right) z} \ \ \left[ \frac{\text{V}}{\text{m}} \right] ,
    \end{equation}
    where the coupling coefficient $\kappa_{\ell}^{j, m}$ for a fixed mode $j$ is 
    \begin{equation}\label{eq:CMTCouplingCoeffs}
        \kappa_{\ell}^{j, m}(z) \defeq \frac{\beta^{m}_{\ell} }{ 2 \beta^{j}_{\ell} } \left( \bar{g}_{\ell}(x, y, z) \varphi^{m}_{\ell}(x, y) , \varphi^{j}_{\ell}(x, y) \right)_{\Omega_{xy}} \ \ \left[ \frac{1}{\text{m}} \right] ,
    \end{equation}
\end{subequations}
where $\bar{g}_{\ell} = \bar{g}_{\ell}(x, y, z)$ is the steady-state active gain function [1/m], which couples all of the propagating mode amplitudes together, $\left( \cdot , \cdot \right)_{\Omega_{xy}}$ is the $L_2$-inner product, i.e., $(u, v)_{\Omega_{xy}} = \int_{\Omega_{xy}} u v^{*} dx dy$ for functions $u$ and $v$ within the fiber's transverse cross-section, and $[\cdot]^{*}$ is the complex conjugate operator.  
Note that the mode data $\big( \varphi^{j}_{\ell}, \beta^{j}_{\ell} \big) \ \forall \ j, \ell$ can always be pre-computed prior to the propagation of the optical field amplitudes, and the mode profiles are orthonormal in the sense that $\big( \varphi^{j}_{\ell} , \varphi^{m}_{\ell} \big)_{\Omega_{xy}} = \delta_{jm}$, where $\delta_{jm}$ is the Kronecker delta function. 

Furthermore, presuming a cladding-pumped configuration, the pump light is reasonably modeled as a planewave uniformly spread throughout the fiber's core and inner cladding regions.
Thus, this planewave can be seen as a single mode ($M_{\pmp} = 1$) of the pump field with a normalized profile $\varphi_{\pmp} = 1 / (\sqrt{\pi} \rcl)$ and a corresponding propagation constant $\beta_{\pmp} \cong \omega_{\pmp} \ncl / c$, where the scripting of the mode index is omitted for notational simplicity. 
As a result of the mode profile having no transverse dependencies, the pump field's governing relation reduces to 
\begin{align}
\label{generic_pump_planewave_ODE}
    \frac{\partial A_{\pmp}}{\partial z} & = \kappa_{\pmp}  A_{\pmp} = A_{\pmp} \underbrace{\frac{ \ncl }{ 2 \pi \rcl^{2} } \int_{\Omega_{xy}} \hspace*{-4mm} \bar{g}_{\pmp} dxdy}_{\kappa_{\pmp}} \ . \ %\left[ \frac{\text{V}}{\text{m}} \right] .
\end{align}
Additionally, without loss of generality, we will consider a fiber design that supports only two guided core modes ($M_{\sig} = 2$), namely the fundamental mode (FM, $j = 1$) and a higher-order mode (HOM, $j = 2$). 
Letting ${\Delta}\beta_{\sig} \defeq \beta_{\sig}^{\FM} - \beta_{\sig}^{\HOM}$, the governing relation for the two modes of the signal field can be expressed as 
\begin{equation}\label{eq:SignalModeODEs}
    \begin{aligned}
        \frac{\partial A_{\sig}^{\FM}}{\partial z}     & =    A_{\sig}^{\FM} \underbrace{ \frac{1}{2} \int_{\Omega_{xy}} \hspace*{-4mm} \bar{g}_{\sig} \big| \varphi_{\sig}^{\FM} \big|^{2} dxdy }_{ \kappa_{\sig}^{\FM , \FM} } + A_{\sig}^{\HOM} \underbrace{ \frac{ \beta_{\sig}^{\HOM}  }{ 2 \beta_{\sig}^{\FM} } \int_{\Omega_{xy}} \hspace*{-2mm} \bar{g}_{\sig} \big( \varphi_{\sig}^{\FM} \big)^{*} \varphi_{\sig}^{\HOM} dxdy }_{ \kappa_{\sig}^{\FM , \HOM}  } e^{-i {\Delta}\beta_{\sig} z}  \hspace{5pt} \text{ and} \\
        \frac{\partial A_{\sig}^{\HOM}}{\partial z}   &  =    A_{\sig}^{\FM} \underbrace{ \frac{ \beta_{\sig}^{\FM} }{ 2 \beta_{\sig}^{\HOM} } \int_{\Omega_{xy}} \hspace*{-2mm} \bar{g}_{\sig} \varphi_{\sig}^{\FM} \big( \varphi_{\sig}^{\HOM} \big)^{*} dxdy  }_{ \kappa_{\sig}^{\HOM , \FM} } e^{i {\Delta}\beta_{\sig} z} + A_{\sig}^{\HOM} \underbrace{ \frac{1}{2} \int_{\Omega_{xy}} \hspace*{-4mm} \bar{g}_{\sig} \big| \varphi_{\sig}^{\HOM} \big|^{2} dxdy }_{ \kappa_{\sig}^{\HOM , \HOM} } \ .
    \end{aligned}
\end{equation}
Note that typically $\beta_{\sig}^{\FM} / \beta_{\sig}^{\HOM} \cong 1 \cong \beta_{\sig}^{\HOM} / \beta_{\sig}^{\FM}$; though we will not enforce this approximation. 
Therefore, the CMT model, relations~\eqref{generic_pump_planewave_ODE}--\eqref{eq:SignalModeODEs}, with initial conditions, can be written more succinctly, where the dependent variables are set into a 3-component vector form: ${\mbf A}(z) = [A_{\pmp}(z), A_{\sig}^{\FM}(z), A_{\sig}^{\HOM}(z)]^{\text{T}}$, as 
\begin{subequations}\label{eq:cmt-3x3}    
    \begin{equation}\label{eq:cmt-3x3-ODE}
            \frac{d{\mbf A}}{dz} = \mathbf{F} \left( z, {\mbf A} \right) \ \ , \quad {\mbf A}\left( 0 \right) = {\mbf A}_{0}  \ \ ,  
    \end{equation}
    where 
    \begin{equation}
        \mathbf{F}\left( z, {\mbf A} \right) =
            \bcalK(z, {\mbf A}) \  {\mbf A}(z)  \ \ ,  \label{eq:cmt-3x3-F}
    \end{equation}
    and 
    \begin{equation}
        \bcalK(z, {\mbf A})  = \begin{bmatrix}
                \kappa_{\pmp}(z) & 0 & 0 \\
                0 & \kappa_{\sig}^{\FM , \FM}(z) & \kappa_{\sig}^{\FM , \HOM}(z) e^{-i{\Delta}\beta_{\sig} z}  \\
                0 & \kappa_{\sig}^{\HOM , \FM}(z) e^{i {\Delta}\beta_{\sig} z} & \kappa_{\sig}^{\HOM, \HOM}(z)
            \end{bmatrix}.  \label{eq:cmt-3x3-K}
    \end{equation}
\end{subequations}

A key parameter in both theoretical modeling and physical experiment is irradiance. For the scalar CMT model, the irradiance of an optical field is 
\begin{equation*}\label{eq:Irradiance}
    I_{\ell}(x, y, z) = \frac{ 1 }{2 \mu_{0} \omega_{\ell} } \sum_{j = 1}^{M_{\ell}} \sum_{m = 1}^{M_{\ell}} \sqrt{ \beta_{\ell}^{j} \beta_{\ell}^{m} } \big( A_{\ell}^{j} \big)^{*} A_{\ell}^{m} \big( \varphi_{\ell}^{j} \big)^{*} \varphi_{\ell}^{m} e^{ i \left( \beta_{\ell}^{m} - \beta_{\ell}^{j} \right) z } \ \ \left[ \frac{\text{W}}{\text{m}^{2}} \right] .
\end{equation*}
For the planewave pump field, this irradiance reduces to 
\begin{equation*}\label{eq:PumpIrradiance}
    I_{\pmp}(z) = \frac{ \ncl }{2 \mu_{0} \pi c \rcl^{2} } \big| A_{\pmp} \big|^{2}, \ \ % \left[ \frac{\text{W}}{\text{m}^{2}} \right] \text{ and,}
\end{equation*}
and for the signal field comprised of two modes, this irradiance reduces to 
\begin{equation*}\label{eq:SignalIrradiance}
    \begin{aligned}
        I_{\sig}(x, y, z) & = \frac{ 1 }{2 \mu_{0} \omega_{\sig} } \Big( \beta_{s}^{\FM} \big| A_{\sig}^{\FM} \big|^{2} \big| \varphi_{\sig}^{\FM} \big|^{2} + \beta_{s}^{\HOM} \big| A_{\sig}^{\HOM} \big|^{2} \big| \varphi_{\sig}^{\HOM} \big|^{2} \ + \\
            & \hspace*{12pt} \sqrt{ \beta_{s}^{\FM} \beta_{s}^{\HOM} } \big( A_{\sig}^{\HOM} \big)^{*} A_{\sig}^{\FM} \varphi_{\sig}^{\FM} \big( \varphi_{\sig}^{\HOM} \big)^{*} e^{ i {\Delta}\beta_{\sig} z } \ + \\
            & \hspace*{12pt} \sqrt{ \beta_{s}^{\FM} \beta_{s}^{\HOM} } \big( A_{\sig}^{\FM} \big)^{*} A_{\sig}^{\HOM} \big( \varphi_{\sig}^{\FM} \big)^{*} \varphi_{\sig}^{\HOM} e^{ -i {\Delta}\beta_{\sig} z }  \Big).
    \end{aligned} %\ \ \left[ \frac{\text{W}}{\text{m}^{2}} \right] .
\end{equation*}
Note that by defining 
\begin{align*}
    I_{\sig 0}(x, y, z) & \defeq \frac{ 1 }{2 \mu_{0} \omega_{\sig} } \left( \beta_{s}^{\FM} \big| A_{\sig}^{\FM} \big|^{2} \big| \varphi_{\sig}^{\FM} \big|^{2} + \beta_{s}^{\HOM} \big| A_{\sig}^{\HOM} \big|^{2} \big| \varphi_{\sig}^{\HOM} \big|^{2} \right) \ , \\
    I_{\sig +}(x, y, z) & \defeq \frac{ 1 }{2 \mu_{0} \omega_{\sig} } \sqrt{ \beta_{s}^{\FM} \beta_{s}^{\HOM} } \big( A_{\sig}^{\HOM} \big)^{*} A_{\sig}^{\FM} \varphi_{\sig}^{\FM} \big( \varphi_{\sig}^{\HOM} \big)^{*} \ , \text{ and} \\
    I_{\sig -}(x, y, z) & \defeq \frac{ 1 }{2 \mu_{0} \omega_{\sig} } \sqrt{ \beta_{s}^{\FM} \beta_{s}^{\HOM} } \big( A_{\sig}^{\FM} \big)^{*} A_{\sig}^{\HOM} \big( \varphi_{\sig}^{\FM} \big)^{*} \varphi_{\sig}^{\HOM} \ ,
\end{align*}
where $I_{\sig -} = I_{\sig +}^{*}$, one can express the signal irradiance more compactly as 
\begin{equation}\label{eq:CompactSignalIrradiance}
        I_{\sig}(x, y, z) = I_{\sig 0} + I_{\sig +} e^{ i {\Delta}\beta_{\sig} z } +  I_{\sig -} e^{ -i {\Delta}\beta_{\sig} z }. \ \ %\left[ \frac{\text{W}}{\text{m}^{2}} \right] .
\end{equation}
Furthermore, since the power is the integrated irradiance over the fiber's transverse domain, then the pump power is
\begin{equation}\label{eq:PumpPower}
    P_{\pmp}(z) = \frac{ \ncl }{2 \mu_{0} c } \big| A_{\pmp} \big|^{2} \ \ \left[ \text{W} \right] 
\end{equation}
and the signal power is
\begin{equation}\label{eq:SignalPower}
        P_{\sig}(z) = P_{\sig}^{\FM}(z) + P_{\sig}^{\HOM}(z) \ \ \left[ \text{W} \right],
\end{equation}
 where, taking advantage of the mode orthonormalities,
\begin{align*}
    P_{\sig}^{\FM}(z) & = \frac{ \beta_{s}^{\FM}  }{2  \mu_{0} \omega_{\sig} } \big| A_{\sig}^{\FM} \big|^{2} \text{ and } 
        P_{\sig}^{\HOM}(z) = \frac{ \beta_{s}^{\HOM}  }{2  \mu_{0} \omega_{\sig} } \big| A_{\sig}^{\HOM} \big|^{2} \ \ . % \left[ \text{W} \right]
\end{align*}
\begin{figure}
    \centering
    \begin{subfigure}{0.4\textwidth}
        \centering
        \includegraphics[width = \textwidth]{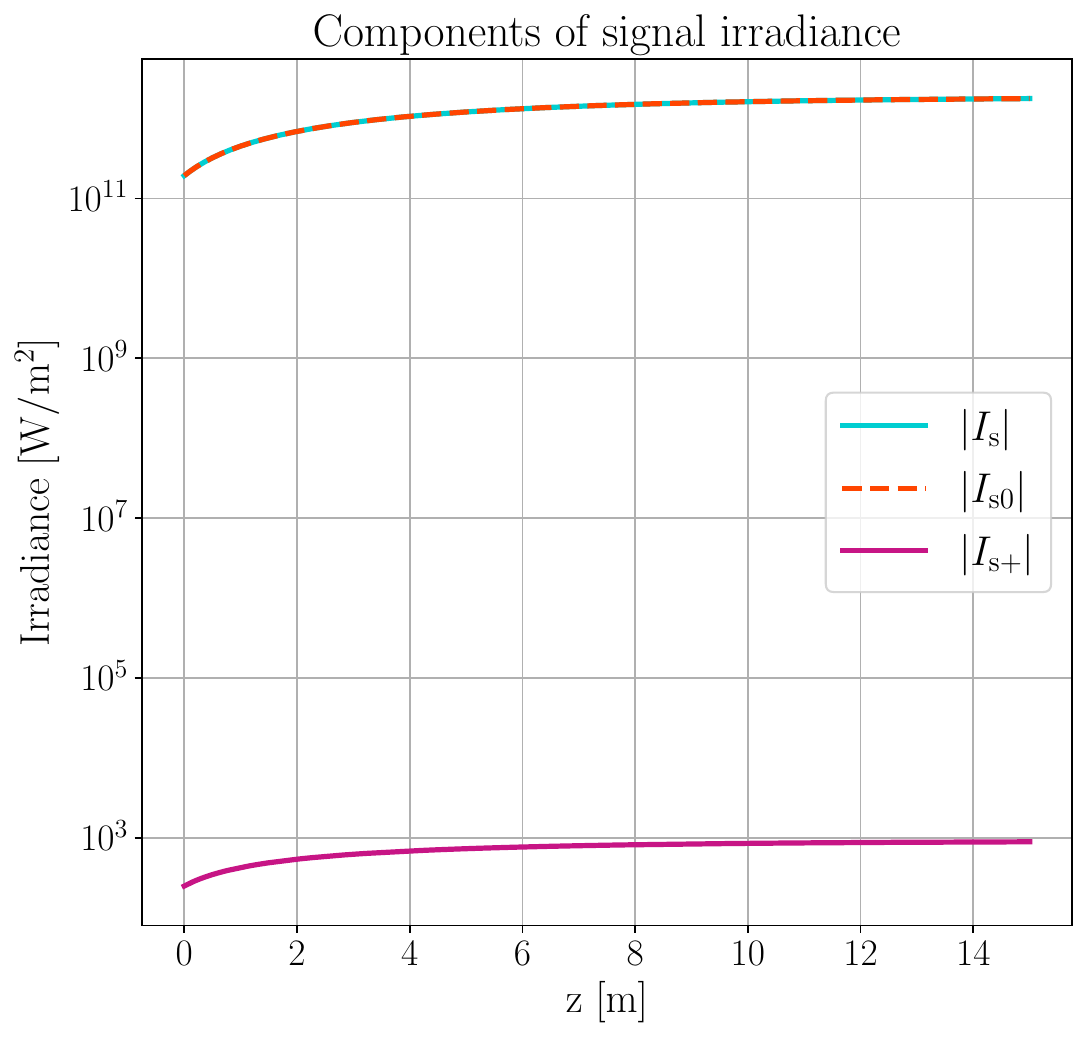}
        \caption{Magnitudes of $I_{{\sig}}$, $I_{{\sig}0}$, and $I_{{\sig}+}$ ($|I_{{\sig} +}| \equiv |I_{{\sig} -}|$).}
        \label{plot:IsComponents}
    \end{subfigure}
    \hspace{0.75cm}
    \begin{subfigure}{0.39\textwidth}
        \centering
        \includegraphics[width = \textwidth]{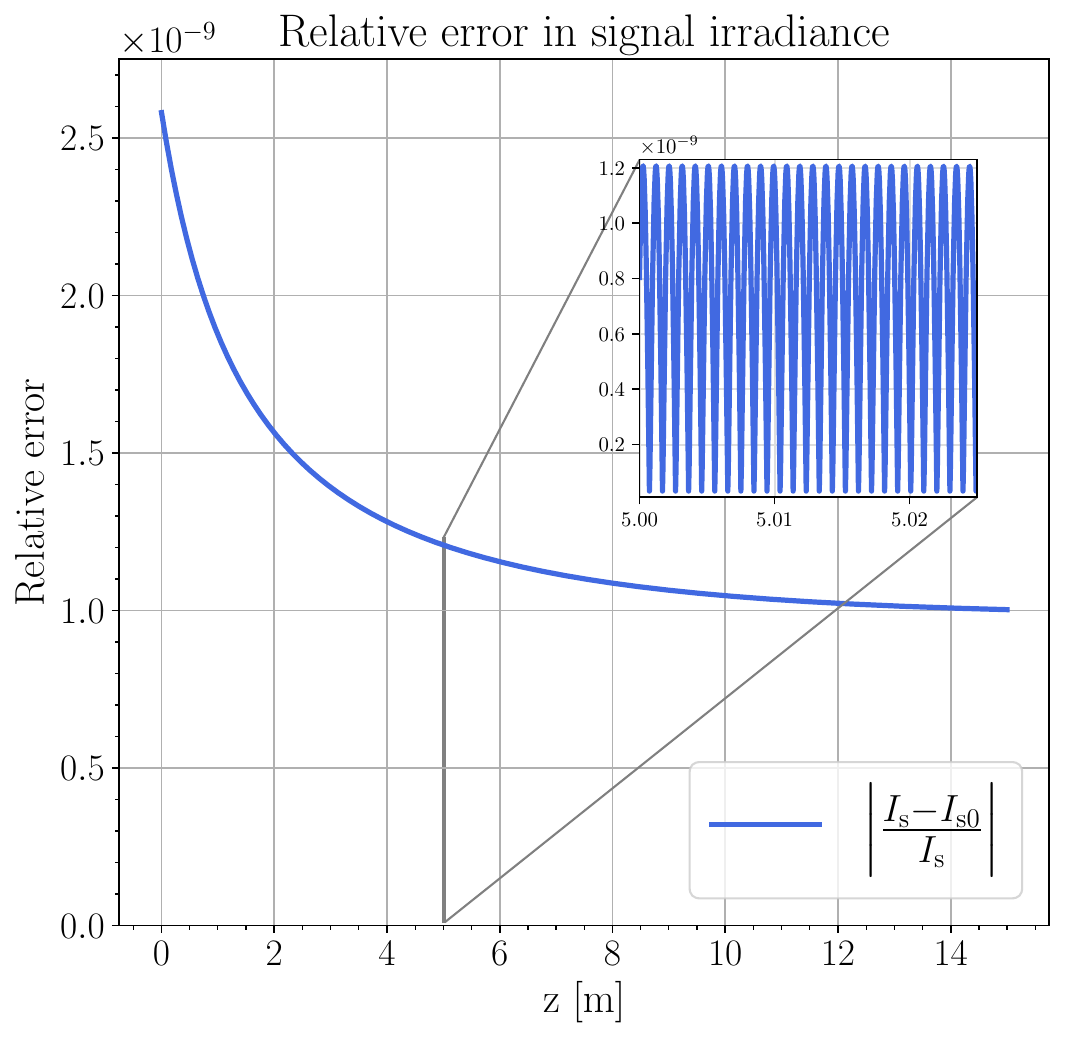}
        \caption{Relative error between $I_{\text{s}}$ and $I_{\text{s}0}$.}
        \label{plot:IsError}
    \end{subfigure} \\
    \vspace{0.75cm}
    \begin{subfigure}{0.4\textwidth}
        \centering 
        \includegraphics[width=\textwidth]{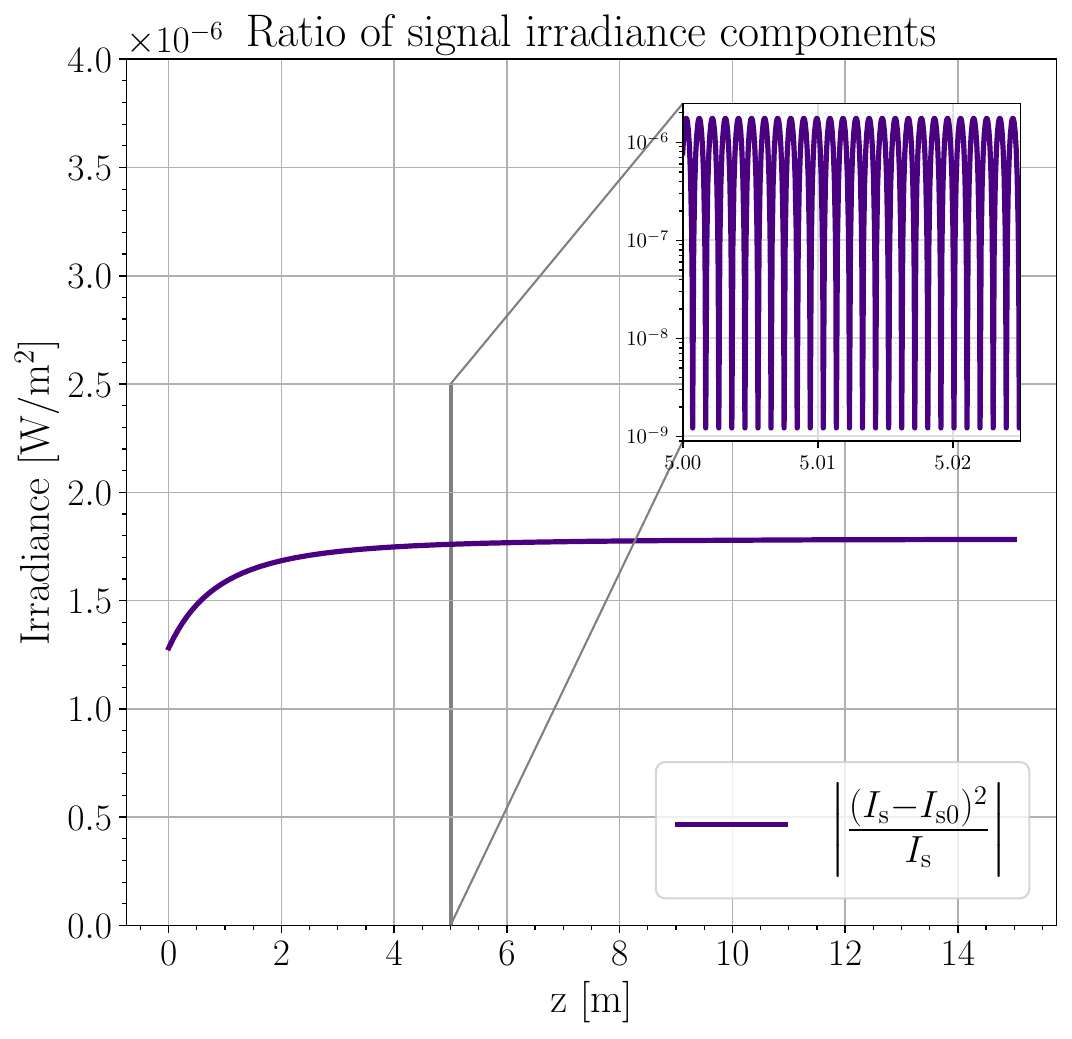}
        \caption{Envelope of a ratio involving $I_{\sig}$ and $I_{\sig 0}$.}
        \label{plot:Is0_ratio}
    \end{subfigure}
    \caption{Comparison of signal irradiance to its components along the fiber centerline $(x,y) = (0,0)$.}
    \label{fig:SignalIrradianceComparedToItsComponents}
\end{figure}

Using the Yb-doped fiber amplifier that is delineated in Section~\ref{sec:comp-results}, the components of the signal irradiance $\big\{ I_{\sig 0}, I_{\sig \pm} \big\}$ are plotted along the fiber centerline, $(x,y) = (0,0)$, in  Figure~\ref{fig:SignalIrradianceComparedToItsComponents}. 
Plot~(\subref{plot:IsComponents}) indicates that the majority of the signal field's energy is in the DC component ($I_{\sig 0}$) and not in the AC components ($I_{\sig \pm}$). 
The relative error between the full irradiance and its DC component ($I_{\sig 0}$) is illustrated in plot~(\subref{plot:IsError}). 
Of course, the key difference between $\Is$ and $I_{\sig 0 }$ is the presence of, or absence of, the periodic factors $\exp\big( \pm i \Delta \beta_{\sig} z \big)$, resulting in a dense set of low-magnitude rapid oscillations in the relative error, as depicted in the inset of plot~(\subref{plot:IsError}). 
In this case of a fairly typical fiber amplifier configuration, over most of the fiber length, one can expect that 
\begin{equation}\label{eq:Is-Iso}
    \big| I_{{\sig}\pm}(x = 0, y = 0, z) \big| \ll I_{{\sig} 0}(x = 0, y = 0, z) \ \text{ and }
    \max_{x, y, z} \Big| \frac{ \Is - I_{{\sig} 0}}{ I_{{\sig}} } \Big| \equiv \delta \ll 1 \ \text{ for } \ 0 \leq z \leq L \ \ ,
\end{equation}
where $\delta \sim \mathcal{O}\big( 10^{-9} \big)$ and $L$ is the length of the fiber.  
Moreover, 
\begin{equation}\label{eq:zeta}
    \max_{x, y, z} \Big| \frac{ \left( \Is - I_{{\sig} 0} \right)^{2}}{ I_{{\sig}} } \Big| \equiv \zeta \ll 1 \ \text{ for } \ 0 \leq z \leq L \ \ ,
\end{equation}
and $\zeta  \sim \mathcal{O}\big( 10^{-6} \big)$ shown in plot~(\subref{plot:Is0_ratio}). 
Later, we will use this information (\ref{eq:Is-Iso}--\ref{eq:zeta}) to help extract the oscillatory components within the governing system.

The steady-state gain function is nonlinearly dependent on the pump and signal irradiances: $\bar{g}_{\ell} = \bar{g}_{\ell}\big( I_{\pmp}, I_{\sig} \big)$ (cf. relation~\ref{eq:YbFixedPoint}). 
The definition of the coupling coefficients ($\kappa_{\ell}^{j , m}$)~\eqref{eq:CMTCouplingCoeffs} demonstrate that they are dependent on the gain functions ($\bar{g}_{\ell}$). 
Since the signal irradiance has factors of $\exp\big( \pm i {\Delta}\beta_{\sig} z \big)$ (see~\ref{eq:CompactSignalIrradiance}), the r.h.s. of the CMT governing relations ($\mathbf{F}( z, {\mbf A} )$) has embedded dependencies on the periodic factors $\exp\big( \pm i {\Delta}\beta_{\sig} z \big)$ within the coupling coefficients ($\kappa_{\ell}^{j , m}$). 
In order to obtain a reduced model by means of the integral averaging technique (Theorem~\ref{thm:avg-Verl}), it will be prudent to first extract the embedded periodic factors $\exp\big( \pm i {\Delta}\beta_{\sig} z \big)$ such that the r.h.s. of the CMT model expresses all of the dependencies on these factors explicitly (cf. Criterion~\ref{asmp:expl_period}). 

\FloatBarrier

%%%%%%%%%%%%%%%%%%%%%%%%%%%%%%%%%%% 
% Section: CMT Acceleration Via Averaged Dynamics: ACM Model
%%%%%%%%%%%%%%%%%%%%%%%%%%%%%%%%%%%

\section{CMT Acceleration Via Averaged Dynamics: ACM Model}\label{sec:acm}

In this section, we demonstrate that a reliable \textit{accelerated coupled mode} (ACM) model can be obtained from the CMT model~\eqref{eq:cmt-3x3} by applying the method of averaging periodicities (Theorem~\ref{thm:avg-Verl}). 
This integral averaging process will make the ACM model an autonomous ODE system for which one can use larger discrete longitudinal step sizes as compared to the original CMT model and yet still well-resolve the solution numerically. 
Deriving the ACM model using Theorem~\ref{thm:avg-Verl} requires the satisfaction of the aforementioned criteria~\ref{asmp:expl_period} and~\ref{asmp:small_param}; namely, that the dependencies on periodicities within the r.h.s. of the ODE system are explicitly expressed, and that there exists a problem-specific parameter $\varepsilon > 0$.

\FloatBarrier
%%%%%%%%%%%%%%%%%%%%%%%%%%%%%%%%%%% 
% Subsection: Explicit Dependence on Periodic Terms
%%%%%%%%%%%%%%%%%%%%%%%%%%%%%%%%%%%

\subsection{Explicit Dependence on Periodic Terms}\label{subsec:ExplicitDependence}

The main computational expense in numerically solving~\eqref{eq:cmt-3x3} is the evaluation of $\mbf F(z, {\mbf A})$ which requires integration over fiber cross-sections at each discrete longitudinal point. 
The number of times ${\mbf F}$ is to be evaluated in a practical simulation is largely determined by the presence of, and need to resolve, the oscillating exponentials within~\eqref{eq:cmt-3x3-K}, $\exp\big( \pm i {\Delta}\beta_{\sig} z \big)$, which have periods (mode beat lengths) of $2 \pi / {\Delta}\beta_{\sig} \sim \mathcal{O}\big( 1 \text{ mm} \big)$. 
The integral averaging method (Theorem~\ref{thm:avg-Verl}) can be used to remove these oscillating terms. 

To this end, consider that the steady-state gain for an active lanthanide dopant molecule is expressed as 
\begin{equation}\label{eq:g-ell}
    \bar{g}_{\ell}(x, y, z) \defeq \sigma_{\ell}^{\text{ems}} \overline{\calN}_{\text{excited}}(x, y, z) - \sigma_{\ell}^{\text{abs}} \overline{\calN}_{\text{ground}}(x, y, z) \ \ \left[ \frac{1}{\text{m}} \right] ,
\end{equation}
where $\sigma_{\ell}^{\text{abs}}$ and $\sigma_{\ell}^{\text{ems}}$ are the measured absorption and emission cross-sections of the active dopant [m$^{2}$/ion], respectively, and $\overline{\calN}_{\text{ground}}$ and $\overline{\calN}_{\text{excited}}$ are the steady-state outer shell electron's ground-state and excited-state energy manifold population concentrations [ions/m$^{3}$], respectively. 
These population concentrations are dictated by the local irradiance levels of the pump and signal fields:
\[
     \bar{g}_{\ell}(I_{\pmp}, I_{\sig}) = \sigma_{\ell}^{\text{ems}} \overline{\calN}_{\text{excited}}(I_{\pmp}, I_{\sig}) - \sigma_{\ell}^{\text{abs}} \overline{\calN}_{\text{ground}}(I_{\pmp}, I_{\sig}) \ \ .
\]

\begin{figure}
    \begin{subfigure}[b!]{0.39\textwidth}
        \centering
        \includegraphics[width=\linewidth]{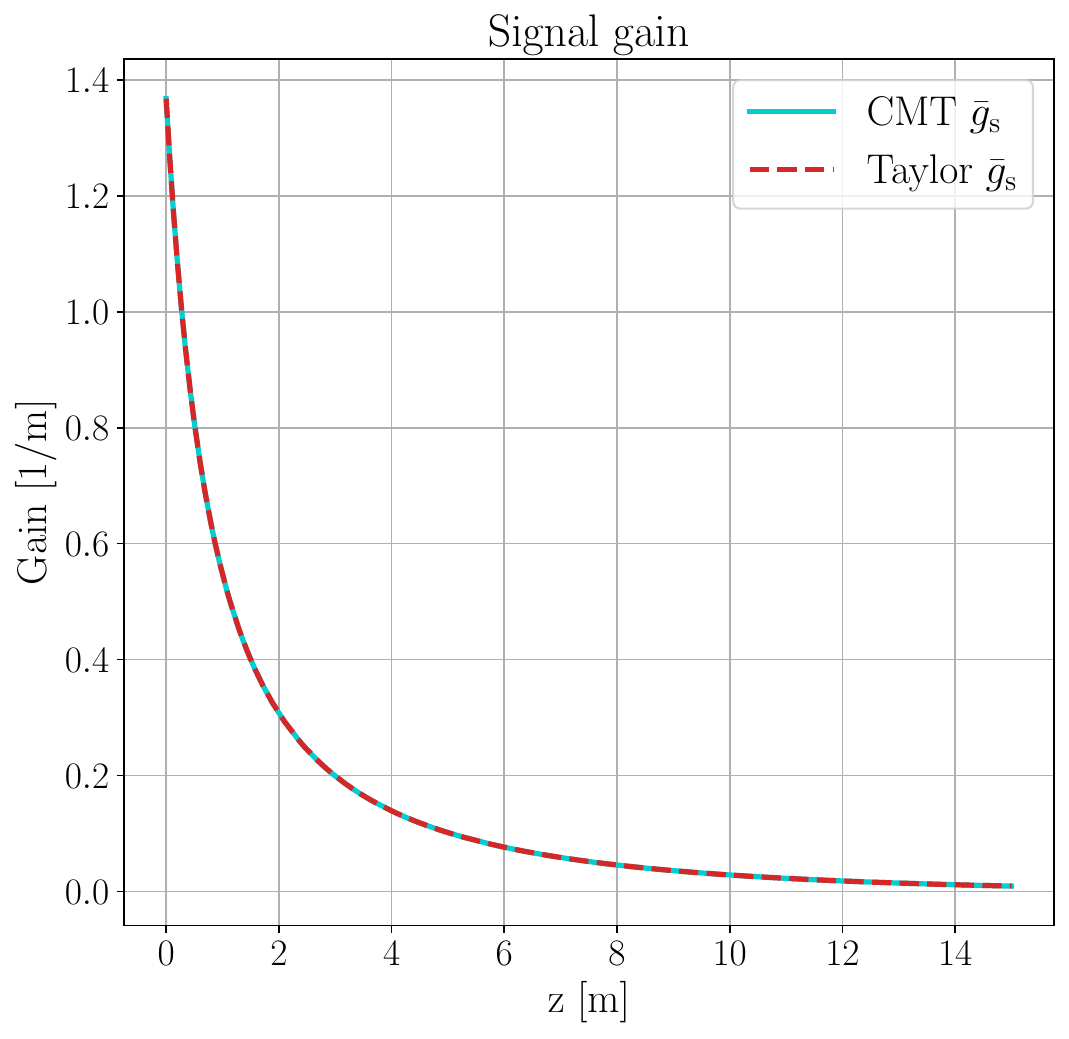}
        \caption{Signal gain and first-order Taylor expansion of signal gain.}
        \label{fig:sgain}
    \end{subfigure}
    \hspace{0.75cm}
    \begin{subfigure}[b!]{0.4\textwidth}
        \centering
        \includegraphics[width=\linewidth]{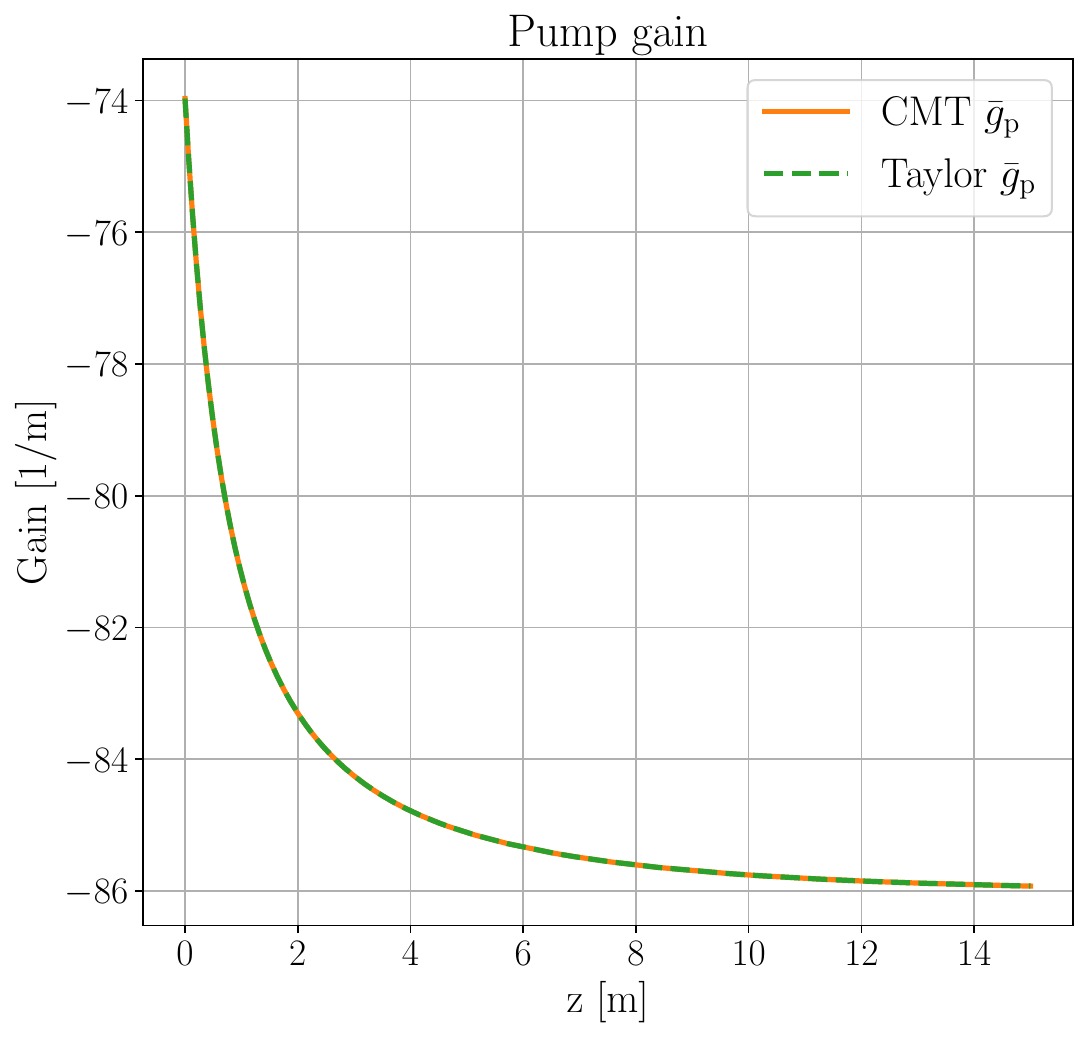}
        \caption{Pump gain and first-order Taylor expansion of pump gain.}
        \label{fig:pgain}
    \end{subfigure} \\
    \vspace{0.75cm}
    \begin{subfigure}{0.39\textwidth}
        \centering
        \includegraphics[width=\linewidth]{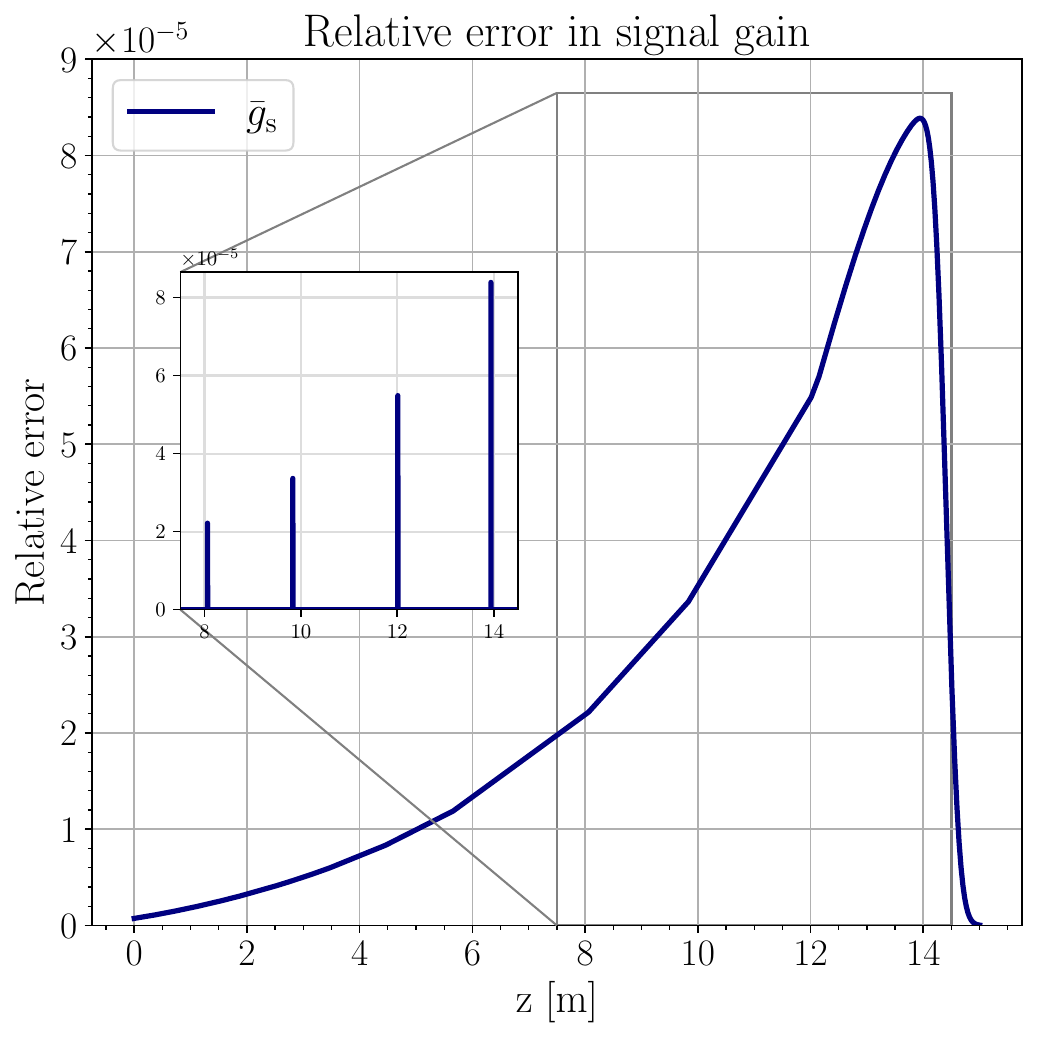}
        \caption{Signal gain Taylor approximation error.}
        \label{fig:sgain_diff}
    \end{subfigure}
    \hspace{0.75cm}
    \begin{subfigure}{0.4\textwidth}
        \centering
        \includegraphics[width=\linewidth]{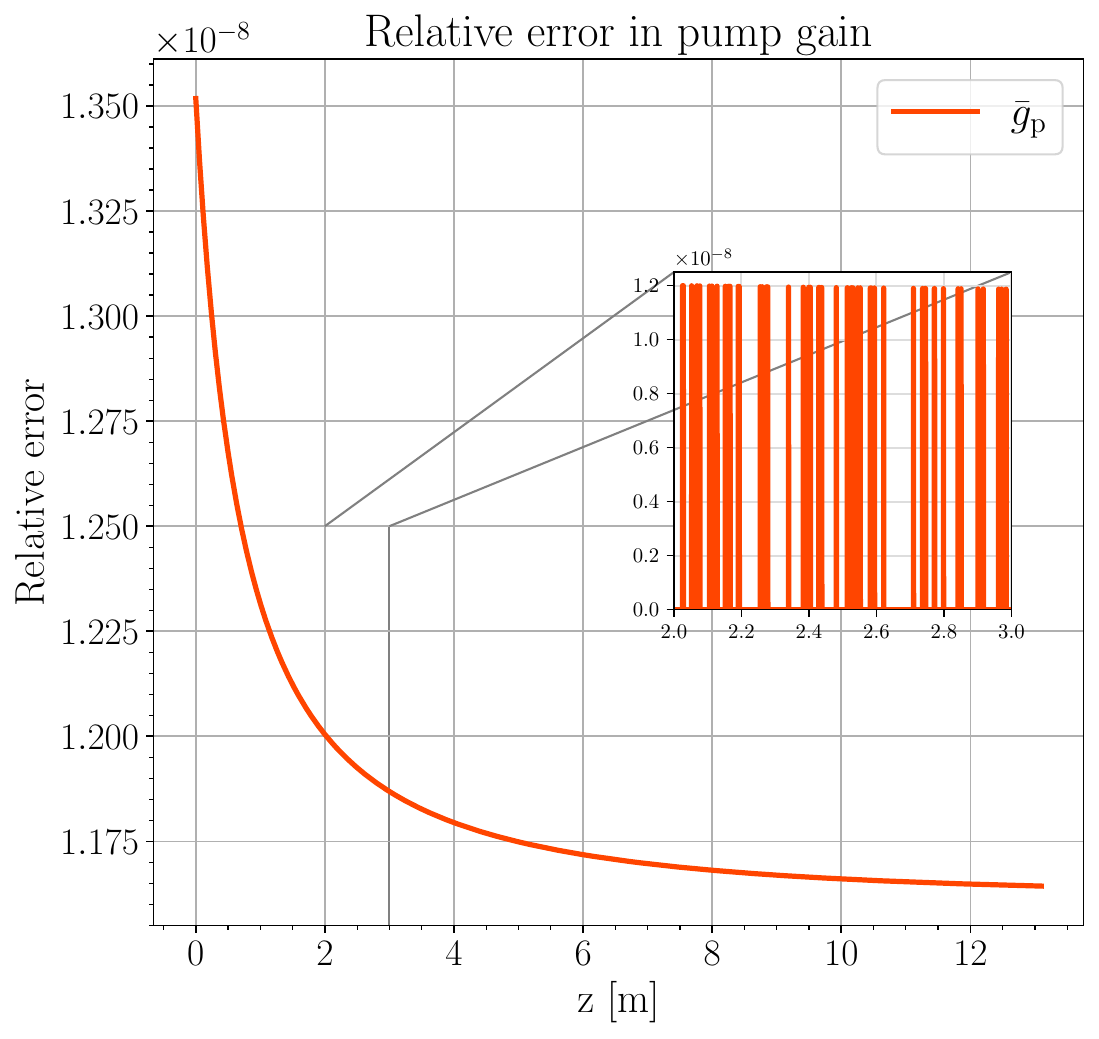}
        \caption{Pump gain Taylor approximation error.}
        \label{fig:pgain_diff}
    \end{subfigure}
    \caption{Full gain definition (solid lines) and first-order Taylor expansion of gain (dashed lines) along the fiber centerline $(x,y) = (0,0) $.}
    \label{fig:comparegain}
\end{figure}

With evidence supporting that $I_{{\sig} 0} \cong I_{\sig}$ in the
sense of~\eqref{eq:Is-Iso}, one may wish to approximate the steady-state gain using $I_{{\sig} 0}$ rather than $I_{\sig}$ as a means to extract additional embedded periodic factors. 
Holding $\Ip$ fixed and performing a first-order Taylor expansion about $I_{\sig} = I_{\sig 0}$ on the steady-state gain, one finds that 
\begin{equation}\label{eq:TaylorExpandedGain}
    \bar{g}_{\ell}\big( \Ip, \Is \big) \cong \bar{g}_{\ell}\big( \Ip, I_{{\sig} 0} \big) 
    + 
    \frac{\partial \bar{g}_{\ell} \big( \Ip, I_{{\sig} 0} \big)}{\partial I_{\sig}} \Big( I_{{\sig}} - I_{{\sig} 0} \Big)   \ \ ,
\end{equation}
where ${\p}\bar{g}_{\ell} / {\p}\Is$  can be ascertained from the known expressions for $\overline{\calN}_{\text{ground}}$ and $\overline{\calN}_{\text{excited}}$ (some partial derivatives are delineated in Appendix Section~\ref{sec:GainFormulas} for the ytterbium (Yb), thulium (Tm), and holmuim (Ho) dopant molecules). 
The error in approximation~\eqref{eq:TaylorExpandedGain} depends on the size of
\begin{align}\label{eq:first_ord_Taylor_error}
    \left| \frac{\partial^{2} \bar{g}_{\ell}}{\partial I_{\sig}^{2}} \left(I_{\sig} - I_{\sig 0} \right)^{2} \right| = \left| I_{\sig}^{2}  \frac{\partial^{2} \bar{g}_{\ell}}{\partial I_{\sig}^{2}} \left( \frac{I_{\sig} - I_{\sig 0} }{I_{\sig}} \right)^{2} \right| \leq \left| I_{\sig}^{2}  \frac{\partial^{2} \bar{g}_{\ell}}{\partial I_{\sig}^{2}} \right| \delta^{2},
\end{align}
where $\delta$ is delineated in~\eqref{eq:Is-Iso}. 
Appendix Section~\ref{subsec:taylor_error} shows that $I_{\sig}^{2} (\partial^{2} g_{\ell} / \partial I_{\sig}^{2})$ can be bounded independently of $I_{\sig}$. With the evidence in~\eqref{eq:Is-Iso}, one can be confident that the upper bound in~\eqref{eq:first_ord_Taylor_error}, i.e. the Taylor expansion error, is small. 
In addition to these arguments, Fig.~\ref{fig:comparegain} provides numerical evidence that this first-order Taylor expansion on both signal and pump steady-state gain about $I_{{\sig}} = I_{{\sig}0}$ is highly accurate to the original formulations of signal and pump gain.  
Since signal irradiance has a nonlinear dependence upon the gain equations, the error accumulates differently in the gain curves, as portrayed in plots~(\subref{fig:sgain_diff}) and (\subref{fig:pgain_diff}), as compared to how it evolved for the DC component of the signal irradiance (see plot~(\subref{plot:IsError}) of Fig.~\ref{fig:SignalIrradianceComparedToItsComponents}). 

Applying the decomposition of the signal irradiance~\eqref{eq:CompactSignalIrradiance}, $\bar{g}_{\ell}$ becomes
\begin{align}
    \bar{g}_{\ell}\big( \Ip, \Is \big) 
        & \cong \bar{g}_{\ell}\big( \Ip, I_{{\sig} 0} \big) + \frac{\partial \bar{g}_{\ell} \big( \Ip, I_{{\sig} 0} \big)}{\partial I_{\sig}} \Big( \cancel{I_{{\sig} 0}} + I_{\sig +} e^{ i {\Delta}\beta_{\sig} z } +  I_{\sig -} e^{ -i {\Delta}\beta_{\sig} z } - \cancel{I_{{\sig} 0}} \Big) \notag \\
    \bar{g}_{\ell}\big( \Ip, \Is \big) 
        & \cong \bar{g}_{\ell 0} + \bar{g}_{\ell +} e^{ i {\Delta}\beta_{\sig} z } + \bar{g}_{\ell -} e^{ -i {\Delta}\beta_{\sig} z } \ \ ,  \label{eq:DecomposedGain} %  \left[ \frac{1}{\text{m}} \right] \notag
\end{align}
where $\bar{g}_{\ell 0}(x, y, z) \defeq \bar{g}_{\ell}\big( \Ip, I_{{\sig} 0} \big)$ and  $\bar{g}_{\ell \pm}(x, y, z) \defeq I_{\sig \pm} \cdot {\p}\bar{g}_{\ell} / {\p}{\Is}\big( \Ip, I_{{\sig} 0} \big)$, respectively, and $\bar{g}_{\ell +} = \bar{g}_{\ell -}^{*}$. 

Recall that the coupling coefficients, $\kappa_{\ell}^{j , m}$~\eqref{eq:CMTCouplingCoeffs}, integrate the product of the steady-state gain function and the guided core modes over the fiber's transverse cross-section. 
Therefore, the periodic factors $\exp\big( \pm i {\Delta}\beta_{\sig} z \big)$ can easily be extracted in a similar format using the linearity of integration: 
\begin{align}
    \kappa_{\ell}^{j , m}\big( \Ip, \Is \big) & = \frac{ \beta_{\ell}^{m} }{ 2 \beta_{\ell}^{j} } \Big( \bar{g}_{\ell} \varphi_{\ell}^{m} , \varphi_{\ell}^{j} \Big)_{\Omega_{xy}} \notag \\
    & \cong \frac{ \beta_{\ell}^{m} }{ 2 \beta_{\ell}^{j} } \int_{\Omega_{xy}} \big[ \bar{g}_{\ell 0} + \bar{g}_{\ell +} e^{ i {\Delta}\beta_{\sig} z } + \bar{g}_{\ell -} e^{ -i {\Delta}\beta_{\sig} z } \big] \varphi_{\ell}^{m} \big( \varphi_{\ell}^{j} \big)^{*} dx dy \notag \\
    \kappa_{\ell}^{j , m}(z) & \cong \kappa_{\ell 0}^{j , m}(z) + \kappa_{\ell +}^{j , m}(z) e^{i {\Delta}\beta_{\sig} z} + \kappa_{\ell -}^{j , m}(z) e^{-i {\Delta}\beta_{\sig} z} \ \ ,  \label{eq:DecomposedCouplingCoeffs} % \left[ \frac{1}{\text{m}} \right] 
\end{align}
 where
 \begin{equation*}
     \kappa_{\ell 0}^{j , m}(z) \defeq \frac{ \beta_{\ell}^{m} }{ 2 \beta_{\ell}^{j} } \Big( \bar{g}_{\ell 0} \varphi_{\ell}^{m} , \varphi_{\ell}^{j} \Big)_{\Omega_{xy}} \hspace{1cm} \text{and} \hspace{1cm} \kappa_{\ell \pm}^{j , m}(z) \defeq \frac{ \beta_{\ell}^{m} }{ 2 \beta_{\ell}^{j} } \Big( \bar{g}_{\ell \pm} \varphi_{\ell}^{m} , \varphi_{\ell}^{j} \Big)_{\Omega_{xy}} \ \ .
 \end{equation*}
With these periodicities within the coupling coefficients explicitly expressed~\eqref{eq:DecomposedCouplingCoeffs}, Criterion~\ref{asmp:expl_period} is satisfied. 
The non-zero r.h.s. terms of the governing system~\eqref{eq:cmt-3x3-K} become 
\begin{equation}\label{eq:cmt-3x3-K-TaylorExpanded}
    \begin{aligned}
        \big[ \mathbf{F} \big]_{11} & = \left( \kappa_{\pmp 0} + \kappa_{\pmp +} e^{i {\Delta}\beta_{\sig} z} + \kappa_{\pmp -} e^{-i {\Delta}\beta_{\sig} z} \right) A_{\pmp} \ , \\
        \big[ \mathbf{F} \big]_{22} & = \left(\kappa_{\sig 0}^{\FM , \FM} + \kappa_{\sig +}^{\FM , \FM} e^{i {\Delta}\beta_{\sig} z} + \kappa_{\sig -}^{\FM , \FM} e^{-i {\Delta}\beta_{\sig} z}\right) A_{\sig}^{\FM} \ , \\ 
        \big[ \mathbf{F} \big]_{23} & = \Big[ \kappa_{\sig 0}^{\FM , \HOM} + \kappa_{\sig +}^{\FM , \HOM} e^{i {\Delta}\beta_{\sig} z} + \kappa_{\sig -}^{\FM , \HOM} e^{-i {\Delta}\beta_{\sig} z} \Big]  A_{\sig}^{\HOM} e^{-i{\Delta}\beta_{\sig} z} \\
            & = \left( \kappa_{\sig 0}^{\FM , \HOM} e^{-i{\Delta}\beta_{\sig} z} + \kappa_{\sig +}^{\FM , \HOM} + \kappa_{\sig -}^{\FM , \HOM} e^{-2 i {\Delta}\beta_{\sig} z} \right) A_{\sig}^{\HOM} \ , \\
        \big[ \mathbf{F} \big]_{32} & = \Big[ \kappa_{\sig 0}^{\HOM , \FM} + \kappa_{\sig +}^{\HOM , \FM} e^{i {\Delta}\beta_{\sig} z} + \kappa_{\sig -}^{\HOM , \FM} e^{-i {\Delta}\beta_{\sig} z} \Big]  A_{\sig}^{\FM} e^{i {\Delta}\beta_{\sig} z} \\
            & = \left( \kappa_{\sig 0}^{\HOM , \FM} e^{i {\Delta}\beta_{\sig} z} + \kappa_{\sig +}^{\HOM , \FM} e^{2 i {\Delta}\beta_{\sig} z} + \kappa_{\sig -}^{\HOM , \FM} \right) A_{\sig}^{\FM} \ , \text{ and} \\
        \big[ \mathbf{F} \big]_{33} & = \left( \kappa_{\sig 0}^{\HOM, \HOM} + \kappa_{\sig +}^{\HOM , \HOM} e^{i {\Delta}\beta_{\sig} z} + \kappa_{\sig -}^{\HOM , \HOM} e^{-i {\Delta}\beta_{\sig} z} \right)  A_{\sig}^{\HOM} \ \ .
    \end{aligned}
\end{equation}

\FloatBarrier
%%%%%%%%%%%%%%%%%%%%%%%%%%%%%%%%%%%
% Subsection: Extracting $\varepsilon$
%%%%%%%%%%%%%%%%%%%%%%%%%%%%%%%%%%%

\subsection{Extracting $\varepsilon$}\label{subsec:Extract}

\begin{figure}%[H]
    \centering
    \includegraphics[width=0.5\linewidth]{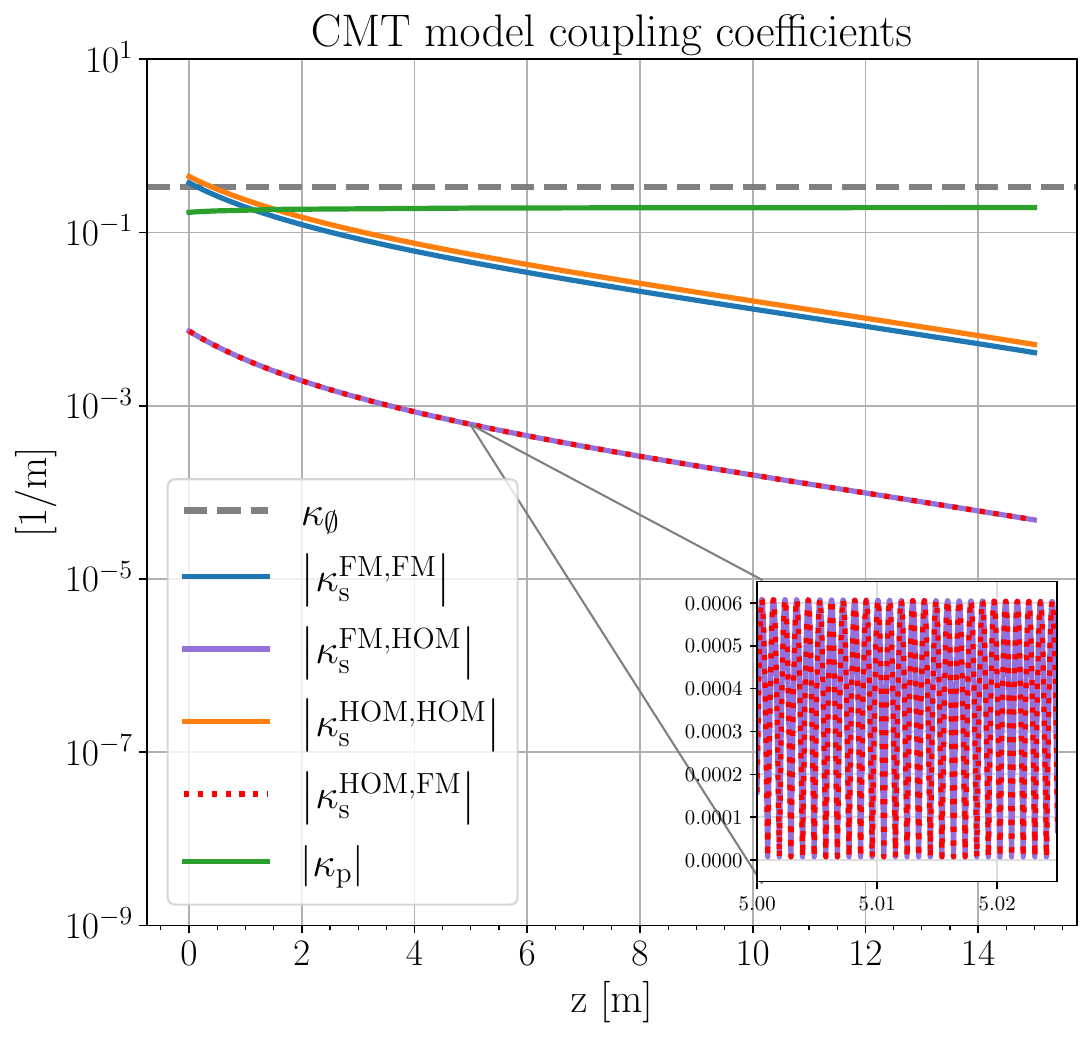}
    \caption{Coupling coefficient magnitudes along with dimensional scaling value $\kappa_{\varnothing}$.}
    \label{fig:kappas}
\end{figure}

Criterion~\ref{asmp:small_param} is met through the nondimensionalization of the CMT model governing equations~\eqref{eq:cmt-3x3}. 
Consider the following new model parameters\dots 
\begin{equation}\label{eq:non_dim}
    \mbf A  = \mbf A_{\varnothing} \hat{\mbf A}, \hspace{1cm} \Delta \beta_{\sig} = \Delta \beta_{\sig \varnothing} \Delta \hat{\beta_{\sig}}, \hspace{1cm}   z = z_{\varnothing} \hat{z},  \hspace{1cm} \text{and } \kappa^{j,m}_{\ell} = \kappa_{\varnothing } \hat{\kappa}^{j,m}_{\ell} ,
\end{equation}
where $[\cdot_{\varnothing}]$ denotes some scaling value and $[\hat{\cdot}]$ denotes the dimensionless parameter. 
The dimensional scaling values are chosen to be 
\begin{equation}\label{eq:scaling_values}
    \begin{aligned}
        \mbf A_{\varnothing} & = 10^{2} \ \left[ \text{V} \right] \ ,  & \Delta \beta_{\sig \varnothing} & = 3.25 \cdot 10^{3} \ \left[\frac{\text{rads}}{\text{m}} \right] \ , \\
        z_{\varnothing} & \defeq \frac{1}{\Delta \beta_{\sig \varnothing}} \ \left[ \text{m} \right] \ , & \text{and } \kappa_{\varnothing } & = 3.4 \cdot 10^{-1} \left[ \frac{1}{\text{m}} \right] \ .
    \end{aligned}
\end{equation}
This rescaling of the governing relations sets $\varepsilon$ to be 
\begin{align}\label{eq:veps_rho}
    \varepsilon \defeq \kappa_{\varnothing} z_{\varnothing} \ \ .
\end{align}
Figure~\ref{fig:kappas} displays the chosen size for $\kappa_{\varnothing}$ in comparison to the magnitudes of the coupling coefficients~\eqref{eq:CMTCouplingCoeffs} for the typical Yb-doped amplifier with the listed properties in Section~\ref{sec:comp-results}. 

Substituting (\ref{eq:non_dim}--\ref{eq:veps_rho}) into~\eqref{eq:cmt-3x3}, and simplifying, results in the nondimensional CMT ODE system 
\begin{subequations}\label{eq:cmt-nondim-3x3}  
    \begin{align}
        \frac{\partial {\hat{\mbf A}}}{\partial \hat{z}} & = \varepsilon {\hat{\mbf F}}( \hat{{\mbf A}}) \ \ , \quad {\hat{\mbf A}}\left( 0 \right) = \hat{{\mbf A}}_{0} \ \ , \text{ where} \label{eq:cmt-nondim-3x3-ODE} \\
        \hat{\mathbf{F}}( \hat{\mbf A}) & = \hat{\bcalK} (\hat {\mbf A}) \ \hat{\mbf A} \ \text{ and } \ 
            \hat{\bcalK} ( \hat {\mbf A}) = 
                \begin{bmatrix}
                    \hat{\kappa}_{\pmp}  & 0 & 0 \\
                    0 &   \hat{\kappa}_{\sig }^{\FM , \FM}  &  \hat{\kappa}_{\sig }^{\FM , \HOM}  e^{-i \Delta \hat{\beta_{\sig}}  \hat{z}}  \\
                    0 &  \hat{\kappa}_{\sig }^{\HOM , \FM}  e^{i  \Delta \hat{\beta_{\sig}} \hat{z}} &   \hat{\kappa}_{\sig }^{\HOM , \HOM}
                \end{bmatrix} \ \ , \label{eq:cmt-nondim-3x3-K}
    \end{align}
    where $\varepsilon \simeq 1 \cdot 10^{-4}$. 
\end{subequations}
For the sake of notational simplicity, we discontinue the use of the $[\hat{\cdot}]$ notation hereafter, and the governing systems are to be considered nondimensionalized. 

With a reasonable $\varepsilon$ extracted, Criterion~\ref{asmp:small_param} is satisfied, and we may apply Theorem~\ref{thm:avg-Verl} to~\eqref{eq:cmt-nondim-3x3} by integral averaging the r.h.s. of the governing ODEs over the period $2 \pi / {\Delta}\beta_{\sig}$. 
This yields our ACM model: 
\begin{subequations}\label{eq:acm-3x3}
    \begin{align}
        \frac{\partial {\mbt A}}{\partial \hat{z}} & = \varepsilon {\mbt F}({\mbt A}) \ \ , \quad {\mbt A}\left( 0 \right) = {\mbf A}_{0} \ \ , \text{ where} \label{eq:acm-3x3-ODE} \\
        \tilde{\mathbf{F}}( \tilde{\mbf A}) & = \tilde{\bcalK} (\tilde {\mbf A}) \ \tilde{\mbf A} \ \text{ and } \ 
            \tilde{\bcalK} ( \tilde {\mbf A}) = 
                \begin{bmatrix}
                    \hat{\kappa}_{\pmp 0}  & 0 & 0 \\
                    0 &   \hat{\kappa}_{\sig 0}^{\FM , \FM}  &  \hat{\kappa}_{\sig +}^{\FM , \HOM}   \\
                    0 &  \hat{\kappa}_{\sig -}^{\HOM , \FM}   &   \hat{\kappa}_{\sig 0}^{\HOM , \HOM}
                \end{bmatrix} \ \ . \label{eq:acm-3x3-Ktilde}
    \end{align}
\end{subequations}
By Theorem~\ref{thm:avg-Verl}, the solutions to the original~\eqref{eq:cmt-nondim-3x3} and averaged~\eqref{eq:acm-3x3} dynamical systems are bounded such that 
\[
    \| {\mbf A}(z) - {\mbt A}(z) \| \le c \veps 
\]
holds with the Euclidean norm $\| \cdot \|$ for $0 \le z \le S / \veps$, and where 
\begin{align*}
    c & = \left( 2 S \lambda_{\mbf F} c_{\mbf A} + \frac{ 2 \pi }{ {\Delta}\beta_{\sig} } c_{{\mbf F} - {\mbt F}} \right) e^{S \lambda_{\mbf F}} \ \ .
\end{align*}
The value of $S / \veps$, the maximum fiber length over which the averaged solution remains accurate, is problem-dependent since it depends on the initial conditions for ${\mbf A}$, and on various other fiber parameters, properties, and configurations. 
However, because this optical fiber amplifier problem conserves the total of photons and excited electrons, one can safely presume ${\mbf A}$ will remain bounded for any finite $z$. 
The Lipschitz constant of $\mbf F$, $\lambda_{\mbf F}$, is defined in terms of the CMT model in \eqref{eq:cmt-nondim-3x3-K} as 
\begin{align*}
   \lambda_{\mbf F} & = \max \| \mbf A \| \ \lambda_{\bcalK} + \max \| \bcalK \| \ \ ,
\end{align*}
where $\lambda_{\bcalK}$ is the Lipschitz constant of $\bcalK$.

\textbf{Remark.}
\emph{A similar type of reduced model was explored by Drake et al.~\cite{drake2020simulation} after they realized that, by removing the factors $\exp\big( \pm i {\Delta}\beta_{\sig} z \big)$ from the governing equations, one could accurately model the original problem over a shorter amplifier length, but with an unaltered longitudinal discrete step size (${\Delta}z$), by rescaling the equations. 
Or, equivalently, use a larger longitudinal discrete step size (${\Delta}z$) with the original fiber length. 
In their case, they did not make any approximations on the steady-state gain functions and were able to show an exact equivalence between the original amplifier problem and their rescaled version, resulting in a substantial computational speed-up in their numerical solve times. 
However, their approach was limited to modeling simpler amplification and loss processes. 
In our future work, we plan to demonstrate that this current method, where the gain functions undergo a Taylor approximation, will allow the model to be augmented with other physical effects, and yet remain an accurate representation of its corresponding original dynamical system while still offering an improvement in computational efficiency. }

\begin{figure}%
    \centering
        \centering
        \includegraphics[width=0.5\linewidth]{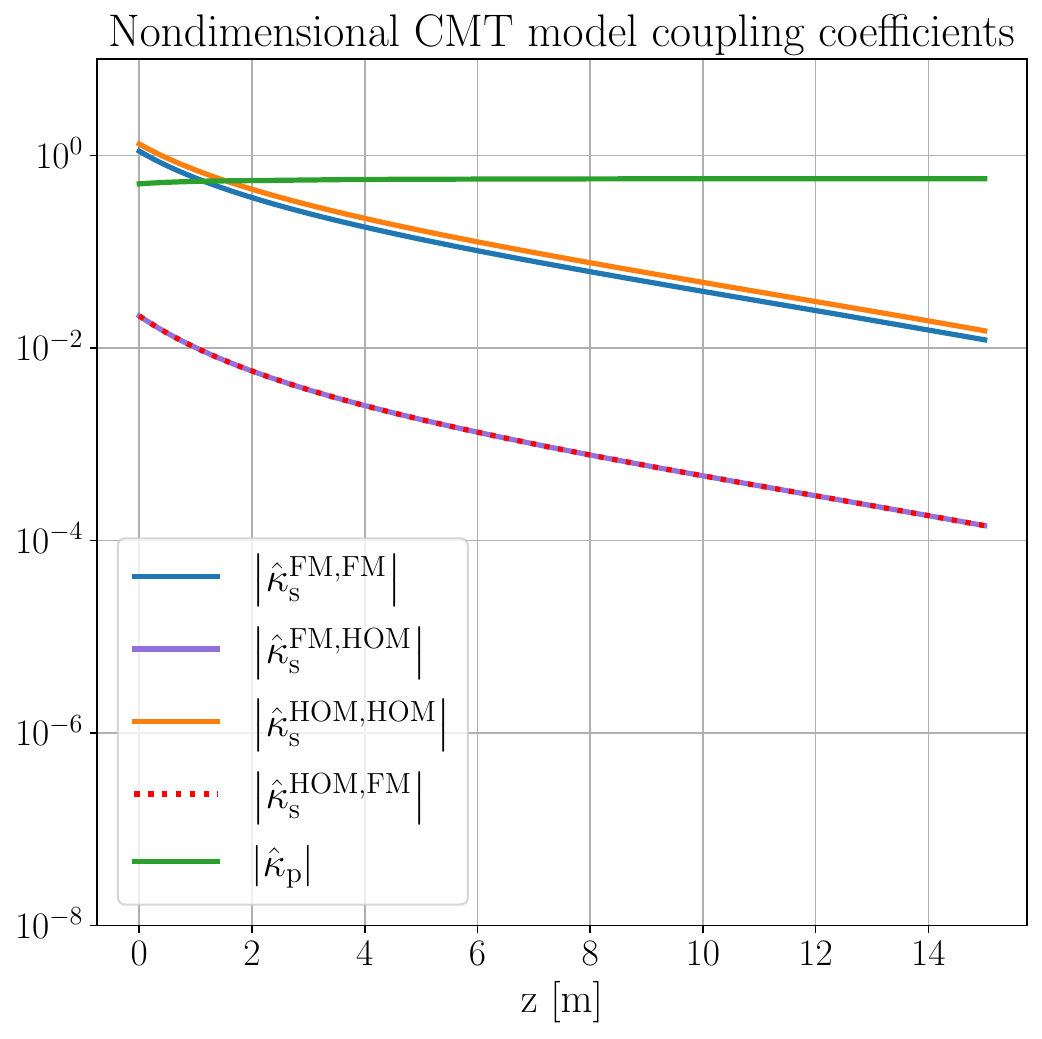}
        \caption{The nondimensional coefficients of the CMT model are approximately $\mathcal{O} (1)$ or smaller.}
        \label{fig:nondim_kappas}
\end{figure}
The nondimensional coupling coefficients plotted in Figure~\ref{fig:nondim_kappas} and can be used to estimate the magnitude of $\lambda _{\mbf F}$. 
The constant $c_{\mbf A}$ is the bound for which $\| {\mbf A} \left( z_{1} \right) - {\mbf A} \left( z_{2} \right) \| \leq \veps c_{\mbf A}$ for any $0 \leq z_{1}, z_{2} \leq L$. 
Finally, the constant $c_{{\mbf F} - {\mbt F}}$ is the maximum of the difference between ${\mbf F}$~\eqref{eq:cmt-nondim-3x3-K} and ${\mbt F}$~\eqref{eq:acm-3x3-Ktilde}, $c_{{\mbf F} - {\mbt F}} = \underset{{\mbf A}, \ z \in \left[ 0, S / \varepsilon \right]}{\max} \| \left( {\mbf F} - {\mbt F} \right) \left( {\mbf A} , z \right) \|$, where the non-zero ${\mbf F} - {\mbt F}$ elements can be written as 
\begin{align*}
    \big[ {\mbf F} - {\mbt F} \big]_{11} & =   \left( \kappa_{\pmp +} e^{i {\Delta}\beta_{\sig} z} + \kappa_{\pmp -} e^{-i {\Delta}\beta_{\sig} z} \right)  A_{\pmp} \ , \\
    \big[ {\mbf F} - {\mbt F} \big]_{22} & =   \left( \kappa_{\sig +}^{\FM , \FM} e^{i {\Delta}\beta_{\sig} z} + \kappa_{\sig -}^{\FM , \FM} e^{-i {\Delta}\beta_{\sig} z} \right)  A_{\sig}^{\FM} \ , \\
    \big[ {\mbf F} - {\mbt F} \big]_{23} & = \left( \kappa_{\sig 0}^{\FM , \HOM} e^{-i {\Delta}\beta_{\sig} z} + \kappa_{\sig -}^{\FM , \HOM} e^{-2 i {\Delta}\beta_{\sig} z} \right)  A_{\sig}^{\HOM} \ , \\
    \big[ {\mbf F} - {\mbt F} \big]_{32} & = \left( \kappa_{\sig 0}^{\HOM , \FM} e^{i {\Delta}\beta_{\sig} z} + \kappa_{\sig +}^{\HOM , \FM} e^{2 i {\Delta}\beta_{\sig} z} \right)  A_{\sig}^{\FM}\ , \text{ and} \\
    \big[ {\mbf F} - {\mbt F} \big]_{33} & =   \left( \kappa_{\sig +}^{\HOM , \HOM} e^{i {\Delta}\beta_{\sig} z} + \kappa_{\sig -}^{\HOM , \HOM} e^{-i {\Delta}\beta_{\sig} z} \right)   A_{\sig}^{\HOM} \ \ .
\end{align*}

\FloatBarrier
%%%%%%%%%%%%%%%%%%%%%%%%%%%%%%%%%%% 
% Subsection: Summary of the CMT and ACM Model Governing Equations
%%%%%%%%%%%%%%%%%%%%%%%%%%%%%%%%%%%

\subsection{Summary of the CMT and ACM Model Governing Equations}\label{subsec:GEsSummary}

In summary, the CMT model solves for $\mbf A = [A_{\pmp}, A_{\sig}^{\FM}, A_{\sig}^{\HOM}]^{\text{T}}$ and is governed by 
\begin{align}
    \frac{\partial \mbf{A}}{\partial z} & = \varepsilon 
        \begin{bmatrix}
            {\kappa}_{\pmp} & 0 & 0 \\
            0 &   {\kappa}_{\sig}^{\FM,\FM} & {\kappa}_{\sig}^{\FM,\HOM} \ e^{-i  {\Delta}\beta_{\sig} z}  \\
            0 & {\kappa}_{\sig}^{\HOM,\FM} \ e^{i  {\Delta}\beta_{\sig} z} &    {\kappa}_{\sig}^{\HOM,\HOM}
        \end{bmatrix}   \mbf{A} , \text{ where } \label{eq:NondimensionalCMTModel} \\
        & \hspace*{12pt} 
            \begin{cases}
                {\kappa}_{\pmp} = \frac{ \ncl }{ 2 \pi \rcl^{2} } \int_{\Omega_{xy}} \hspace*{-2.5mm} \bar{g}_{\pmp} dxdy \vspace{1.5mm} \\ 
                {\kappa}_{{\sig}}^{j, m} = \frac{ \beta^{m}_{\sig} }{ 2 \beta^{j}_{\sig}  } \left( \bar{g}_{{\sig}} \varphi_{m}, \varphi_{j} \right)_{\Omega_{xy}}
            \end{cases} \ , \notag \\
    \intertext{while the ACM model solves for $\mbt A = [\tilde A_0^\pmp, \tilde A_{\FM}^\sig, \tilde A_{\HOM}^\sig]^{\text{T}}$ and follows}
    \frac{\partial \mbt{A}}{\partial z} & = \varepsilon 
        \begin{bmatrix}
            {\kappa}_{\pmp 0} & 0 & 0 \\
            0 & {\kappa}_{\sig 0}^{\FM,\FM} & {\kappa}_{\sig +}^{\FM,\HOM} \\
            0 & {\kappa}_{\sig -}^{\HOM,\FM}  & {\kappa}_{\sig 0}^{\HOM,\HOM}
        \end{bmatrix}   \mbt{A} , \text{ where } 
        \begin{cases}
            {\kappa}_{\pmp 0} = \frac{ \ncl }{ 2 \pi \rcl^{2}  } \int_{\Omega_{xy}} \hspace*{-2.5mm} \bar{g}_{\pmp 0} dxdy \vspace{1.5mm} \\
            {\kappa}_{{\sig} 0}^{j, m} = \frac{ \beta^{m}_{\sig} }{ 2 \beta^{j}_{\sig}  } \left( \bar{g}_{{\sig}0} \varphi_{m}, \varphi_{j} \right)_{\Omega_{xy}} \vspace{1.5mm} \\
            {\kappa}_{{\sig} \pm}^{j, m} = \frac{ \beta^{m}_{\sig} }{ 2 \beta^{j}_{\sig}  } \left( \bar{g}_{{\sig}\pm} \varphi_{m}, \varphi_{j} \right)_{\Omega_{xy}}
        \end{cases} \ . \label{eq:NondimensionalACMModel}
\end{align}
The two key differences between these models are that (1) the explicit periodic terms $\exp \big(\pm i {\Delta}\beta_{\sig} z \big)$ have been removed from the ACM model, and that (2) the ACM model only uses the first-order Taylor expansion of the steady-state gain, whereas the full CMT model does not approximate this gain relation. 
Thus, for the CMT model, $\bar{g}_{\sig}$ and $\bar{g}_{\pmp}$ are formulated in~\eqref{eq:g-ell}, while, for the ACM model, $\bar{g}_{\sig 0}$, $\bar{g}_{\sig\pm}$, $\bar{g}_{\pmp 0}$, and $\bar{g}_{\pmp\pm}$ are expressed in~\eqref{eq:DecomposedGain}. 
Otherwise, all parameters, initial conditions, and scalings between the two models are the kept identical for comparison purposes. 
In the next section, we demonstrate how these differences translate into computational accuracy and performance. 

\FloatBarrier

\FloatBarrier
%%%%%%%%%%%%%%%%%%%%%%%%%%%%%%%%%%%
% Section: Model Comparison Results
%%%%%%%%%%%%%%%%%%%%%%%%%%%%%%%%%%%

\section{Model Comparison Results}\label{sec:comp-results}

For the purpose of comparing the CMT and ACM models, consider an ytterbium-doped, cladding-pumped, fiber amplifier with the design parameters that are listed in Table~\ref{fiberParams}.   
\begin{table}% [H]
    \centering
    \begin{tabular}{|l|l|l|}
        \hline
        Symbol & Description & Value     \\
        \hline
        \hline
        $r_{\text{core}}$       & outer extent of the fiber's core region                   & 9.5 $\mu$m \\
        $r_{\text{clad}}$       & outer extent of the fiber's inner cladding region         & 200 $\mu$m \\
        $r_{\text{fiber}}$      & outer extent of the fiber (with polymer jackets)          & 260 $\mu$m \\
        $L$                     & total fiber amplifier length  & 15 m    \\
        $n_{\text{core}}$       & core refractive index         & 1.4500 \\
        $n_{\text{clad}}$       & cladding refractive index     & 1.4485\\
        NA$_{\text{core}}$      & core numerical aperture       & 0.065 \\
        $\lambda_{\text{s}}$    & laser signal wavelength       & 1064 nm \\
        $\lambda_{\text{p}}$    & pump field wavelength         & 976 nm \\
        $V$                             & fiber configuration $V$-number    & 3.6465 \\
        $\sigma_{\sig}^{\text{abs}}$    & signal absorption cross-section   & $6 \cdot 10^{-27} \ \frac{\text{m}^{2}}{\text{ion}}$ \\
        $\sigma_{\sig}^{\text{ems}}$    & signal emission cross-section     & $3.58 \cdot 10^{-25} \ \frac{\text{m}^{2}}{\text{ion}}$ \\
        $\sigma_{\pmp}^{\text{abs}}$    & pump absorption cross-section     & $1.429 \cdot 10^{-24} \ \frac{\text{m}^{2}}{\text{ion}}$ \\
        $\sigma_{\pmp}^{\text{ems}}$    & pump emission cross-section       & $1.776 \cdot 10^{-24} \ \frac{\text{m}^{2}}{\text{ion}}$ \\
        $\calN_{\text{total}}$  & total dopant population concentration     & $6.25 \cdot 10^{25} \ \frac{\text{ion}}{\text{m}^{3}}$ \\
        $\tau^{\text{Yb}}$      & upper level radiative lifetime for Yb$_{2}$O$_{3}$    & $8.014 \cdot 10^{-4} \ \text{s}$ \\
        $P_{\sig}(0)$           & total signal seed power       & 50 W \\
        $P_{\pmp}(0)$           & launched pump power           & 500 W \\
        $\kappa_{\text{thermal}}$       & thermal conductivity of glass     & $1.38 \ \frac{\text{W}}{\text{m} \cdot \text{K}}$ \\
        \hline
    \end{tabular}
    \caption{Fiber laser amplifier simulation parameters, properties, and initial conditions.}
    \label{fiberParams}
\end{table}
\FloatBarrier
For practical fiber amplifier applications, experimentalists are interested in the real-valued quantities of power and amplification efficiency, which require some post-processing of the CMT and ACM amplitude solutions. 
The pump and signal powers are defined in equations~\eqref{eq:PumpPower} and~\eqref{eq:SignalPower}, respectively. 
The amplification efficiency, $\eta$, is defined as
\begin{equation*}
    \eta(z) = \frac{P_{\sig}(z) - P_{\sig}(z = 0)}{P_{\pmp}(z = 0) - P_{\pmp}(z)} \ \ .
\end{equation*}
Note that this amplification efficiency will always be less than the ideal efficiency of $\eta_{\text{ideal}} = \hbar \omega_{\sig} / (\hbar \omega_{\pmp}) \cong \lp / \ls \simeq 91.7\%$, which would only be achieved if every pump photon were converted into a laser signal photon. 
Due to the quantum defect ($1 - \hbar \omega_{\sig} / (\hbar \omega_{\pmp})$) between the pump and signal frequencies, some energy is deposited as heat along the length of the fiber. 

\begin{figure} %[H]
    \centering
    \begin{subfigure}[b]{0.38\textwidth}
        \includegraphics[width=\linewidth]{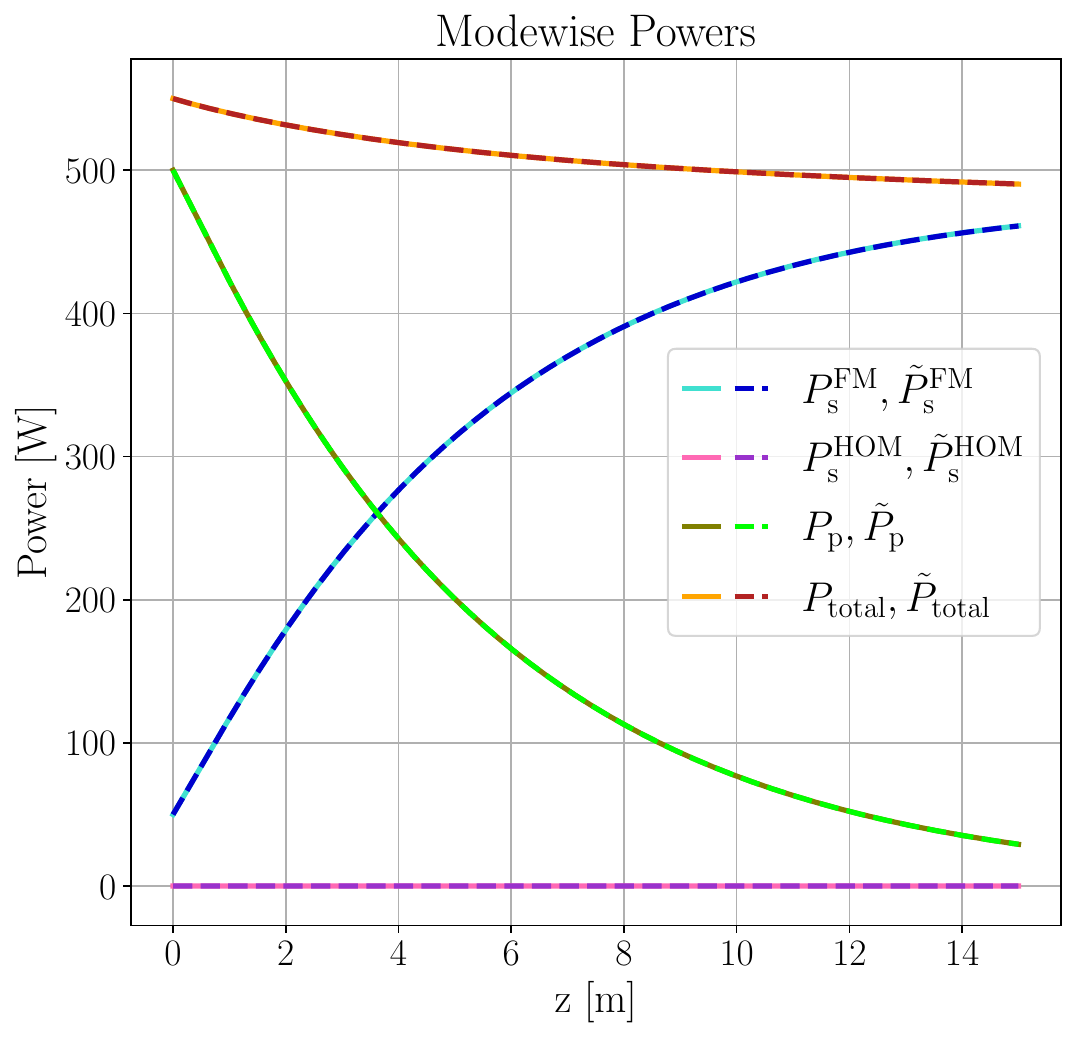}
        \caption{Power: $P = P(z)$.}
        \label{fig:overPower}
    \end{subfigure}
    \hspace{0.75cm}
    \begin{subfigure}[b]{0.4\textwidth}
        \includegraphics[width=\linewidth]{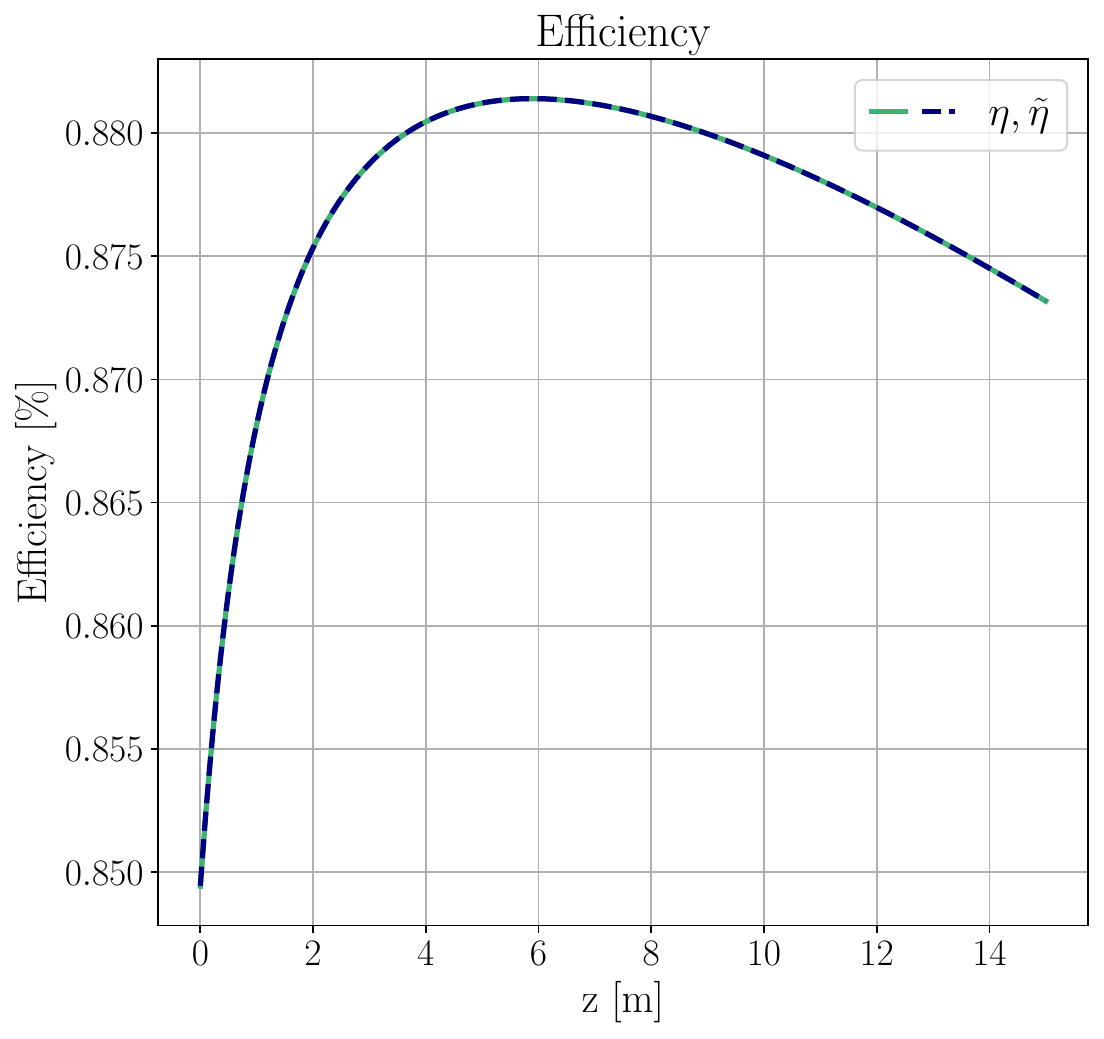}
        \caption{Amplification efficiency: $\eta = \eta(z)$.}
        \label{fig:overEff}
    \end{subfigure} \\
    \vspace{0.75cm}
    \begin{subfigure}[b]{0.4\textwidth}
        \includegraphics[width=\linewidth]{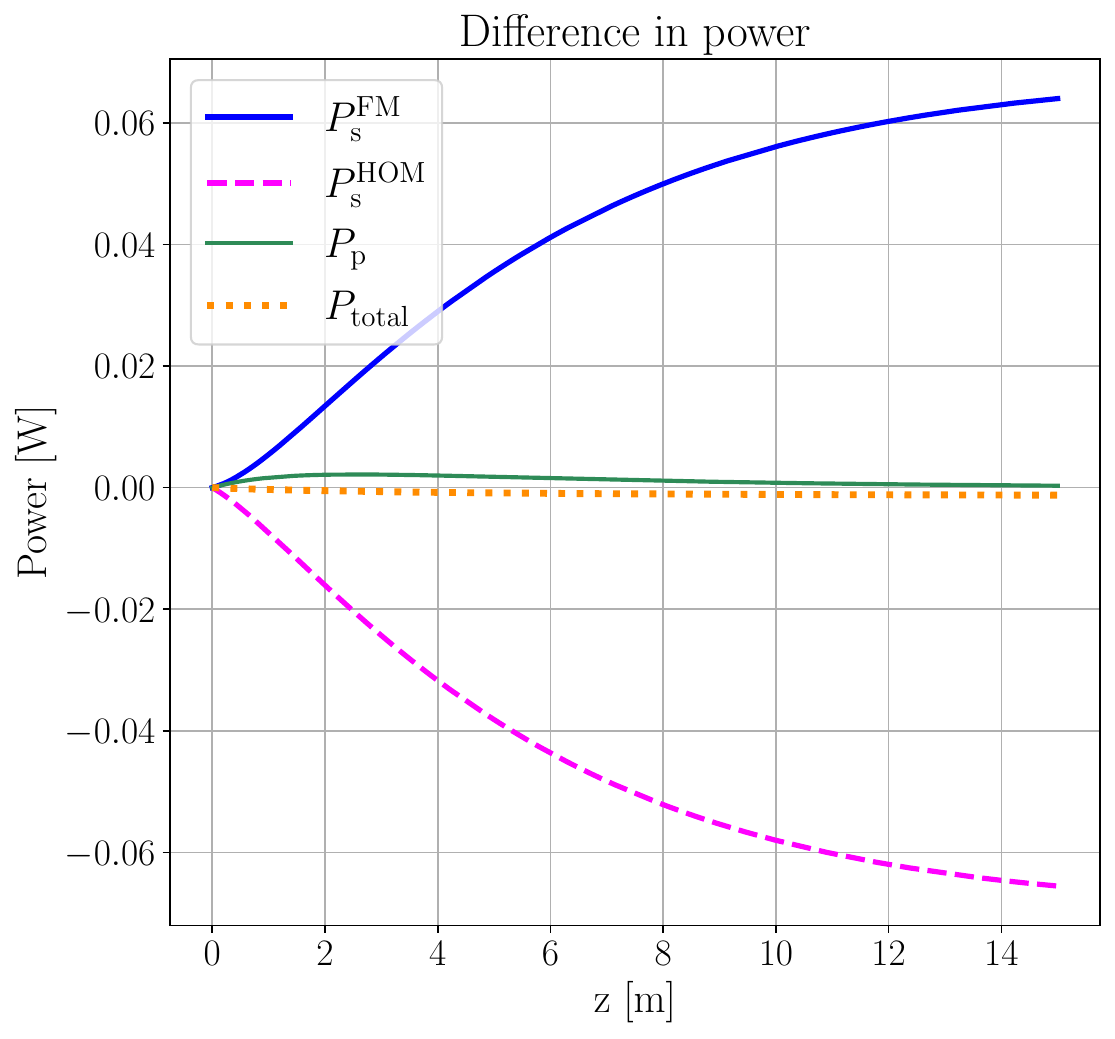}
        \caption{$P_{\text{CMT}}(z) - P_{\text{ACM}}(z)$.}
        \label{fig:powerDiff}
    \end{subfigure}
    \hspace{0.75cm}
    \begin{subfigure}[b]{0.38\textwidth}
        \includegraphics[width=\linewidth]{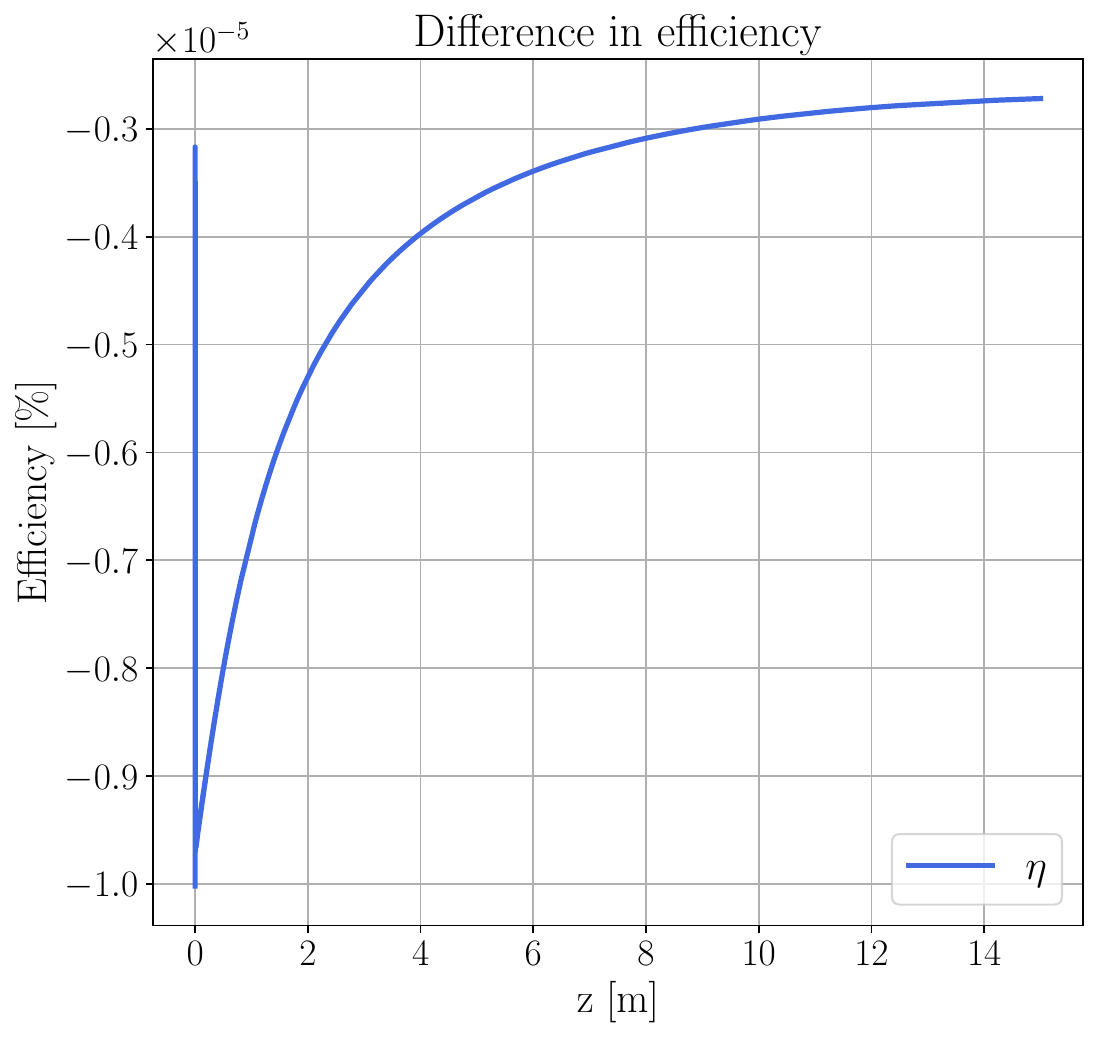}
        \caption{$\eta_{\text{CMT}}(z) - \eta_{\text{ACM}}(z)$.}
        \label{fig:EffDiff}
    \end{subfigure}
    \caption{The CMT (solid lines) and ACM (dashed lines) model results for power and amplification efficiency levels along the amplifier length.}
    \label{fig:two_models}
\end{figure}

Injecting $99.99\%$ of the seed power into the FM and $0.01\%$ into the HOM, Fig.~\ref{fig:two_models} shows the power evolution (plot~\subref{fig:overPower}), and the amplification efficiency evolution (plot~\subref{fig:overEff}), along the amplifier length for the solutions to the CMT (represented by solid lines) and ACM (represented by dashed lines) models. 
The use of solid lines for the full CMT model, and the use dashed lines for the ACM model, will be maintained throughout the communication of this section. 
The fact that the power levels for the pump fields and individual signal modes (plot~\subref{fig:overPower}), and the fact that the amplification efficiency levels (plot~\subref{fig:overEff}), are visually indistinguishable between the two models are strong indications that there is an exceptionally low error between the two models' solutions. 
This qualitative high-accuracy is better supported in plot~(\subref{fig:powerDiff}), which depicts the evolution of the differences between the two models' predicted power levels, and in plot~(\subref{fig:EffDiff}), which depicts the difference between these two models' predicted amplification efficiencies. 
Note that the ACM model overpredicts the HOM power by nearly the same amount that it underpredicts the FM power. 
This is somewhat surprising, considering that the majority of the power is seeded into the FM; so relatively speaking, the error in the HOM power is considerably greater than that of the FM. 
Regardless, the relative error in the pump and total signal field is exceptionally low, less than $0.002\%$ for either field. 
This small error in the pump and signal fields contributes to the high precision of the amplification efficiency observed in plot~(\subref{fig:EffDiff}). 
Since the ACM model simulates amplification efficiency so accurately, the authors are hopeful that the ACM model will also accurately simulate time-dependent thermal effects in future work. 
This idea is further supported by steady-state temperature results presented later in Section~\ref{subsec:Q_and_steady_state_T}, specifically in Figs.~\ref{fig:heat_temp_full_Q} and~\ref{fig:heat_temp_Taylor_Q}. 
\FloatBarrier
\begin{figure} %[H]
    \centering
    \begin{subfigure}[b]{0.49\textwidth}
        \centering
        \includegraphics[width = \linewidth]{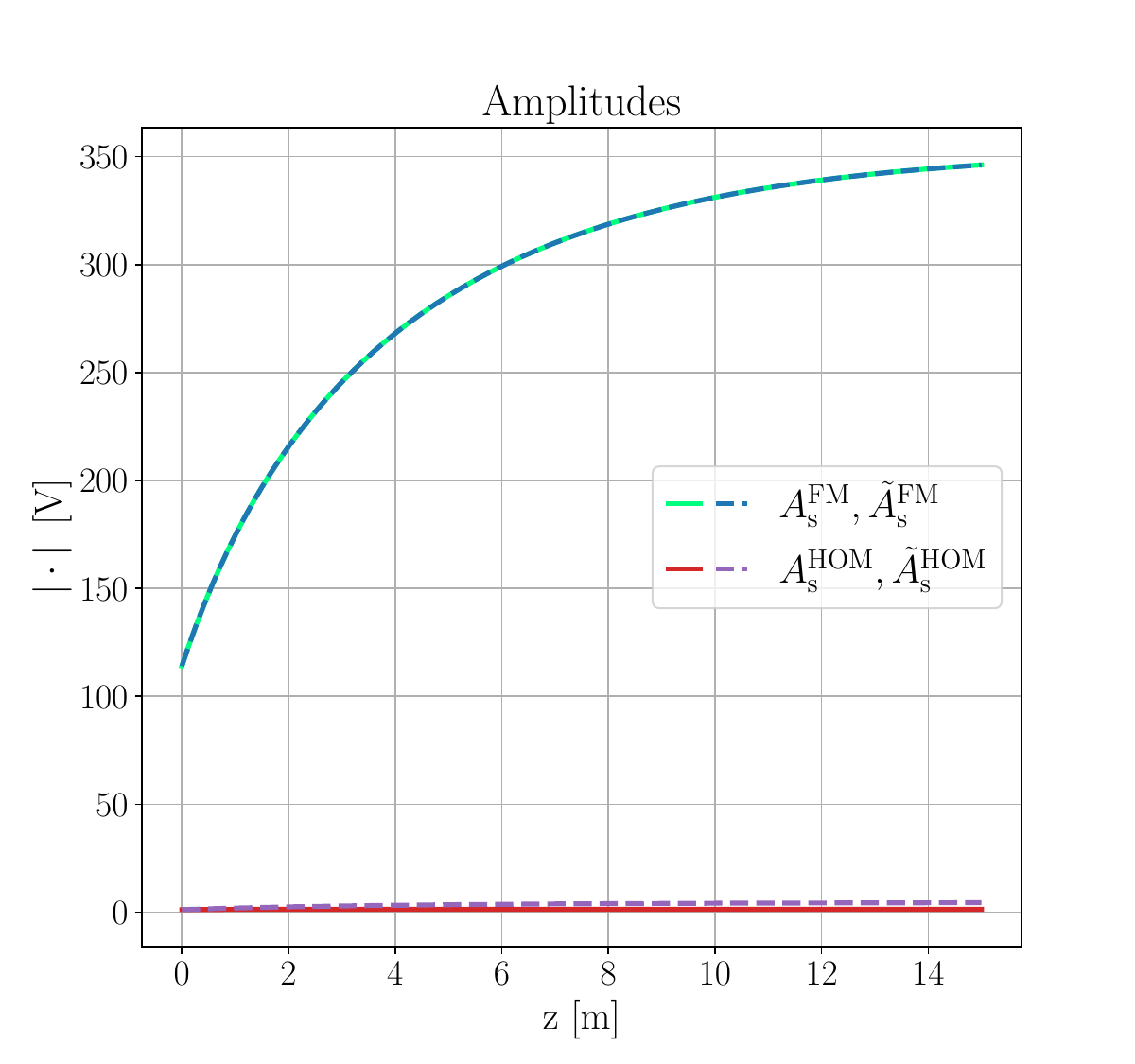}
        \caption{Magnitude of the signal mode amplitudes. \\ \centering \textcolor{white}{$\left| \hspace{2pt} \fft{z}{k}{A = A(z)} \right|$} }
        \label{fig:AMPS}
    \end{subfigure}
    \hfill
    \begin{subfigure}[b]{0.49\textwidth}
        \centering 
        \includegraphics[width = \linewidth]{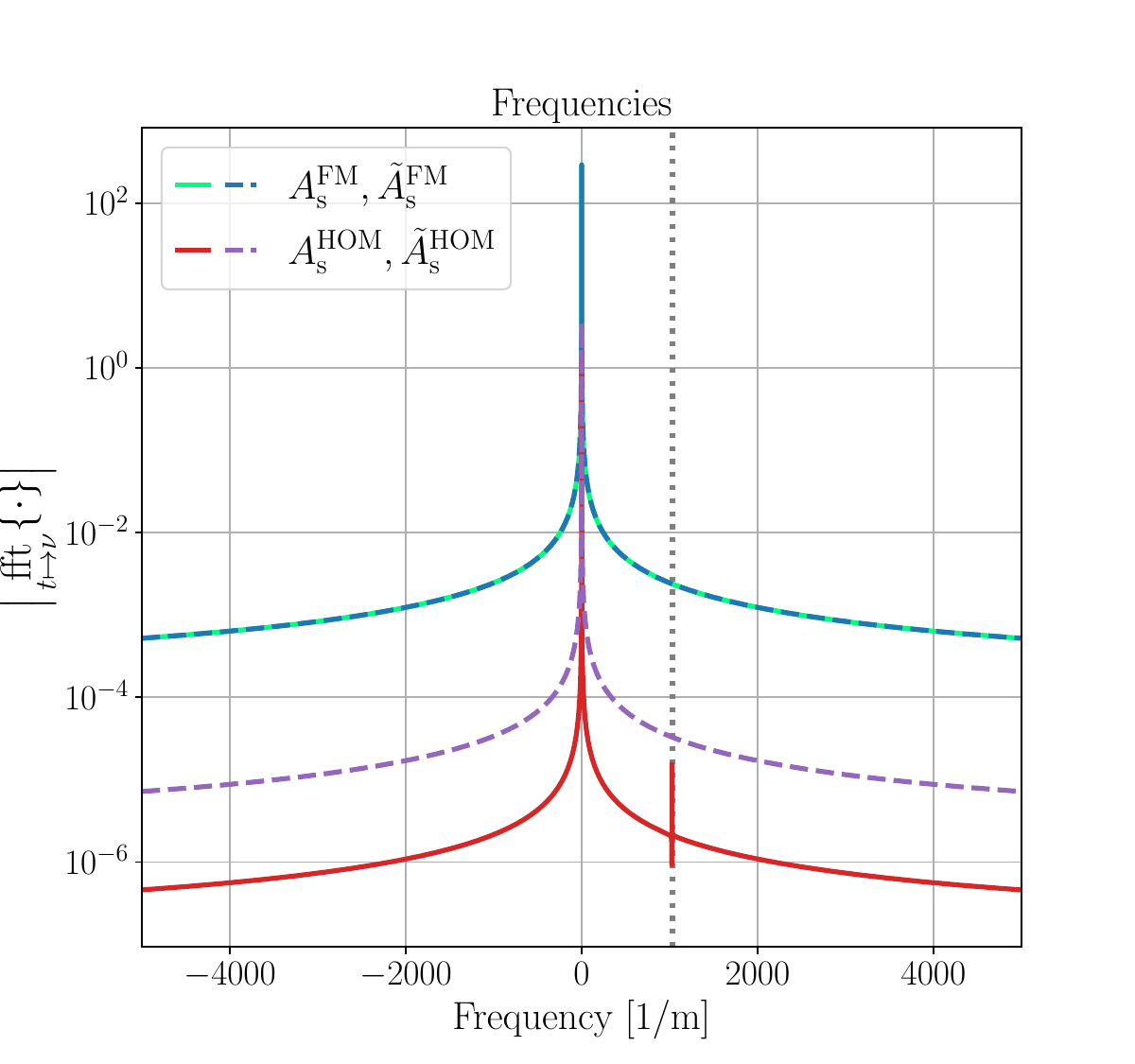}
        \caption{Signal mode amplitude spectra: \\ $\left| \hspace{2pt} \fft{z}{k}{A = A(z)} \right|$.}
        \label{fig:FREQS}
    \end{subfigure}
    \caption{The CMT (solid lines) and ACM (dashed lines) model mode amplitude results.}
    \label{fig:amplitudes}
\end{figure}
Although power is a physically relevant figure-of-merit to assess the amplifier performance, it does not quite represent the model performance. 
The true error in the ACM model accumulates in the complex-valued amplitudes (i.e. the dependent variable). 
Since power is proportional to square of the amplitude's magnitude, not surprisingly plot~(\subref{fig:AMPS}) of Fig.~\ref{fig:amplitudes} illustrates that the two model predictions for $|A_{\sig}^{\text{FM/HOM}}(z)|$ are nearly identical, just as they were for the power levels (cf. plot~(\subref{fig:overPower}) of Fig.~\ref{fig:two_models}). 
Similar to the averaging example in Section~\ref{subsec:avg-ex}, evidence from plot~(\subref{fig:FREQS}) shows that averaging removes negligible high-frequency information from the ACM amplitude solutions, in this case, specifically filtering at $k = {\Delta}\beta_{\sig} / \pi$. 

\begin{figure} %[H] 
    \begin{subfigure}[b]{0.49\textwidth}
        \centering
        \includegraphics[width = \linewidth]{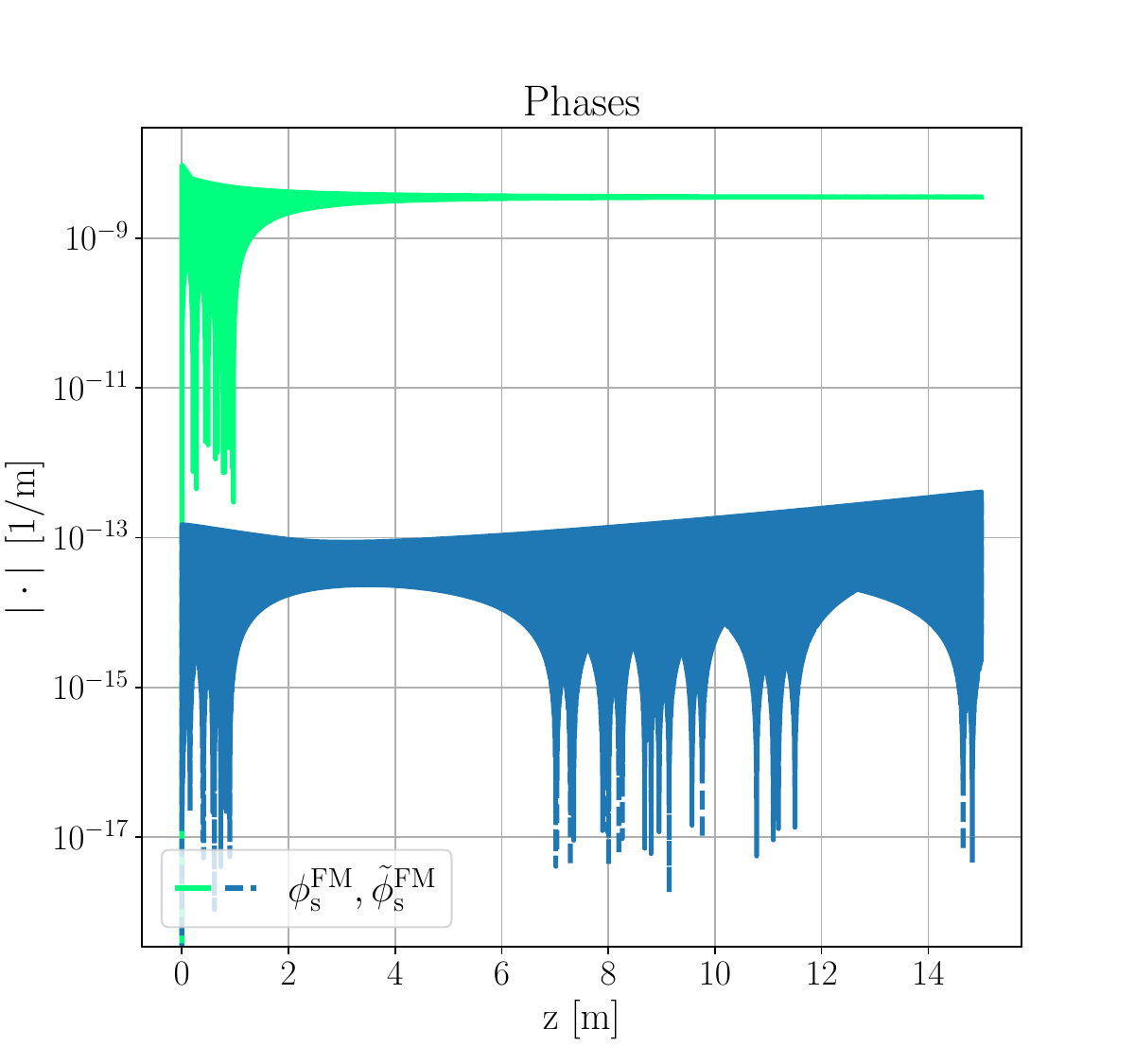}
        \caption{$\FM$ phase: $\big| \phi^{\FM}_{\sig} \big| = \big| \phi^{\FM}_{\sig} (z) \big|$.}
        \label{fig:FM_phases}
    \end{subfigure}
    \hfill
    \begin{subfigure}[b]{0.49\textwidth}
        \centering
        \includegraphics[width = \linewidth]{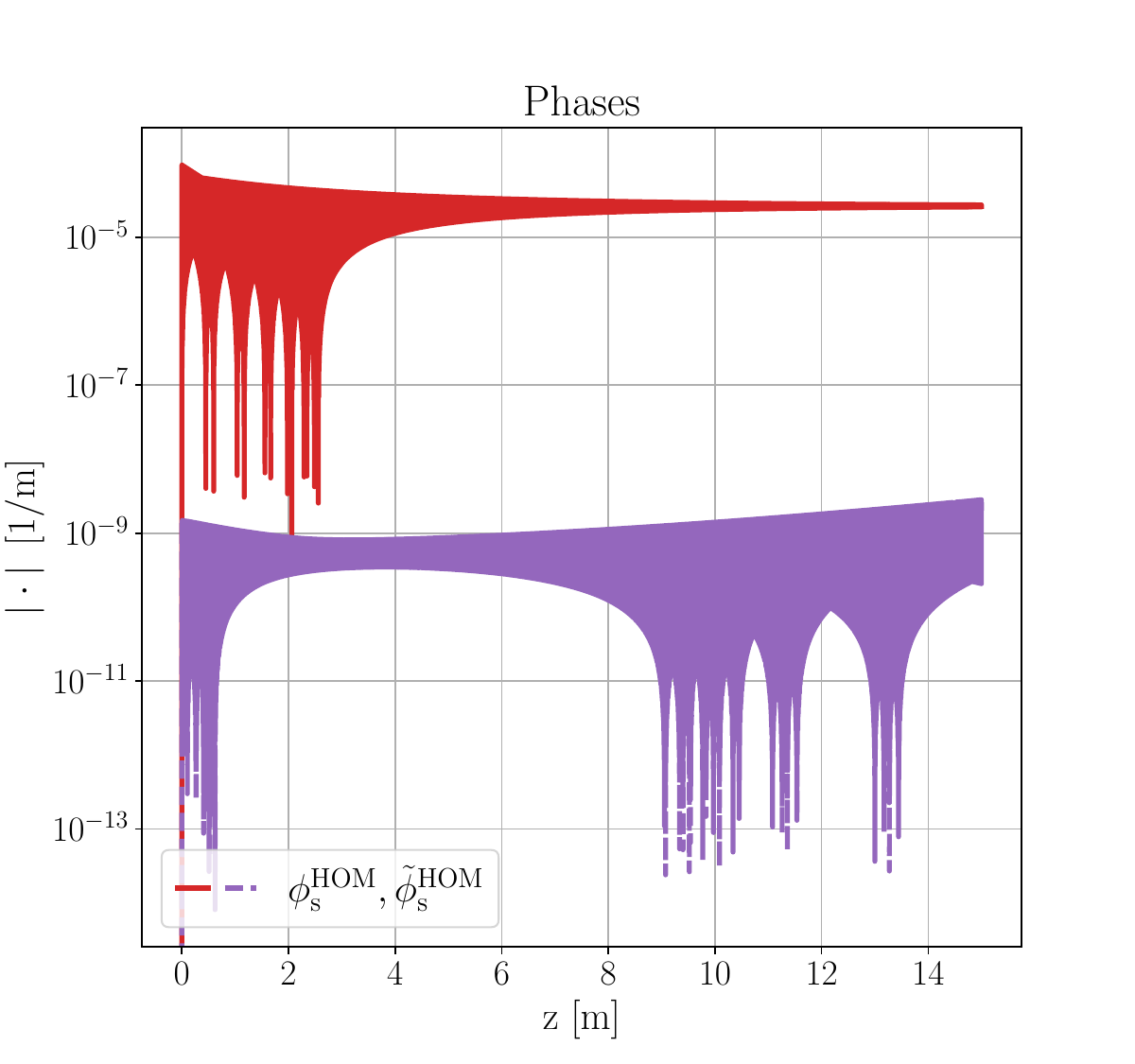}
        \caption{$\HOM$ phase: $\big| \phi^{\HOM}_{\sig} \big| = \big| \phi^{\HOM}_{\sig} (z) \big|$.}
        \label{fig:HOM_phases}
    \end{subfigure}
    \caption{The CMT (solid lines) and ACM (dashed lines) model results for mode amplitudes' magnitude of the phase along the amplifier length.}
    \label{fig:phase_info}
\end{figure}
One can isolate each amplitude's phase information by expressing these amplitudes in its phasor form: 
\begin{align}
    \mathcal{A}^{j}_{\text{s}}  = \left| \mathcal{A}^{j}_{\text{s}} \right| e^{i \Phi^{j}_{\text{s}} \left( \mathcal{A}^{j}_{\text{s}} \right)} \ , \text{ where } \Phi^{j}_{\text{s}}\left( \mathcal{A}^{j}_{\text{s}} \right) = \arctan\left( \frac{\text{Imag} \left( \mathcal{A}^{j}_{\text{s}} \right)}{\text{Real} \left( \mathcal{A}^{j}_{\text{s}} \right)} \right) \ \ ,
\end{align}
and where $\mathcal{A}^{j}_{\text{s}} \in \{A^{j}_{\text{s}}, \tilde{A}^{j}_{\text{s}} \}$ and $\Phi^{j}_{\text{s}} \in \{ \phi^{j}_{\text{s}} , \tilde{\phi }^{j}_{\text{s}} \}$ for $j \in \{ \FM, \HOM \}$. 
Note that, in practice, the $\arctan(\cdot)$ function wraps the phase between $-\pi$ and $\pi$. 
Figure~\ref{fig:phase_info} shows that there are phase differences between the CMT and ACM models that span several orders of magnitude, but are still overall quite small. Specifically, the phase error accounts for less than 1\% of the error in the amplitude. 
Note that the phase mixing between the two signal modes occurs in the cross-coupling factors ($\big\{ \kappa_{\sig}^{\text{FM,HOM}} \exp(-i {\Delta}\beta_{\sig} z), \kappa_{\sig}^{\text{HOM,FM}} \exp(i {\Delta}\beta_{\sig} z) \big\}$ within relation~\eqref{eq:NondimensionalCMTModel} for the CMT model and $\big\{ \kappa_{\sig +}^{\text{FM,HOM}}, \kappa_{\sig -}^{\text{HOM,FM}} \big\}$ within relation~\eqref{eq:NondimensionalACMModel} for the ACM model), which are combating mode orthogonality in order to be non-zero -- this is why the mode phases are small in magnitude. 
This accumulation of phase error will likely have a greater impact on the ACM model's performance when phase-dependent nonlinear effects are added to the model (in future work). 

\begin{figure}
    \centering
    \includegraphics[width=0.5\linewidth]{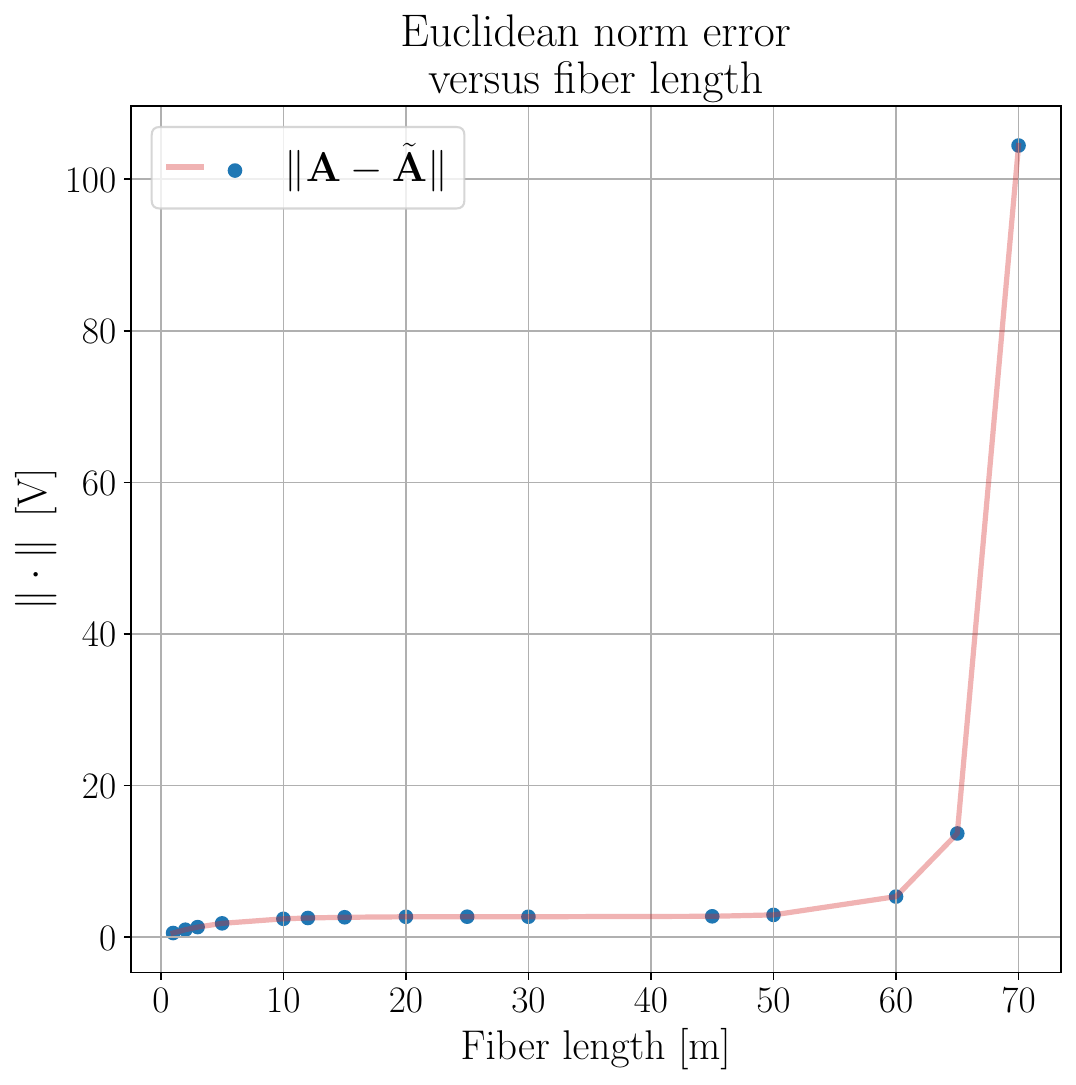}
    \caption{The error~\eqref{A_L_error} between the CMT and ACM model solutions as a function of amplifier length.}
    \label{fig:L_against_error}
\end{figure}

One may also observe the span over which averaging holds for this amplifier problem. 
The span, $S$, from Theorem~\ref{thm:avg-Verl}, which dictates the length over which the averaged model remains accurate to its original counterpart, can be ascertained by imposing an acceptable error tolerance, as was demonstrated for the example problem in Section~\ref{subsec:avg-ex}. 
Figure~\ref{fig:L_against_error} depicts the maximum Euclidean norm error between the solutions $\mbf A$ and $\mbt A$ over the last mode beat length,
\begin{align}\label{A_L_error}
        \text{error}_{\mbf A, \mbt A} \left( L \right)  = \underset{z \in \left[ L - \frac{2 \pi}{\Delta \beta_{\text{s}}}, L \right]}{\max} \| \mbf A \left( z \right) - \mbt A \left( z \right) \| \ \ ,
\end{align}
for various fiber lengths. 
Considering that a typical Yb-doped fiber amplifier ranges from 5 to 20 m in length, the fact that error remains so low for fibers as long as 50 m, indicates that averaging is a reliable model reduction technique for this application. 
Note that even a well-calibrated power meter in an experiment can have about a $\pm 5\%$ uncertainty in its measurements, and this amplifier has an output power of about $P_{\sig}(L) \simeq 460$ W at $L = 15$ m, which corresponds to an uncertainty of ${\delta}P_{\sig}(L) \simeq {\pm}23$ W.

\FloatBarrier
%%%%%%%%%%%%%%%%%%%%%%%%%%%%%%%%%%% 
% Subsection: Steady-State Temperature Solution
%%%%%%%%%%%%%%%%%%%%%%%%%%%%%%%%%%%

\subsection{Steady-State Temperature Solution}\label{subsec:Q_and_steady_state_T}

Another two quantities-of-interest that can be extracted from a fiber laser amplifier model are the heat generation ($Q$) and the corresponding steady-state change in temperature ($\dT$). 
The heat generation in the fiber is computed at each discrete point along the fiber length:  
\begin{align}\label{eq:heat_dep}
    Q(x, y; z) & = -\sum_{\ell}\left( \bar{g}_{\ell}(x, y; z) I_{\ell}(x, y; z) \right) \ \ \Bigg[\frac{\text{W}}{\text{m}^{3}}\Bigg]
\end{align}
and it drives a change in temperature, $\dT \ \left[ \mbox{}^{\circ}\text{C} \right]$, within the fiber: 
\begin{align}\label{eq:steady_temp}
    -\kappa_{\text{thermal}} {\Delta}_{xy} \dT(x, y; z) & = Q(x, y; z) \ \ ,
\end{align}
where $\kappa_{\text{thermal}}$ [W/(m$\cdot$K)] is the thermal conductivity of glass and $\Delta_{xy} = {\p}^{2}/{\p}x^{2} + {\p}^{2}/{\p}y^{2}$ denotes the transverse Laplacian operator. 

Let $Q$, as defined in~\eqref{eq:heat_dep}, denote the heat generation extracted from the CMT model and $\tilde{Q}$ denote the heat generation computed by ACM model. 
A notable difference between $Q$ and $\tilde{Q}$ is that $\tilde{Q}$ uses the first-order Taylor expansion on the gain functions, i.e. 
\begin{align}\label{eq:avg_heat_dep}
    \tilde{Q} & \defeq -\left[ \left( \bar{g}_{{\sig}0} +  \bar{g}_{{\sig}+} e^{i  {\Delta}\beta_{\sig}  z} +  \bar{g}_{{\sig}-} e^{-i  {\Delta}\beta_{\sig}  z} \right) I_{{\sig}} + \left(  \bar{g}_{{\pmp}0} +  \bar{g}_{{\pmp}+} e^{i {\Delta}\beta_{\sig} z} +  \bar{g}_{{\pmp}-} e^{-i {\Delta}\beta_{\sig} z} \right) I_{{\pmp}} \right] \ \  .
\end{align}
The ACM model's heat generation drives to its own steady-state temperature, $\widetilde{\dT}$, that obeys 
\begin{align}\label{eq:avg_steady_temp}
    -\kappa_{\text{thermal}} {\Delta}_{xy} \widetilde{\dT}(x, y; z) & = \tilde{Q}(x, y; z) \ \ .
\end{align}

\begin{figure}
    \begin{subfigure}[b!]{0.39\textwidth}
        \centering
        \includegraphics[width=\linewidth]{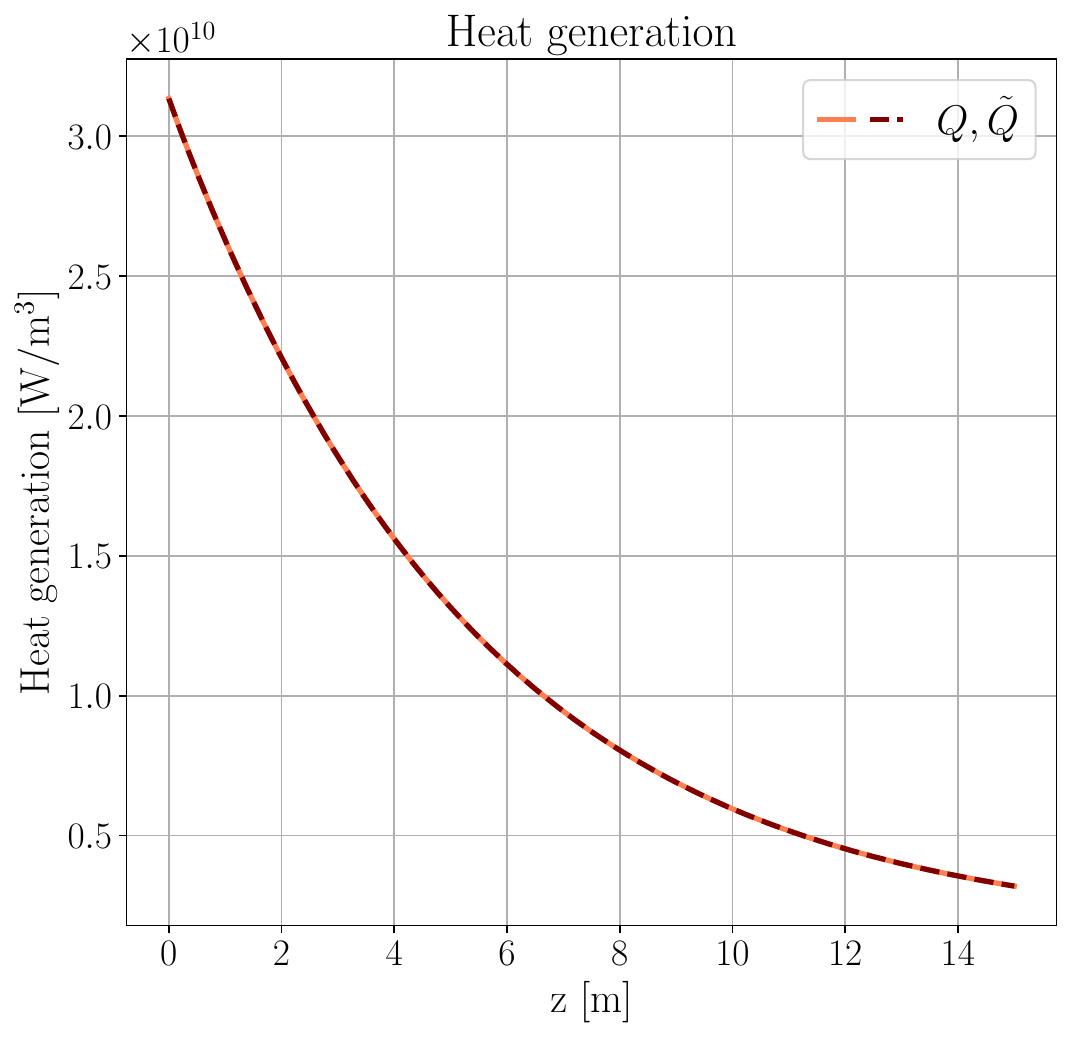}
        \caption{Heat generation: $Q = Q(z)$.}
        \label{fig:heatLoad}
    \end{subfigure}
    \hspace{0.75cm}
    \begin{subfigure}[b!]{0.4\textwidth}
        \centering
        \includegraphics[width=\linewidth]{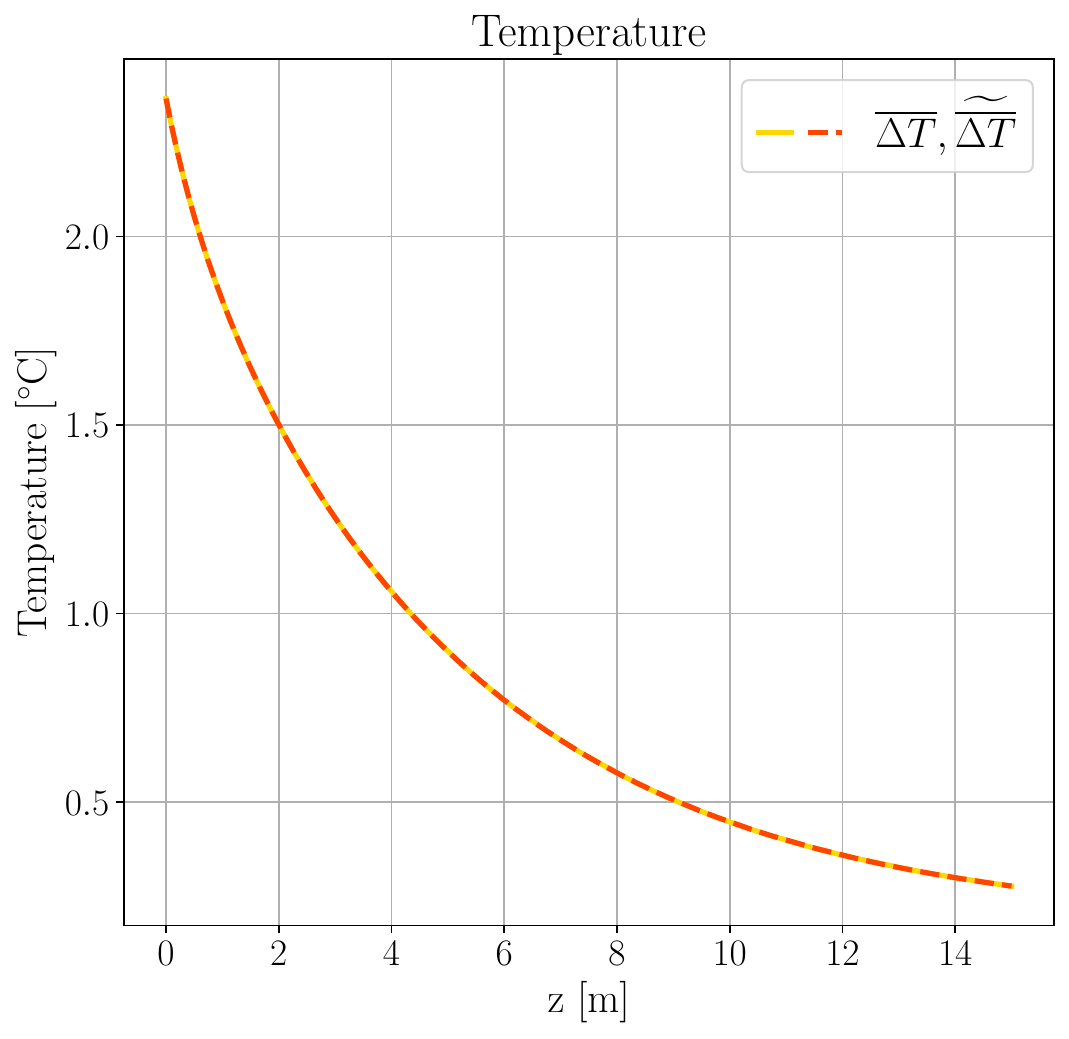}
        \caption{Change in temperature: $\dT = \dT(z)$.}
        \label{fig:temp}
    \end{subfigure} \\
    \vspace{0.75cm}
    \begin{subfigure}[b!]{0.405\textwidth}
        \centering
        \includegraphics[width=\linewidth]{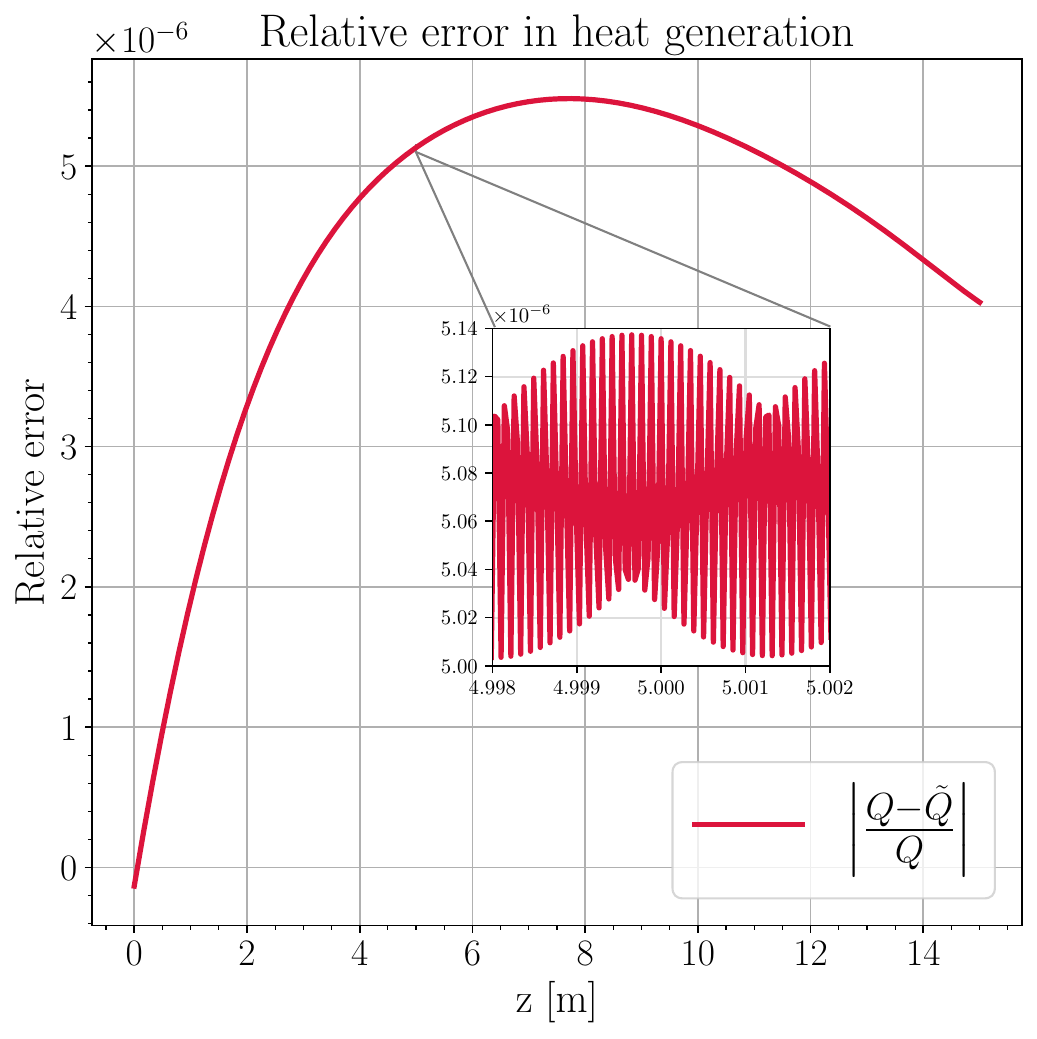}
        \caption{Relative error: $\left|(Q - \tilde{Q}) / Q \right|$.}
        \label{fig:heatLoadError}
    \end{subfigure}
    \hspace{0.75cm}
    \begin{subfigure}[b!]{0.39\textwidth}
        \centering
        \includegraphics[width=\linewidth]{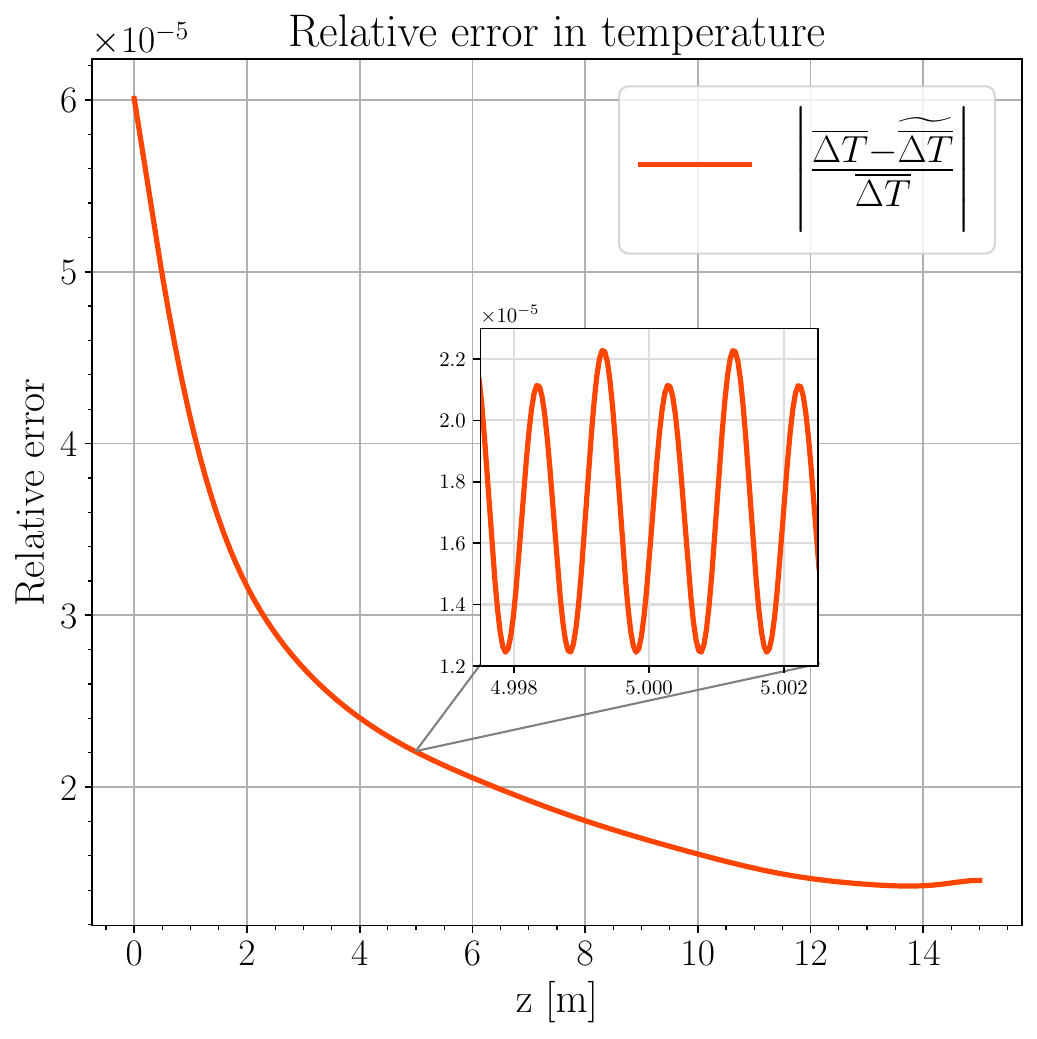}
        \caption{Relative error: $\left| (\dT - \widetilde{\dT}) / \dT \right|$.}
        \label{fig:tempError}
    \end{subfigure}
    \caption{The CMT and ACM model predictions along the fiber centerline $(x,y) = (0,0) $ for the heat generation~\eqref{eq:heat_dep} is seen in plot~(\subref{fig:heatLoad}), and for the steady-state change in temperature~\eqref{eq:steady_temp} is depicted in plot~(\subref{fig:temp}). The corresponding relative errors are in plot~(\subref{fig:heatLoadError}) for the heat generation, and are in plot~(\subref{fig:tempError}) for the change in temperatures.}
    \label{fig:heat_temp_full_Q}
\end{figure}

The close agreement between the CMT and ACM predicted heat generation quantities are portrayed in plots~(\subref{fig:heatLoad}) and~(\subref{fig:heatLoadError}) of Fig.~\ref{fig:heat_temp_full_Q}. 
Plot~(\subref{fig:heatLoadError}) of Fig.~\ref{fig:heat_temp_full_Q} demonstrates that the relative heat generation error along the fiber (given as a fraction) remains $\mathcal{O}(10^{-4}\%)$ and indicates that the ACM model is maintaining a high degree of accuracy compared to the CMT model. 
Additionally, after solving the heat equations for both $Q$~\eqref{eq:steady_temp} and $\tilde{Q}$~\eqref{eq:avg_steady_temp} and comparing the results, the relative error between the two models' steady-state change in temperature solutions remain less than $0.006\%$ along the fiber, as portrayed in plot~(\subref{fig:tempError}) of Fig.~\ref{fig:heat_temp_full_Q}.

\begin{figure}
    \begin{subfigure}[b!]{0.39\textwidth}
        \centering
        \includegraphics[width=\linewidth]{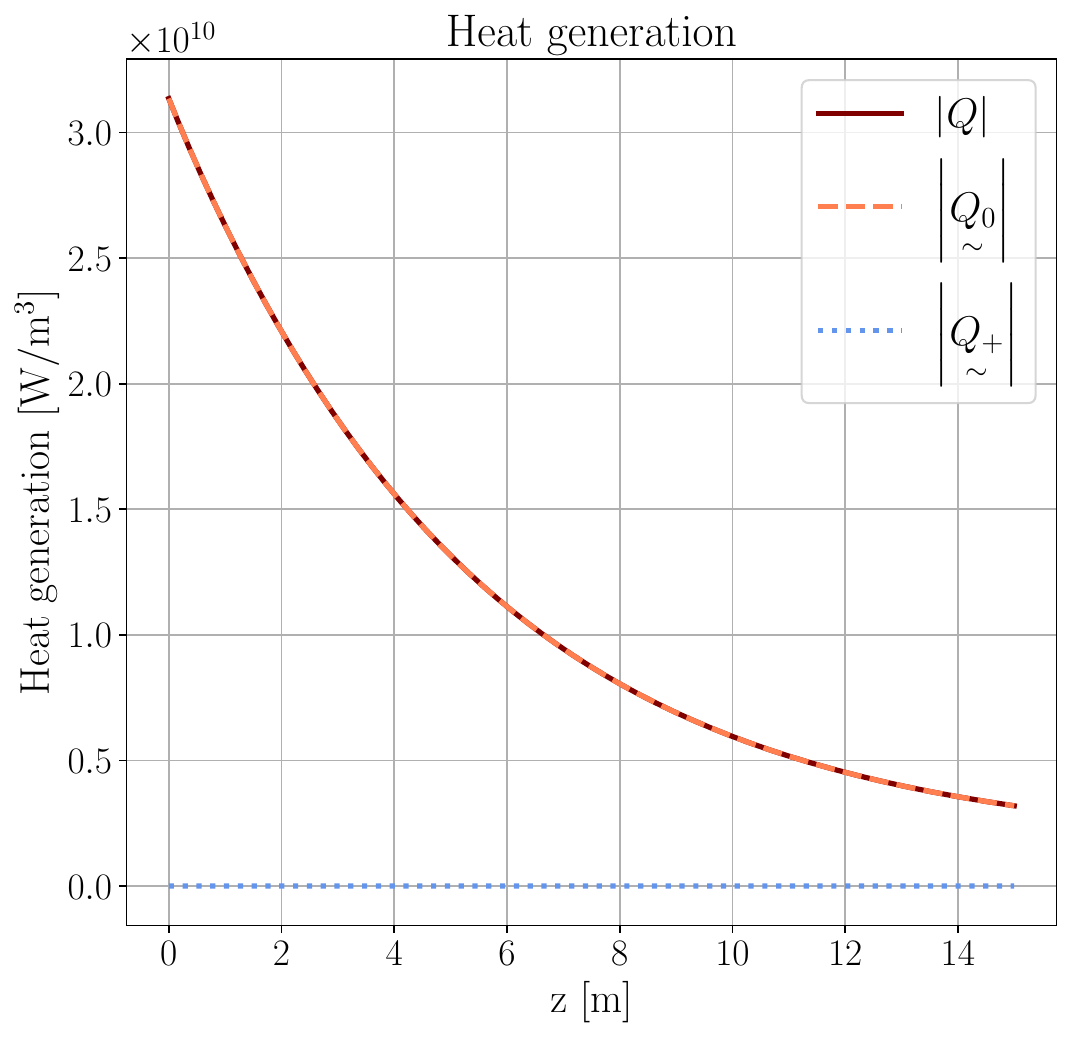}
        \caption{Heat generation: $Q = Q(z)$.}
        \label{fig:heatLoadcomps}
    \end{subfigure}
    \hspace{0.75cm}
    \begin{subfigure}[b!]{0.4\textwidth}
        \centering
        \includegraphics[width=\linewidth]{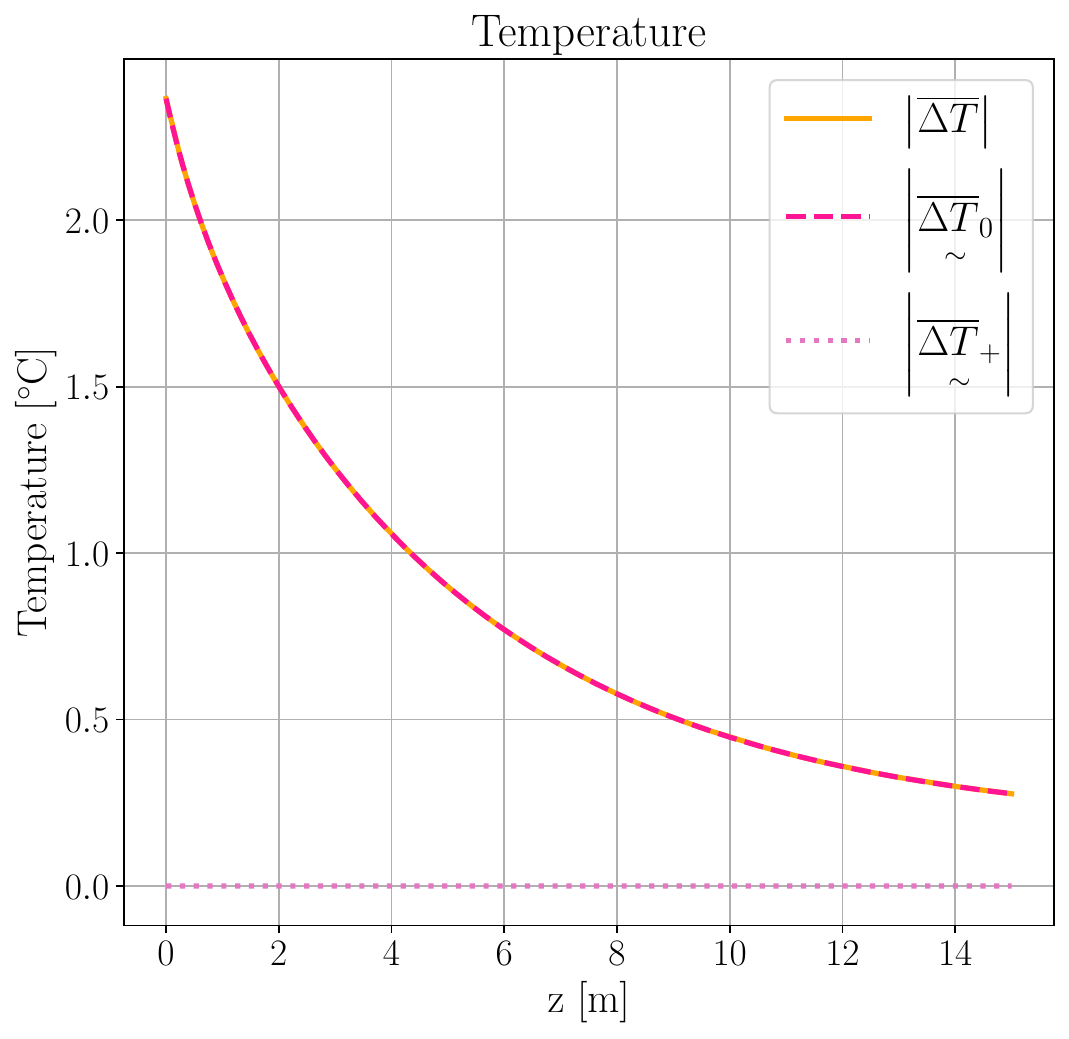}
        \caption{Change in temperature: $\dT = \dT(z)$.}
        \label{fig:tempcomps}
    \end{subfigure} \\
    \vspace{0.75cm}
    \begin{subfigure}[b!]{0.41\textwidth}
        \centering
        \includegraphics[width=\linewidth]{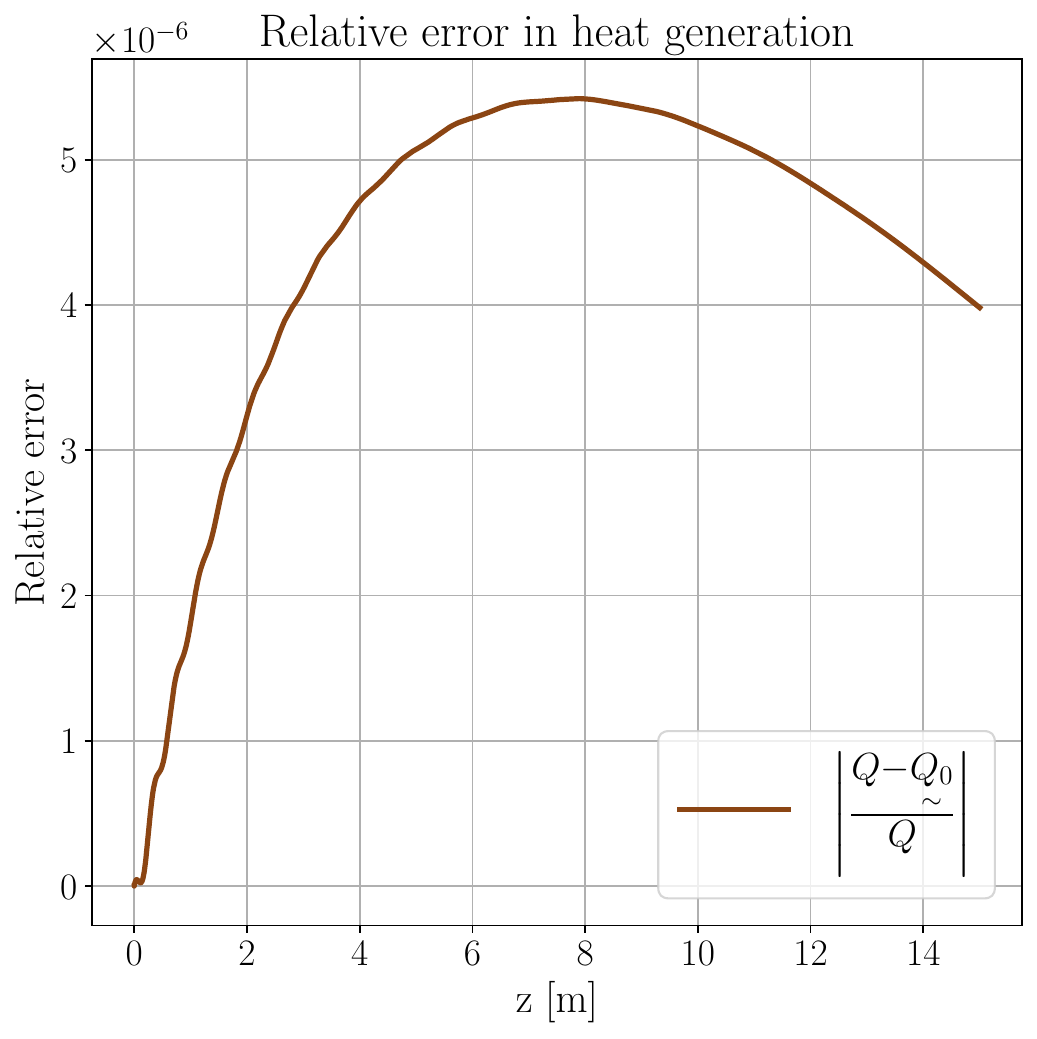}
        \caption{Relative error: $\left| (Q - \utilde{Q}_{0}) / Q \right|$.}
        \label{fig:heatLoadcompsError}
    \end{subfigure}
    \hspace{0.75cm}
    \begin{subfigure}[b!]{0.39\textwidth}
        \centering
        \includegraphics[width=\linewidth]{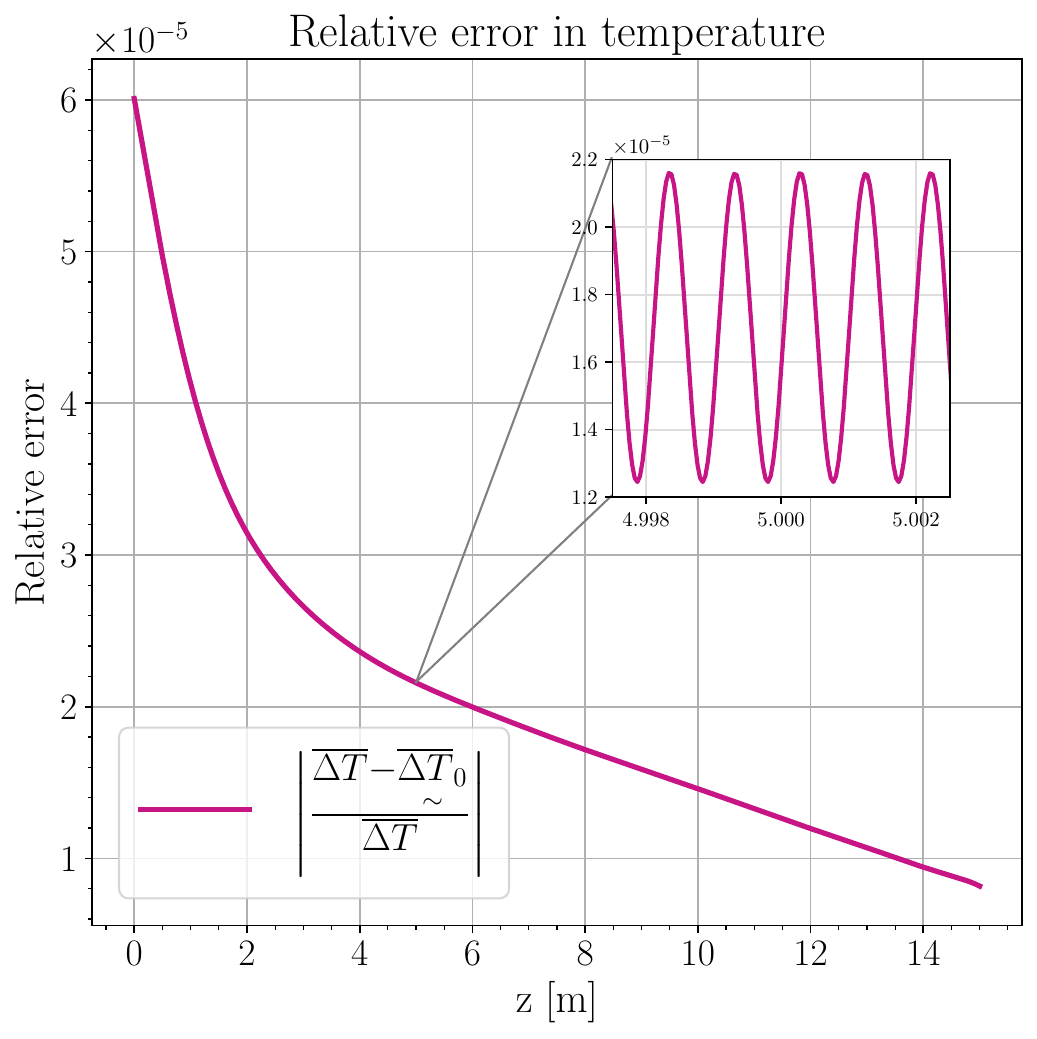}
        \caption{Relative error: $ \left|(\dT - \utilde{\dT}_{0}) / \dT \right| $.}
        \label{fig:tempcompsError}
    \end{subfigure}
    \caption{The CMT (solid lines) model's predicted heat generation~\eqref{eq:heat_dep} levels and steady-state change in temperatures~\eqref{eq:steady_temp}, and, likewise, the ACM (dashed lines) model's computed heat generation~\eqref{eq:Q-3-term-expansion} levels and steady-state change in temperatures~\eqref{eq:T-3-term-expansion} along the fiber centerline $(x,y) = (0,0) $.}
    \label{fig:heat_temp_Taylor_Q}
\end{figure}

Instead of using the first-order gain Taylor expansion \eqref{eq:TaylorExpandedGain} in the heat generation equation~\eqref{eq:heat_dep}, one may rather conduct a first-order Taylor expansion about $I_{\sig} = I_{\sig 0}$ on $Q$ directly: 
\begin{align}\label{eq:Q-Taylor}
    Q(\Ip, \Is) & \cong Q(\Ip, I_{{\sig}0}) + \pp{Q (\Ip, I_{{\sig} 0})}{\Is} \;\left( \Is - I_{{\sig}0} \right) \ \ .
\end{align}
The error in this approximation involves the second derivative of $Q$ with respect to $I_{\sig}$, which is
\begin{align*}
    \frac{\partial^{2} Q}{\partial I_{\sig}^{2}} & = \frac{\partial^{2} \bar{g}_{\sig} }{\partial I^{2}_{\sig}} +  \frac{\partial \bar{g}_{\sig} }{\partial I_{\sig}}  +  \frac{\partial^{2} \bar{g}_{\pmp} }{\partial I^{2}_{\sig}},
\end{align*}
and depends on the size of 
\begin{align*}
    \left| \frac{\partial^2 Q  }{\partial I^{2}_{{\sig}}}\big( I_{\sig} -  I_{{\sig} 0} \big)^{2} \right| & \leq \left| 
  \left( {I_{\sig}^2}  \frac{\partial^{2} \bar{g}_{\sig} }{\partial I^{2}_{\sig}} +  {I_{\sig}^2} \frac{\partial^{2} \bar{g}_{\pmp} }{\partial I^{2}_{\sig}}  \right) \left(\frac{ I_{{\sig}} - I_{{\sig}0}  }{ I_{\sig} }  \right)^{2} \right| + 
  \left| {I_{\sig}} \frac{\partial \bar{g}_{\sig} }{\partial I_{\sig}} \left(\frac{ \left( I_{{\sig}} - I_{{\sig}0} \right)^{2}  }{ I_{\sig} } \right) \right| \\
  & \leq \left| {I_{\sig}^2}  \frac{\partial^{2} \bar{g}_{\sig} }{\partial I^{2}_{\sig}} +  {I_{\sig}^2} \frac{\partial^{2} \bar{g}_{\pmp} }{\partial I^{2}_{\sig}}  \right| \delta + \Bigg| {I_{\sig}} \frac{\partial \bar{g}_{\sig} }{\partial I_{\sig}} \Bigg| \zeta \ \ ,
\end{align*}
where $\delta$ is from~\eqref{eq:Is-Iso} and $\zeta$ is from~\eqref{eq:zeta}. 
The other terms on the r.h.s.~of this inequality are bounded by the analysis in {\S}\ref{subsec:taylor_error}, ultimately showing that this Taylor expansion error is small. 
Substituting the three-term decomposition of $I_{{\sig}}$ from~\eqref{eq:CompactSignalIrradiance} into the r.h.s. of~\eqref{eq:Q-Taylor}, one obtains 
\begin{equation}\label{eq:Q-3-term-expansion}
    Q \cong Q_{0} + Q_{+} e^{i  \Delta \beta_{\sig}  z} + Q_{-} e^{-i  \Delta \beta_{\sig} z} \ \ ,
\end{equation}
where $Q_{0}$, $Q_{+}$, and $Q_{-}$ are defined by
\begin{align*}
    Q_{0} & = Q \left( I_{{\sig}0}, \Ip \right) \ , & 
    Q_{+} & = \frac{\partial Q \left( I_{{\sig}0}, \Ip \right)}{\partial I_{{\sig}}} \, I_{{\sig}+} \ , & \text{ and }
    Q_{-} & = \frac{\partial Q \left( I_{{\sig}0}, \Ip \right)}{\partial I_{{\sig}}} \, I_{{\sig}-} \ .
\end{align*}
We let $\utilde{Q}$ denote the ACM model's evaluation of the heat generation first-order Taylor expansion~\eqref{eq:Q-Taylor}. 
Even for this alternative approach, $\utilde{Q}_{0}$ well-approximates $Q$, as is illustrated in Fig.~\ref{fig:heat_temp_Taylor_Q}. 
Specifically, plot~(\subref{fig:heatLoadcomps}) shows that $Q$ and $\utilde{Q}_{0}$ coincide nicely, with the minor difference of the periodic oscillations embedded within $Q$.
Plot~(\subref{fig:heatLoadcompsError}) displays that $\utilde{Q}_{0}$ has a maximum relative error of $5 \cdot 10^{-4}\%$ when compared to $Q$. 

The heat deposition $\utilde{Q}$ can also be used to predict a steady-state change in temperature by solving a similar heat equation to~\eqref{eq:avg_steady_temp}. 
Since $\utilde{Q}$ can be expressed in terms of $\utilde{Q}_{0}$ and $\utilde{Q}_{\pm}$, satisfying~\eqref{eq:Q-3-term-expansion}, then, by comparison of coefficients, heat equation can be split into 3 distinct relations such that 
\begin{align*}
    -\kappa_{\text{thermal}} \Delta_{xy} \utilde{\dT}_{0} & = \utilde{Q}_{0} \ , \\
    -\kappa_{\text{thermal}} \Delta_{xy} \utilde{\dT}_{+} & = \utilde{Q}_{+} \ , \text{ and} \\
    -\kappa_{\text{thermal}} \Delta_{xy} \utilde{\dT}_{-} & = \utilde{Q}_{-} \ ,
\end{align*}
where
\begin{equation}\label{eq:T-3-term-expansion}
    \utilde{\dT} = \utilde{\dT}_{0} + \utilde{\dT}_{+} e^{i {\Delta}\beta_{\sig} z} + \utilde{\dT}_{-} e^{-i {\Delta}\beta_{\sig} z} \ \ .
\end{equation}
Numerical simulations find that the DC component of the ACM model's prediction of steady-state change in temperature ($\utilde{\dT}_{0}$) matches exceptionally well with the CMT model's assessment ($\dT$), as is demonstrated in plots~(\subref{fig:tempcomps}) and (\subref{fig:tempcompsError}) of Fig.~\ref{fig:heat_temp_Taylor_Q}, resulting, again, in a relative error less than $0.006\%$. 

Whether one leverages the Taylor expansion of the steady-state gains to compute the heat load~\eqref{eq:avg_heat_dep} or uses a direct Taylor expansion on the heat generation formula~\eqref{eq:Q-Taylor}, there is a path to forcing the heat equation to explicitly express its dependence on the periodic $\exp\big( \pm i {\Delta}\beta_{\sig} z \big)$ factors, and yet still allow the ACM model to accurately estimate the change in temperatures throughout the amplifier. 
This means that, for other applications, there is good reason to believe that this integral averaging technique (Theorem~\ref{thm:avg-Verl}) can be successfully applied to the time dependent heating in the fiber amplifier, allowing for a more complete model that captures the ongoing thermal effects within an operating laser amplifier. 
Indeed, this is what the authors intend to do in their future work. 
Furthermore, by using the Taylor expansion approximations, as compared to the approximation within the ytterbium gain relations that was presented in Menyuk et al.'s methodology~\cite{menyuk2021accurate}, which allowed them to explicitly express the dependencies on $\exp\big( \pm i {\Delta}\beta_{\sig} z \big)$ in their governing relations, we propose that this model reduction approach can be applied to a wider variety of fiber laser amplifier types and configurations. 
For example, the Taylor expansion approach is indifferent to the active dopant of the amplifier, whereas Menyuk et al.'s approximation is only valid for the ytterbium gain relation. 
Also, by making a bivariate (or multivariate) Taylor expansion about $I_{\ell} = I_{{\ell} 0}$, for each core-guided laser field present in the fiber, one would be able to apply this model reduction technique to a two-tone (or multitoned/laser gain competition) amplifier configuration, as the authors intend to demonstrate in their future efforts. 

\FloatBarrier
%%%%%%%%%%%%%%%%%%%%%%%%%%%%%%%%%%% 
% Subsection: Points Per Mode Beat Length Study
%%%%%%%%%%%%%%%%%%%%%%%%%%%%%%%%%%%

\subsection{Points Per Mode Beat Length Study}\label{subsec:points_study}

\begin{figure} %[H]
    \centering
    \begin{subfigure}{0.48\textwidth}
        \includegraphics[width=\linewidth]{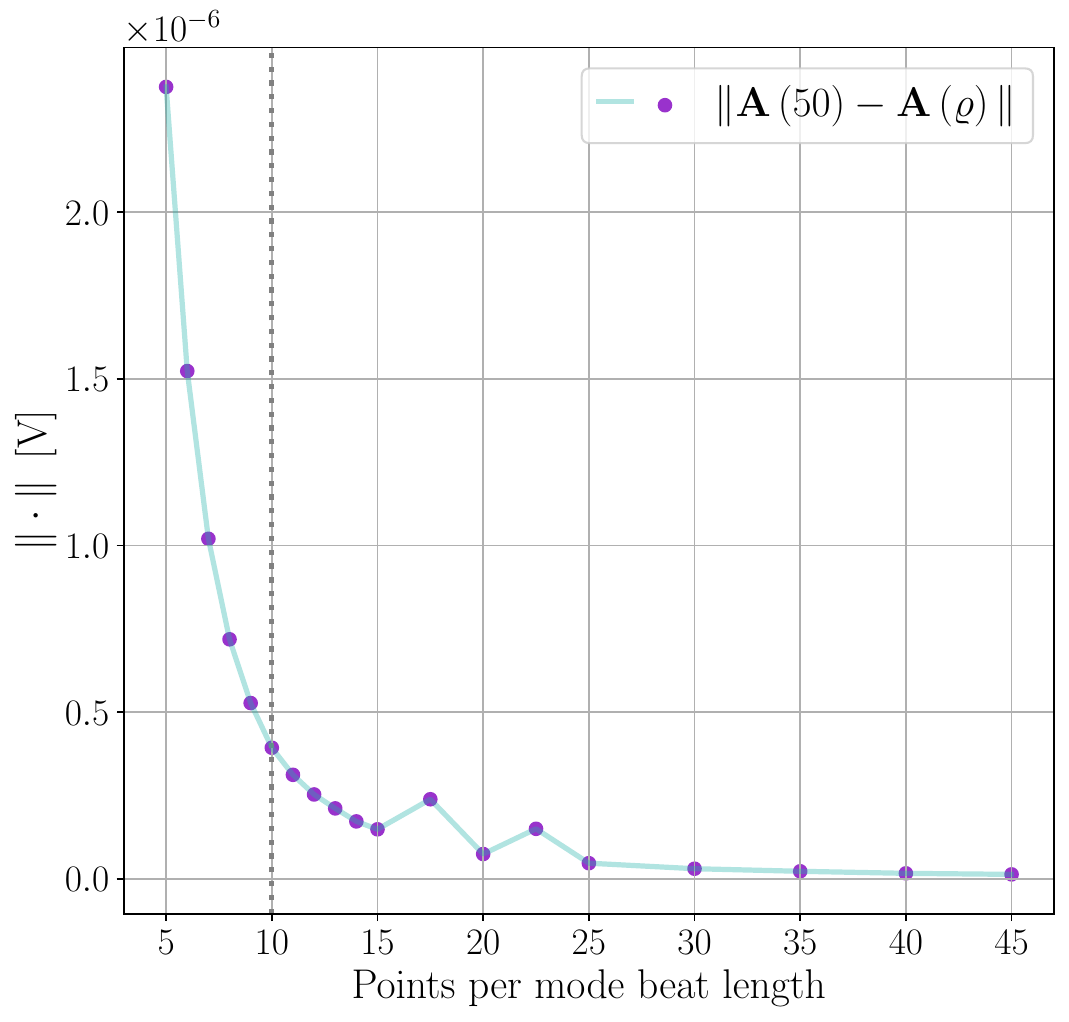}
        \caption{The error~\eqref{CMT_A_rho_error} in the CMT model.}
        \label{fig:CMTpb}
    \end{subfigure}
    \hfill
    \begin{subfigure}{0.48\textwidth}
        \includegraphics[width=\linewidth]{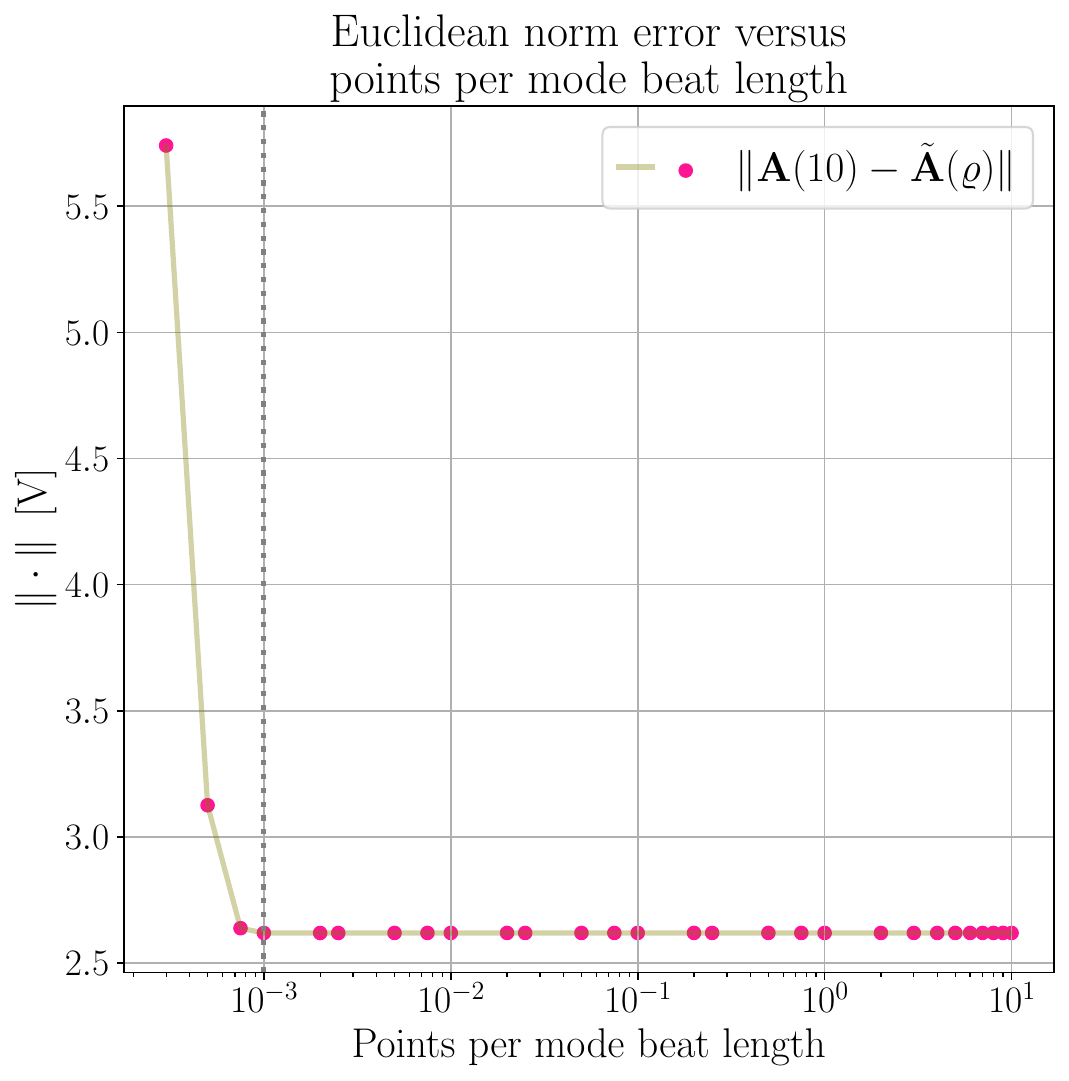}
        \caption{The error~\eqref{A_rho_error} in the ACM model.}
        \label{fig:ACMpb}
    \end{subfigure}
    \caption{The error in amplitude over the last mode beat length for varying values of $\varrho$.}
    \label{fig:ppbError}
\end{figure}

We conclude our studies on the numerical performance of the ACM model by quantifying the increase in this model's computational speed compared to that of the CMT model. 
The lack of high-frequency oscillations in the ACM model allows for a coarser model discretization in $z$. 
Thus far, both the CMT and ACM models have used the same discretization of 50 points per mode beat length, which is much finer than necessary to resolve the oscillations of the optical field. 
For the sake of finding a fair comparison, we first determine a reasonably low number of points per mode beat length to use in the CMT model before error accumulates. 
Let $\mbf A = \mbf A \left( \varrho \right)$ be the CMT model amplitude for $\varrho$ points per mode beat length. 
The model error is defined to be the maximum Euclidean norm error between $\mbf A \left( 50 \right)$ and $\mbf A \left( \varrho \right)$ over the last mode beat length, where $\varrho < 50$: 
\begin{align}\label{CMT_A_rho_error}
    \text{error}_{\mbf A} \left( \varrho \right) \defeq \underset{z \in \left[ L - \frac{2 \pi}{\Delta \beta_{\text{s}}}, L \right]}{\max} \| \mbf A \left( 50 \right) - \mbf A \left( \varrho \right) \|.
\end{align}
The solution with $\varrho$ points per mode beat length is linearly interpolated as if there is an equal total number of discrete points in each solution, and ensuring that one can find the error within the last mode beat length no matter how large ${\Delta}z$ becomes. 
For $\varrho = 10$, we find that $\text{error}_{\mbf A}\left( \varrho = 10 \right) \approx 3.932 \cdot 10^{-7} < 5 \cdot 10^{-7}$, and we use this case to set the default total number of discrete longitudinal points to be 77497, which is depicted by the vertical dotted line in plot~(\subref{fig:CMTpb}) of Fig.~\ref{fig:ppbError}. 
We chose $\varrho = 10$ to be the standard because one would expect that 10 discrete points per mode beat length ought to be sufficient to capture the mode beating oscillations that are expected to occur in the solution; and, indeed, our computational results (plot~\subref{fig:CMTpb}) indicate that the solution is still well-resolved for this discretization, but is also near the turning point at which any fewer points per mode beat length starts to rapidly accumulate error. The CMT model ODE system solved at 77497 discrete points in $z$ results in a runtime of approximately 2.571 seconds.

Next, we compute the error between the CMT model's solution using this 10 points per mode beat length and the ACM model's solution with decreasing values for $\varrho$: 
\begin{align}\label{A_rho_error}
    \text{error}_{\mbf A, \mbt A} \left( \varrho \right) \defeq \underset{z \in \left[ L - \frac{2 \pi}{\Delta \beta_{\text{s}}}, L \right]}{\max} \| \mbf A \left( 10 \right) - \mbt A \left( \varrho \right) \| \ \ .
\end{align}
Similarly, the ACM model's solution is linearly interpolated to have the same number of points per beat length as the CMT model solution. 
Varying the ACM model's number of points per mode beat length, $\varrho$, within $3 \cdot 10^{-4} \leq \varrho \leq 10$, results in the errors portrayed in plot~(\subref{fig:ACMpb}) of Fig.~\ref{fig:ppbError}. 
Decreasing the value of $\varrho$, the error remains small and stable until about $10^{-3}$ points per mode beat length, which corresponds to having a total of 9 discrete $z$-steps in the model, resulting in a runtime of approximately $6.690 \cdot 10^{-4}$ seconds. 
This strongly indicates that the ACM model can take much larger discrete step sizes while maintaining a satisfactory accuracy compared to the CMT model. 

\begin{figure} 
    \centering
    \begin{subfigure}{0.4\textwidth}
        \includegraphics[width=\linewidth]{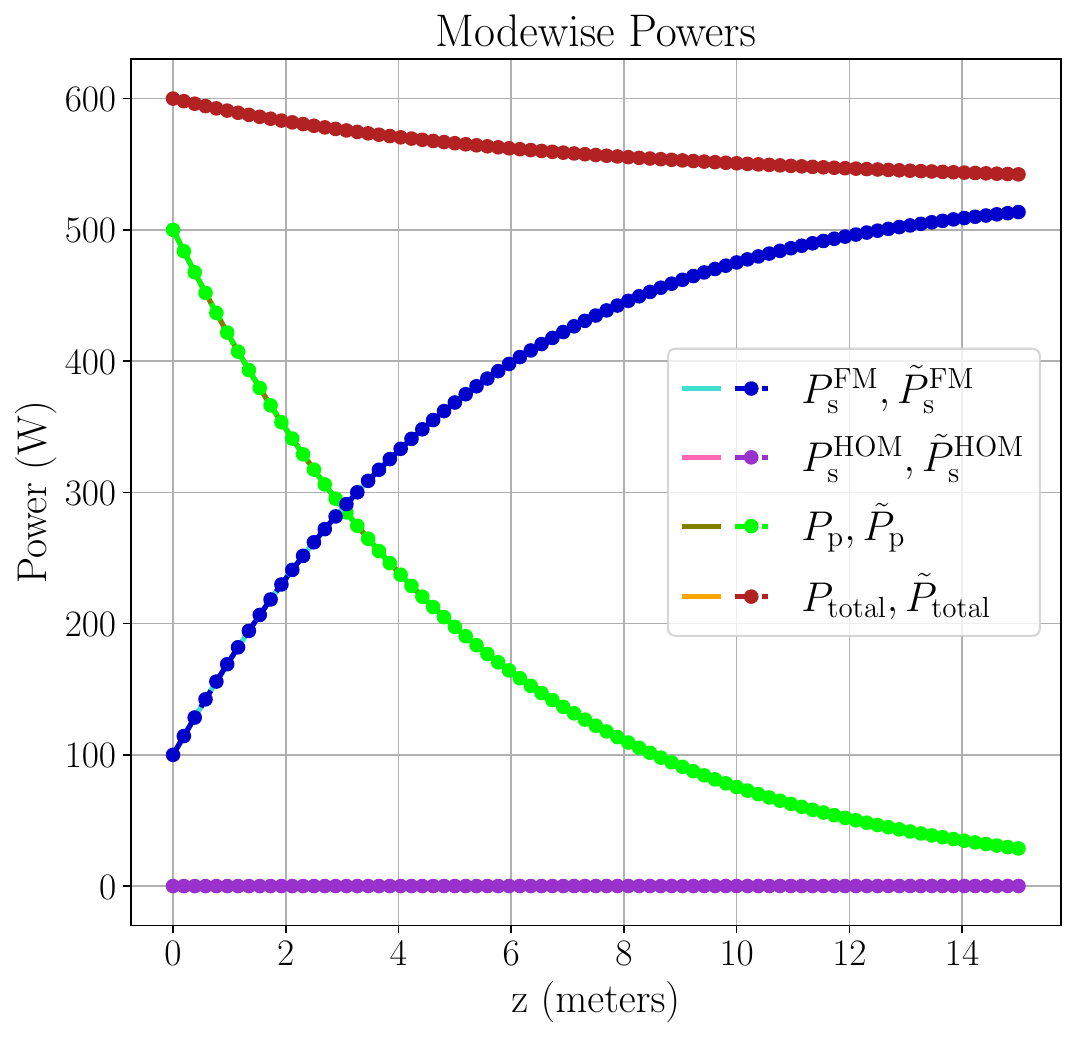}
        \caption{$\varrho = 0.01$ (79 $z$-steps).}
        \label{fig:pb0.01}
    \end{subfigure}
    \hspace{0.75cm}
    \begin{subfigure}{0.4\textwidth}
        \includegraphics[width=\linewidth]{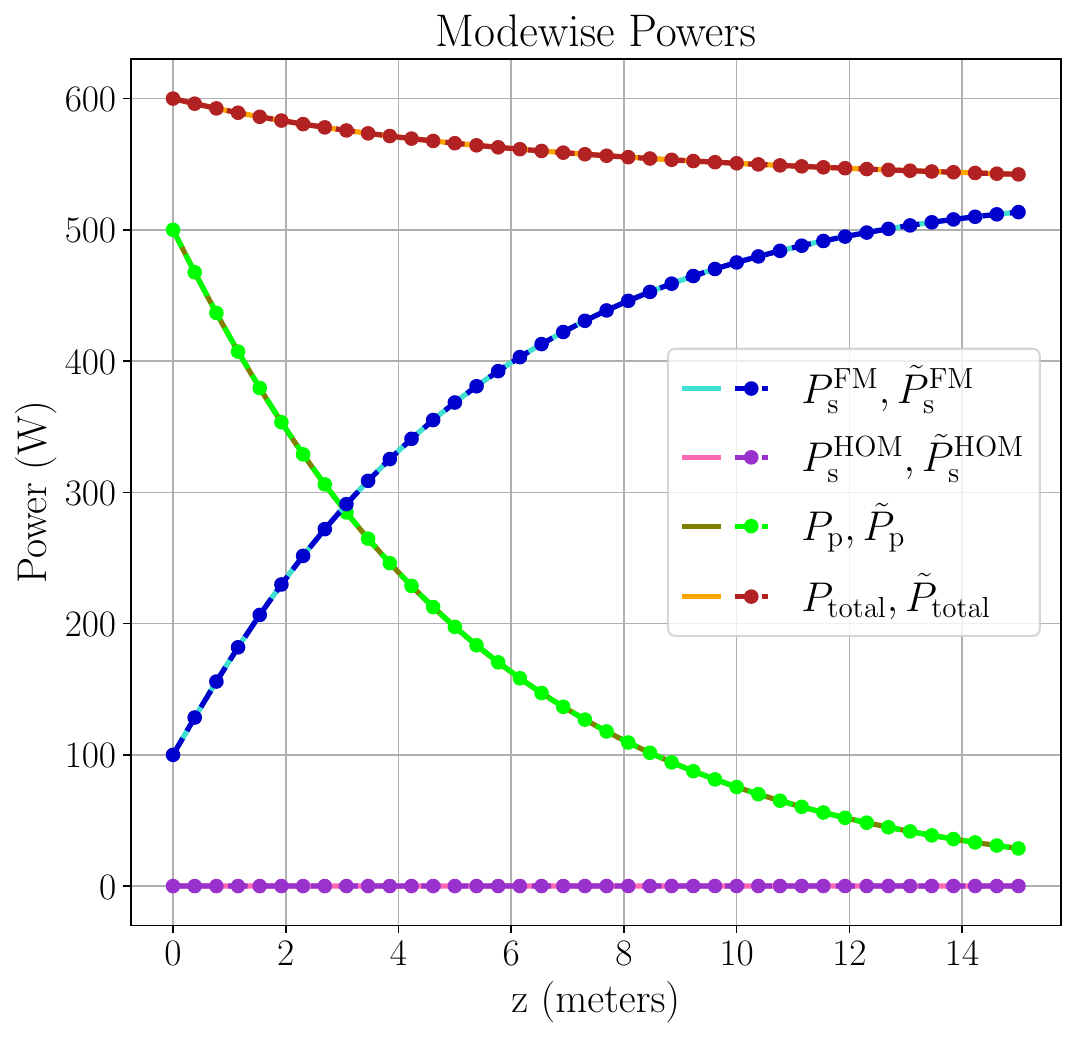}
        \caption{$\varrho = 0.005$ (40 $z$-steps).}
        \label{fig:pb0.005}
    \end{subfigure} \\
    \vspace{0.75cm}
    \begin{subfigure}{0.4\textwidth}
        \includegraphics[width=\linewidth]{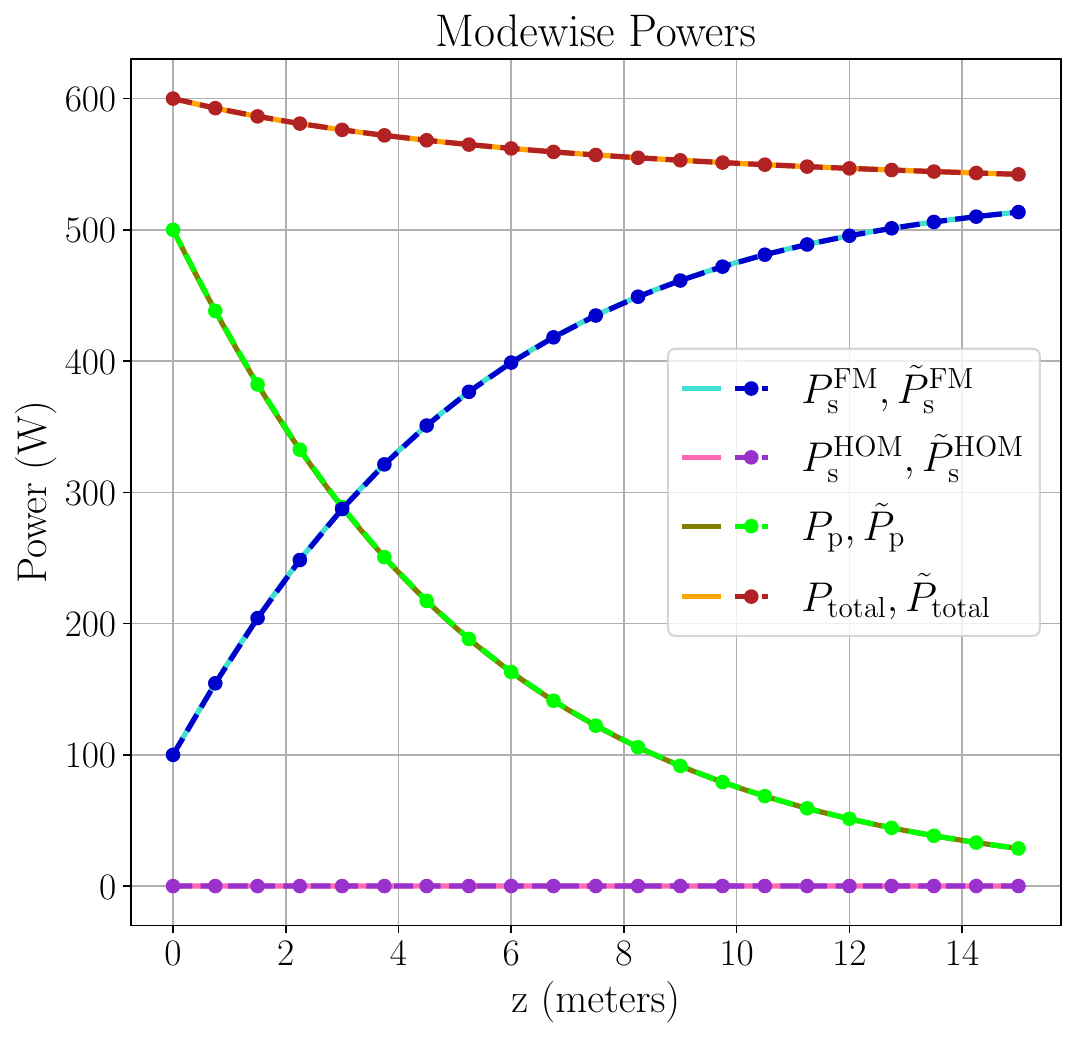}
        \caption{$\varrho = 0.0025$ (21 $z$-steps).}
        \label{fig:pb0.0025}
    \end{subfigure}
    \hspace{0.75cm}
    \begin{subfigure}{0.4\textwidth}
        \includegraphics[width=\linewidth]{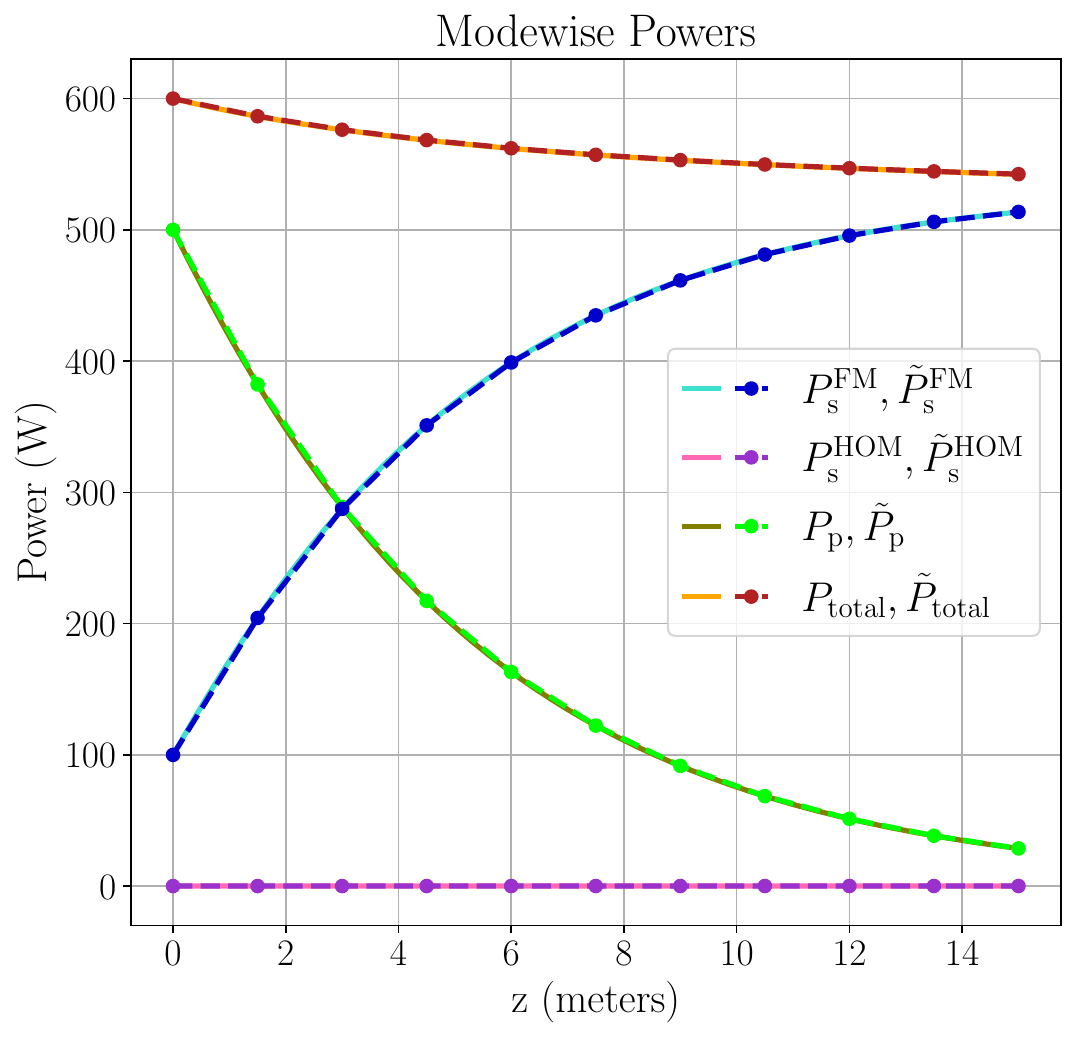}
        \caption{$\varrho = 0.00125$ (11 $z$-steps).}
        \label{fig:pb0.001}
    \end{subfigure}
    \caption{The CMT (solid lines) and ACM (dashed lines) model results for power along the amplifier length, where the ACM model is solved with coarser longitudinal discretizations determined by $\varrho$.}
    \label{fig:low_ppb}
\end{figure}

This high accuracy of the ACM model with considerably fewer points per mode beat length compared to the CMT model is also depicted in the power evolution plots of Figure~\ref{fig:low_ppb}. 
The CMT model power predictions along the amplifier length, computed for ${\mbf A}(\varrho = 10 \text{ or } N_{z} = 77497)$, are plotted concurrently with the ACM model's power levels, computed for ${\mbt A}(\varrho = 10^{-2} \text{ or } N_{z} = 79)$ (plot~\subref{fig:pb0.01}), ${\mbt A}(\varrho = 5 \cdot 10^{-3} \text{ or } N_{z} = 40)$ (plot~\subref{fig:pb0.005}), ${\mbt A}(\varrho = 2.5 \cdot 10^{-3} \text{ or } N_{z} = 21)$ (plot~\subref{fig:pb0.0025}), and ${\mbt A}(\varrho = 1.25 \cdot 10^{-3} \text{ or } N_{z} = 11)$ (plot~\subref{fig:pb0.001}), respectively. 
In all four of these cases, the powers predicted by the ACM model remain visually indistinguishable from those predicted by the CMT model. 
Note that the gain (stimulated emission process) is a relatively benign nonlinearity that monotonically gets weaker down the length of the amplifier. 
Also, below $\varrho = 10^{-3}$ points per mode beat length it becomes unclear as to how much error is due to the approximations that distinguish the ACM model from the CMT model verses how much error is on account of the numerical integration method (in this case, the fourth-order Runge-Kutta method), that is being used to propagate the optical field amplitudes along the fiber, becoming unstable due to the excessively large discrete step sizes. 

Even still, in practice, we estimate that the ACM model is achieving an approximate $3840$x computational expense improvement over the CMT model based on runtime, while maintaining an acceptable accuracy. 
Of course, we suspect that if the CMT model were to be augmented with other physical effects, e.g. the Kerr nonlinearity, thermal effects, stimulated Brillouin/Raman scattering, then this model reduction technique would not likely achieve such a large computational improvement, but would be still computationally advantageous nonetheless. 

\FloatBarrier
%%%%%%%%%%%%%%%%%%%%%%%%%%%%%%%%%%%%%%%%%%%%%%%%%%%%
% Conclusion
%%%%%%%%%%%%%%%%%%%%%%%%%%%%%%%%%%%%%%%%%%%%%%%%%%%%

\section{Conclusion}\label{sec:conclusion}

This work offers a new, mathematically rigorous, perspective on the approximations that are commonly cited for the derivation of many trusted optical models; namely, the slowly varying envelope approximations and/or phase-matching arguments. 
By acknowledging that certain high-frequency terms are removed from the governing system by means of integral averaging (according to Theorem~\ref{thm:avg-Verl}), one is then assured that the reduced system's solution has a finite bounded error over some span of the independent variable. 
For a simple and common fiber laser amplifier configuration, we have numerically demonstrated that our ACM model achieves excellent computational performance (${\sim}4000$x improvement) with suitable accuracy when compared to the CMT model. 
Some portion of the model's error is wrapped up in the phases of the optical fields' amplitudes, and not just in the associated predicted power levels. 
The success of other similar reduced optical models that are reliant upon the SVEAs and/or phase-matching justifications, along with the insight from Theorem~\ref{thm:avg-Verl}, indicates that these models likely preserve their high accuracies over physically plausible lengths for their intended applications. 

This effort sets the stage for our future efforts that will build on this model reduction methodology by integral averaging (motivated by physical intuitions behind the SVEAs and/or phase-matching arguments) and by the Taylor expansions about the DC component of the irradiance. 
Effectively, our plan is to expand upon the work of Menyuk et al.~\cite{menyuk2021accurate}, who built a computationally efficient transverse mode instability model. 
By, instead, implementing our (albeit similar) approach, we intend to demonstrate that the reduced model can be applied to a wider variety of fiber amplifier configurations, types, and designs than could Menyuk et al.'s model, including the two-tone configuration, Raman gain, in-band- and out-of-band-pumped thulium, and/or holmium fiber amplifiers. 
These model extensions would otherwise be impossible without the ease-of-application and generality of the aforementioned Taylor expansion technique, which readily helps extract, and explicitly express, the unwanted periodic (high-frequency) terms in the governing dynamical system.

\FloatBarrier
%%%%%%%%%%%%%%%%%%%%%%%%%%%%%%%%%%%%%%%%%%%%%%%%%%%%
% References
%%%%%%%%%%%%%%%%%%%%%%%%%%%%%%%%%%%%%%%%%%%%%%%%%%%%

\section{References}\label{sec:refs}

\renewcommand\refname{} \vspace*{-4mm}
\bibliography{averaging.bib}

%\newpage
%%%%%%%%%%%%%%%%%%%%%%%%%%%%%%%%%%%%%%%%%%%%%%%%%%%%
% Appendix
%%%%%%%%%%%%%%%%%%%%%%%%%%%%%%%%%%%%%%%%%%%%%%%%%%%%
\appendix
{\noindent}\textbf{\Large{Appendix}}

%%%%%%%%%%%%%%%%%%%%%%%%%%%%%%%%%%%%%%%%%%%%%%%%%%%%
% Appendix Section: Steady-State Gain Formulas
%%%%%%%%%%%%%%%%%%%%%%%%%%%%%%%%%%%%%%%%%%%%%%%%%%%%

\section{Steady-State Gain Formulas}\label{sec:GainFormulas}

Laser amplification occurs by stimulated emission using an active dopant, which is typically a lanthanide metal oxide. 
For in-band-pumping configurations, the absorbed pump light excites the outer-shell electrons into the same energy manifold from which the stimulated emission induces electrons down back to their ground-state manifold. 
For out-of-band-pumping configurations, the absorbed pump light excites the outer-shell electrons into an energy manifold that is higher than the excited-state manifold from which the stimulated emission induces electrons down back to their ground-state manifold. 
These pump-excited electrons must undergo one or more radiative and/or non-radiative transitions to lower energy manifolds before the stimulated emission process can occur. 
Typically, stimulated emission occurs from the lowest excited-state manifold, which implies that the rate of gain is given by 
\[
    g_{\ell}(x, y, z, t) = \sigma_{\ell}^{\text{ems}} \calN_{\text{excited}}(x, y, z, t) - \sigma_{\ell}^{\text{abs}} \calN_{\text{ground}}(x, y, z, t) \ \ \left[ \frac{1}{\text{m}} \right] ,
\]
where $\sigma_{\ell}^{\text{abs}}$ and $\sigma_{\ell}^{\text{ems}}$ [m$^{2}$/ion] are respectively the measured absorption and emission cross-sections of the given dopant, and $\calN_{\text{ground}}$ and $\calN_{\text{excited}}$ [ions/m$^{3}$] are respectively the ground-state and lowest excited-state dopant concentration levels in the fiber's gain region. 
Here, we will summarize the steady-state solutions to the active dopant concentration kinetics relations for ytterbium (Yb), thulium (Tm), and holmium (Ho) oxides in fused silica glass.

\FloatBarrier
%%%%%%%%%%%%%%%%%%%%%%%%%%%
% Appendix Subsection: Gain Kinetics of Ytterbium Oxide
%%%%%%%%%%%%%%%%%%%%%%%%%%%

\subsection{Gain Kinetics of Ytterbium Oxide}\label{subsec:YbDopant}

Yb-doped amplifiers are usually in-band-pumped at $\lp \simeq 976$ nm, exciting the outer-shell electron from the ground-state manifold $\prescript{2}{}{\mathbf{F}}_{7/2}$ (associated with $\calN_{0} \equiv \calN_{\text{ground}}$) to the excited-state manifold $\prescript{2}{}{\mathbf{F}}_{5/2}$ (associated with $\calN_{1} \equiv \calN_{\text{excited}}$)~\cite{pask1995ytterbium}. 
The total dopant concentration is considered a known measured quantity such that 
\begin{equation}\label{eq:ConservationYbTotalDopant}
    \calN_{\text{total}}^{\text{Yb}} = \calN_{\text{ground}}^{\text{Yb}}(x, y, z, t) + \calN_{\text{excited}}^{\text{Yb}}(x, y, z, t) \ \ \left[ \frac{\text{ions}}{\text{m}^{3}} \right] .
\end{equation}
The absorption and emission rates of ytterbium oxide are $\psi_{\ell}^{c} = \sigma_{\ell}^{c} I_{\ell} / (\hbar \omega_\ell) \geq 0$ [1/s] for $\ell \in \{ \pmp, \sig\} $ and $c \in \{ \text{abs} , \text{ems} \}$.
Given specific pump and laser signal irradiance levels ($\{ \Ip, \Is \}$), in a temporal steady-state, the gain kinetics satisfy 
\begin{equation}\label{eq:YbFixedPoint}
    \begin{aligned}
        \bs{\bar{\calN}}^{\text{Yb}} = 
		\bmat
			\bar{\calN}_{\text{ground}}^{\text{Yb}} \\
			\bar{\calN}_{\text{excited}}^{\text{Yb}}
		\ebmat & = 
		\bmat
			\calN_{\text{total}}^{\text{Yb}} - \bar{\calN}_{\text{excited}}^{\text{Yb}} \\
			\bar{\calN}_{\text{excited}}^{\text{Yb}}
		\ebmat =
		\bmat
			\frac{ \calN_{\text{total}}^{\text{Yb}} \left( \psi_{\pmp}^{\text{ems}} + \psi_{\sig}^{\text{ems}} + \frac{1}{\tau^{\text{Yb}}} \right) }{ \psi_{\pmp}^{\text{abs}} + \psi_{\pmp}^{\text{ems}} +  \psi_{\sig}^{\text{abs}} + \psi_{\sig}^{\text{ems}} + \frac{1}{\tau^{\text{Yb}}} } \\
			\frac{ \calN_{\text{total}}^{\text{Yb}} \left( \psi_{\pmp}^{\text{abs}} + \psi_{\sig}^{\text{abs}} \right) }{ \psi_{\pmp}^{\text{abs}} + \psi_{\pmp}^{\text{ems}} +  \psi_{\sig}^{\text{abs}} + \psi_{\sig}^{\text{ems}} + \frac{1}{\tau^{\text{Yb}}} } 
		\ebmat
    \end{aligned} \ \ \left[ \frac{\text{ions}}{\text{m}^{3}} \right] ,
\end{equation}
where $\tau^{\text{Yb}} \ [\text{s}]$ is the measured mean radiative lifetime of the excited electron in the $\prescript{2}{}{\mathbf{F}}_{5/2}$ manifold. 

The steady-state gain relation~\eqref{eq:g-ell} can be re-written as 
\begin{equation}\label{eq:InBandPumpedSteadyStateGain}
    \bar{g}_{\ell}\big( \Ip, \Is \big) = \big( \sigma_{\ell}^{\text{abs}} + \sigma_{\ell}^{\text{ems}} \big) \overline{\calN}_{\text{excited}}^{\text{Yb}}\big( \Ip, \Is \big) - \sigma_{\ell}^{\text{abs}} \calN_{\text{total}}^{\text{Yb}}
\end{equation}
for the Yb dopant because, using~\eqref{eq:ConservationYbTotalDopant}, $\overline{\calN}_{\text{ground}}^{\text{Yb}} = \calN_{\text{total}}^{\text{Yb}} - \overline{\calN}_{\text{excited}}^{\text{Yb}}$. 
Using~\eqref{eq:YbFixedPoint} with $\ell \in \{ \pmp, \sig  \}$, the partial derivative of the steady-state excited-state dopant concentration ($\overline{\calN}_{\text{excited}}^{\text{Yb}}$) with respect to the signal irradiance is given by 
\begin{equation}\label{eq:YbFixedPointDerivative}
    \begin{aligned}
        \pp{\overline{\calN}_{\text{excited}}^{\text{Yb}}}{\Is} & = \pp{}{\Is}\left[ \frac{ \calN_{\text{total}}^{\text{Yb}} \left( \psi_{\pmp}^{\text{abs}} + \psi_{\sig}^{\text{abs}} \right) }{ \psi_{\pmp}^{\text{abs}} + \psi_{\pmp}^{\text{ems}} +  \psi_{\sig}^{\text{abs}} + \psi_{\sig}^{\text{ems}} + \frac{1}{\tau^{\text{Yb}}}  } \right] \\
        & = -\frac{ \frac{ \bar{g}_{\sig} }{ \hbar \omega_{\sig} } }{ \psi_{\pmp}^{\text{abs}} + \psi_{\pmp}^{\text{ems}} +  \psi_{\sig}^{\text{abs}} + \psi_{\sig}^{\text{ems}} + \frac{1}{\tau^{\text{Yb}}}  } \ \ .
    \end{aligned}
\end{equation}
With this information, the partial derivatives of the steady-state gain functions ($\{ \bar{g}_{\pmp}, \bar{g}_{\sig} \}$) with respect to the signal irradiance are determined to be 
\begin{align}
    \pp{\bar{g}_{\pmp}}{\Is} & = \big( \sigma_{\pmp}^{\text{abs}} + \sigma_{\pmp}^{\text{ems}} \big) \pp{\overline{\calN}_{\text{excited}}^{\text{Yb}}}{\Is} \notag \\
    \pp{\bar{g}_{\pmp}}{\Is} & = -\frac{ \frac{ \bar{g}_{\sig} \left( \sigma_{\pmp}^{\text{abs}} + \sigma_{\pmp}^{\text{ems}} \right) }{ \hbar \omega_{\sig} } }{\psi_{\pmp}^{\text{abs}} + \psi_{\pmp}^{\text{ems}} +  \psi_{\sig}^{\text{abs}} + \psi_{\sig}^{\text{ems}} + \frac{1}{\tau^{\text{Yb}}} } \label{eq:Yb-dgp-dIs} \ \ \text{ and} \\
    \pp{\bar{g}_{\sig}}{\Is} & = \big( \sigma_{\sig}^{\text{abs}} + \sigma_{\sig}^{\text{ems}} \big) \pp{\overline{\calN}_{\text{excited}}^{\text{Yb}}}{\Is} \notag \\
    \pp{\bar{g}_{\sig}}{\Is} & = -\frac{ \frac{ \bar{g}_{\sig} \left( \sigma_{\sig}^{\text{abs}} + \sigma_{\sig}^{\text{ems}} \right) }{ \hbar \omega_{\sig} } }{ \psi_{\pmp}^{\text{abs}} + \psi_{\pmp}^{\text{ems}} +  \psi_{\sig}^{\text{abs}} + \psi_{\sig}^{\text{ems}} + \frac{1}{\tau^{\text{Yb}}}  } \ \  . \label{eq:Yb-dgs-dIs}
\end{align}

\FloatBarrier
%%%%%%%%%%%%%%%%%%%%%%%%%%
% Appendix Subsubsection: First-order Taylor Expansion Error Bounds for Ytterbium Gain
%%%%%%%%%%%%%%%%%%%%%%%%%%

\subsubsection{First-order Taylor Expansion Error Bounds for Ytterbium Gain}\label{subsec:taylor_error}

The first~\eqref{eq:Yb-dgs-dIs} and second derivatives of ytterbium steady-state gain are used in Sections~\ref{sec:acm}--\ref{sec:comp-results} to derive three-term approximations of the steady-state gain using a first-order Taylor expansion (see relations~\ref{eq:TaylorExpandedGain}--\ref{eq:DecomposedGain}), and, likewise, to derive the three-term decompositions of the heat generation (see relations~\ref{eq:Q-Taylor}--\ref{eq:Q-3-term-expansion}). 
Starting from relation~\eqref{eq:Yb-dgs-dIs}, one finds that 
\begin{align}\label{eq:dg/Is}
     {\Is} \, \frac{\partial \bar{g}_{{\ell}}\big( \Ip, I_{{\sig}} \big)}{\partial I_{{\sig}}} & = \calN^{\text{Yb}}_{\text{total}} \left( \sigma_{\ell}^{\text{ems}} + \sigma_{\ell}^{\text{abs}} \right) \frac{ \psi^{{\text{abs}}}_{\sig} \psi^{{\text{ems}}}_{\pmp} + \frac{\psi^{{\text{abs}}}_{\sig}}{\tau^{\text{Yb}}} - \psi^{{\text{ems}}}_{\sig} \psi^{{\text{abs}}}_{\pmp} }{ \left(\psi^{{\text{abs}}}_{\pmp} + \psi^{\text{ems}}_{\pmp} + \psi^{{\text{abs}}}_{\sig} + \psi^{\text{ems}}_{\sig} + \frac{1}{\tau^{\text{Yb}}}\right)^{2} } \ .
\end{align}
For notational simplicity, let
\begin{align*}
    \text{denom}_{1} & \defeq \psi^{{\text{abs}}}_{\pmp} + \psi^{\text{ems}}_{\pmp} + \psi^{{\text{abs}}}_{\sig} + \psi^{\text{ems}}_{\sig} + \frac{1}{\tau^{\text{Yb}}} \ , \\
    \text{denom}_{2} & \defeq \left( \psi^{\text{abs}}_{\pmp} \right)^2 + \left( \psi^{\text{ems}}_{\pmp} \right)^2 + \left( \psi^{\text{abs}}_{\sig} \right)^2 + \left( \psi^{\text{ems}}_{\sig} \right)^2 + \left( \frac{1}{\tau^{\text{Yb}}} \right)^{2} \ , \\
    \text{numer}_{1} & \defeq \psi^{{\text{abs}}}_{\sig} \psi^{{\text{ems}}}_{\pmp} + \frac{\psi^{{\text{abs}}}_{\sig}}{\tau^{\text{Yb}}} - \psi^{{\text{ems}}}_{\sig} \psi^{{\text{abs}}}_{\pmp} \ , \text{ and} \\
    \text{numer}_{2} & \defeq \psi^{{\text{abs}}}_{\sig} \psi^{{\text{ems}}}_{\pmp} + \frac{\psi^{{\text{abs}}}_{\sig}}{\tau^{\text{Yb}}} \ .
\end{align*}
Hence, 
\[
    {\Is} \, \frac{\partial \bar{g}_{{\ell}}\big( \Ip, I_{{\sig}} \big)}{\partial I_{{\sig}}} = \calN^{\text{Yb}}_{\text{total}} \left( \sigma_{\ell}^{\text{ems}} + \sigma_{\ell}^{\text{abs}} \right) \frac{ \text{numer}_{1} }{ \text{denom}_{1}^{2} } \ .
\]
Since all of the absorption and emission the rates are non-negative: $\psi_{\ell}^{c} \geq 0$, then $\text{numer}_{1} \leq \text{numer}_{2}$. 
So, 
\begin{align*}
    \text{numer}_{1} & \leq \text{numer}_{2} = \psi^{{\text{abs}}}_{\sig} \left( \psi^{{\text{ems}}}_{\pmp} + \frac{1}{\tau^{\text{Yb}}} \right) \leq 2 \psi^{{\text{abs}}}_{\sig} \psi^{{\text{ems}}}_{\pmp} + 2 \frac{ \psi^{{\text{abs}}}_{\sig} }{ \tau^{\text{Yb}} } \leq {\ldots} \\
    \mbox{} & \ \downarrow \ \text{adding in non-negative terms} \\
    \mbox{} & \hspace*{12pt} {\ldots}\tcfg{\left( \psi^{{\text{abs}}}_{\pmp} \right)^{2}} + \tcfg{2 \psi^{{\text{abs}}}_{\pmp} \psi^{{\text{ems}}}_{\pmp}} + \tcfg{2 \psi^{{\text{abs}}}_{\pmp} \psi^{{\text{abs}}}_{\sig}} + \tcfg{2 \psi^{{\text{abs}}}_{\pmp} \psi^{{\text{ems}}}_{\sig}} + \tcfg{2 \frac{ \psi^{{\text{abs}}}_{\pmp} }{ \tau^{\text{Yb}} }} \ + \\
    \mbox{} & \hspace*{24pt} \tcfg{\left( \psi^{\text{ems}}_{\pmp} \right)^{2}} + 2 \psi^{\text{ems}}_{\pmp} \psi^{\text{abs}}_{\sig} + \tcfg{2 \psi^{\text{ems}}_{\pmp} \psi^{\text{ems}}_{\sig}} + \tcfg{2 \frac{ \psi^{\text{ems}}_{\pmp} }{ \tau^{\text{Yb}} }} + \tcfg{\left( \psi^{\text{abs}}_{\sig} \right)^{2}} \ + \\
    \mbox{} & \hspace*{24pt} \tcfg{2 \psi^{{\text{abs}}}_{\sig} \psi^{{\text{ems}}}_{\sig}} + 2 \frac{ \psi^{{\text{abs}}}_{\sig} }{ \tau^{\text{Yb}} } + \tcfg{\left( \psi^{{\text{ems}}}_{\sig} \right)^{2}} + \tcfg{ 2 \frac{ \psi^{{\text{ems}}}_{\sig} }{ \tau^{\text{Yb}} } } + \tcfg{\left( \frac{1}{\tau^{\text{Yb}}} \right)^{2}} = {\ldots} \\
    \mbox{} & \ \downarrow \ \text{Noting that } \left( x_{1} + x_{2} + x_{3}  + x_{4} + x_{5} \right)^{2} = x_{1}^{2} + 2 x_{1} x_{2} + 2 x_{1} x_{3} + 2 x_{1} x_{4} + 2 x_{1} x_{5} \ + \\
    \mbox{} & \hspace*{172pt} x_{2}^{2} + 2 x_{2} x_{3} + x_{2} x_{4} + x_{2} x_{5} + x_{3}^{2} \ + \\
    \mbox{} & \hspace*{172pt} 2 x_{3} x_{4} + 2 x_{3} x_{5} + x_{4}^{2} + 2 x_{4} x_{5} + x_{5}^{2} \\
    \mbox{} & \hspace*{12pt} {\ldots}\left( \psi^{{\text{abs}}}_{\pmp} + \psi^{\text{ems}}_{\pmp} + \psi^{{\text{abs}}}_{\sig} + \psi^{\text{ems}}_{\sig} + \frac{1}{\tau^{\text{Yb}}} \right)^{2} = \text{denom}_{1}^{2} \ . \\
    \therefore \ \text{numer}_{1} & \leq \text{denom}_{1}^{2} \ .
\end{align*}
Therefore, 
\[
    \frac{ \text{numer}_{1} }{ \text{denom}_{1}^{2} } \leq \frac{ \text{denom}_{1}^{2} }{ \text{denom}_{1}^{2} } = 1 \ .
\]
Thus, one ascertains that relation~\eqref{eq:dg/Is} is bounded entirely by 
\begin{align}\label{eq:der-bound}
    \left| {\Is}\, \frac{\partial \bar{g}_{{\ell}} \big( \Ip, I_{{\sig}} \big)}{\partial I_{{\sig}}} \right| & \leq \calN^{\text{Yb}}_{\text{total}} \left( \sigma_{\ell}^{\text{ems}} + \sigma_{\ell}^{\text{abs}} \right)
\end{align}
at any fixed $\Is$. 

Taking the second derivative of the steady-state gain relation~\eqref{eq:InBandPumpedSteadyStateGain} with respect to $\Is$, yields  
\begin{align}\label{eq:dg2/Is2}
    { \Is^2}\, \frac{\partial^2 \bar{g}_{{\ell}} \big( \Ip, I_{{\sig}} \big)}{\partial I^2_{{\sig}}} & = 2 \calN^{\text{Yb}}_{\text{total}} \left( \sigma_{\ell}^{\text{ems}} + \sigma_{\ell}^{\text{abs}} \right) \frac{ \left( \psi^{{\text{abs}}}_{\sig} + \psi^{\text{ems}}_{\sig} \right) \left( \psi^{{\text{abs}}}_{\pmp} \psi^{\text{ems}}_{\sig} - \psi^{{\text{abs}}}_{\sig} \psi^{\text{ems}}_{\pmp} - \frac{\psi^{{\text{abs}}}_{\sig}}{\tau^{\text{Yb}}} \right) }{ \left(\psi^{{\text{abs}}}_{\pmp} + \psi^{\text{ems}}_{\pmp} + \psi^{{\text{abs}}}_{\sig} + \psi^{\text{ems}}_{\sig} + \frac{1}{\tau^{\text{Yb}}}\right)^{3} } \ .
\end{align}
Let 
\begin{align*}
    \text{numer}_{3} & \defeq \psi_{\sig}^{\text{abs}} + \psi_{\sig}^{\text{ems}} \ , & 
    \text{numer}_{4} & \defeq \psi_{\sig}^{\text{ems}} \psi_{\pmp}^{\text{abs}} - \psi_{\sig}^{\text{abs}} \psi_{\pmp}^{\text{ems}} -  \frac{ \psi_{\sig}^{\text{abs}} }{ \tau^{\text{Yb}} } \ , \\
    \bs{w}_{1} & \defeq 
        \begin{bmatrix}
            1 \\
            1
        \end{bmatrix} \ , & 
    \bs{w}_{2} & \defeq 
        \begin{bmatrix}
            \psi_{\sig}^{\text{abs}} \\
            \psi_{\sig}^{\text{ems}}
        \end{bmatrix} \ , \\
    \bs{w}_{3} & \defeq 
        \begin{bmatrix}
            \psi_{\pmp}^{\text{abs}} \\
            -\psi_{\sig}^{\text{abs}} \\
            -\frac{1}{\tau^{\text{Yb}}}
        \end{bmatrix} \ , & \text{and }
    \bs{w}_{4} & \defeq 
        \begin{bmatrix}
            \psi_{\sig}^{\text{ems}} \\
            \psi_{\pmp}^{\text{ems}} \\
            \psi_{\sig}^{\text{abs}}
        \end{bmatrix} \ .
\end{align*}
Note that 
\begin{align*}
    \text{numer}_{3} & = \bs{w}_{1} \bs{\cdot} \bs{w}_{2} \\
    & \ \downarrow \ \text{by the \textit{Cauchy-Schwarz inequality}} \\
    \text{numer}_{3} & \leq \| \bs{w}_{1} \| \cdot \| \bs{w}_{2} \| \\
    \text{numer}_{3} & \leq \sqrt{1^{2} + 1^{2}} \sqrt{ \left( \psi_{\sig}^{\text{abs}} \right)^{2} + \left( \psi_{\sig}^{\text{ems}} \right)^{2} } \\
    & \ \downarrow \ \text{by adding in non-negative terms} \\
    \text{numer}_{3} & \leq \sqrt{2} \sqrt{ \tcfg{\left( \psi^{\text{abs}}_{\pmp} \right)^{2}} + \tcfg{\left( \psi^{\text{ems}}_{\pmp} \right)^{2}} + \left( \psi^{\text{abs}}_{\sig} \right)^{2} + \left( \psi^{\text{ems}}_{\sig} \right)^{2} + \tcfg{\left( \frac{1}{\tau^{\text{Yb}}} \right)^{2}} } \\
    \text{numer}_{3} & \leq \sqrt{2} \sqrt{ \text{denom}_{2} } \ .
\end{align*}
Likewise,  
\begin{align*}
    \text{numer}_{4} & = \bs{w}_{3} \bs{\cdot} \bs{w}_{4} \\
    & \ \downarrow \ \text{by the \textit{Cauchy-Schwarz inequality}} \\
    \text{numer}_{4} & \leq \| \bs{w}_{3} \| \cdot \| \bs{w}_{4} \| = \sqrt{ \left( \psi_{\pmp}^{\text{abs}} \right)^{2} + \left( \psi_{\sig}^{\text{abs}} \right)^{2} + \left( \frac{1}{\tau^{\text{Yb}}} \right)^{2} } \sqrt{ \left( \psi_{\sig}^{\text{ems}} \right)^{2} + \left( \psi_{\pmp}^{\text{ems}} \right)^{2} + \left( \psi_{\sig}^{\text{abs}} \right)^{2} } \\
    & \ \downarrow \ \text{by adding in non-negative quantities} \\
    \text{numer}_{4} & \leq \sqrt{ \left( \psi^{\text{abs}}_{\pmp} \right)^{2} + \tcfg{\left( \psi^{\text{ems}}_{\pmp} \right)^{2}} + \left( \psi^{\text{abs}}_{\sig} \right)^{2} + \tcfg{\left( \psi^{\text{ems}}_{\sig} \right)^{2}} + \left( \frac{1}{\tau^{\text{Yb}}} \right)^{2} } \ \cdot \\
    & \hspace*{12pt} \sqrt{ \tcfg{\left( \psi^{\text{abs}}_{\pmp} \right)^{2}} + \left( \psi^{\text{ems}}_{\pmp} \right)^{2} + \left( \psi^{\text{abs}}_{\sig} \right)^{2} + \left( \psi^{\text{ems}}_{\sig} \right)^{2} + \tcfg{\left( \frac{1}{\tau^{\text{Yb}}} \right)^{2}} } \\
    \text{numer}_{4} & \leq \text{denom}_{2}
\end{align*}
Since all terms of $\text{denom}_{1}$ are non-negative, it follows that $\text{denom}_{2} \leq \text{denom}_{1}^{2}$, and therefore $\text{denom}_{2}^{\frac{3}{2}} \leq \text{denom}_{1}^{3}$. 
When these observations are used to bound \eqref{eq:dg2/Is2}, we find that
\begin{equation}\label{eq:der2-bound}
    \left| \Is^2\, \frac{\partial^2 \bar{g}_{{\ell}}\big( \Ip, I_{{\sig}} \big)}{\partial I^2_{{\sig}}} \right| \leq 2 \calN^{\text{Yb}}_{\text{total}} \left( \sigma_{\ell}^{\text{ems}} + \sigma_{\ell}^{\text{abs}} \right) \frac{ \sqrt{2} \text{denom}_{2}^{\frac{3}{2}} }{ \text{denom}_{2}^{\frac{3}{2}} } = 2 \sqrt{2} \calN^{\text{Yb}}_{\text{total}} \left( \sigma_{\ell}^{\text{ems}} + \sigma_{\ell}^{\text{abs}} \right)
\end{equation}
for any fixed $\Is$. 
Note that, typically, $10^{-2} \lesssim \calN^{\text{Yb}}_{\text{total}} \left( \sigma_{\ell}^{\text{ems}} + \sigma_{\ell}^{\text{abs}} \right) \lesssim 10^{2}$ for $\ell \in \{ \pmp, \sig \}$ and for most actively doped fiber amplifiers (see Table~\ref{fiberParams}). 

By \textit{Taylor's theorem}, the error associated with the first-order Taylor expansion of steady-state gain is
\begin{align*}
    \text{error}_{g_{\ell}} & \defeq \Big| \bar{g}_{\ell}\big( \Ip, \Is \big) - \bar{g}_{\ell}\big( \Ip, I_{{\sig} 0} \big) - \frac{\partial \bar{g}_{\ell} \big( \Ip, I_{{\sig} 0} \big)}{\partial I_{\sig}} \Big( I_{{\sig}} - I_{{\sig} 0} \Big) \Big| \\
    & = \left| \frac{1}{2} \frac{\partial^2 \bar{g}_{\ell} \big( \Ip, \mathcal{I}_{\sig} \big) }{\partial I^{2}_{{\sig}}}\Big( \mathcal{I}_{\sig} -  I_{{\sig} 0} \Big) \Big( I_{\sig} - I_{\sig 0} \Big) \right| \ \ ,
\end{align*}
which is bounded by
\begin{align}\label{eq:g_ell_err_bound}
    \text{error}_{g_{\ell}} \le \frac{1}{2} \left| {\mathcal{I}_{\sig}^2} \frac{ \partial^2 \bar{g}_{{\ell}} \big( \Ip, \mathcal{I}_{\sig} \big) }
  { \partial I^2_{{\sig}} } \right| \, \left| \frac{\Iso}{\mathcal{I}_{\sig}} \right|^2 \, \left| \frac{ \mathcal{I}_{{\sig}} - I_{{\sig}0}  }{ \Iso } \right|^{2} \ ,
\end{align}
% where the fundamental theorem of calculus demands that there is an 
where $\mathcal{I}_{\sig}$ is some irradiance value. 
As shown in plot~(\subref{plot:IsError}) of Fig.~\ref{fig:SignalIrradianceComparedToItsComponents}, $\max_{x, y, z} \left| \left(I_{\sig} - I_{\sig 0} \right) / I_{\sig 0} \right| \equiv \delta \sim \mathcal{O}\big( 10^{-9} \big)$ (cf. claim in~\ref{eq:Is-Iso}). 
Using this information in conjunction with the bound on the second derivative of the gain function~\eqref{eq:der2-bound} reveals that the r.h.s. to this error is negligible, justifying the first-order Taylor expansion as a trustworthy approximation. 

Likewise, the error associated with the heat generation first-order Taylor expansion is 
\begin{align*}
    \text{error}_{Q} & \defeq \Big| Q \big( \Ip, \Is \big) - Q\big( \Ip, I_{{\sig} 0} \big) - \frac{\partial Q \big( \Ip, I_{{\sig} 0} \big)}{\partial I_{\sig}} \Big( I_{{\sig}} - I_{{\sig} 0} \Big) \Big| \\
    & = \Big| \frac{1}{2} \frac{\partial^2 Q \big( \Ip, \mathcal{I}_{\sig} \big) }{\partial I^{2}_{{\sig}}} \Big( \mathcal{I}_{\sig} -  I_{{\sig} 0} \Big) \Big( I_{\sig} - I_{\sig 0} \Big) \Big| \\
   & = \Big| \frac{1}{2} \left( \frac{\partial^{2} \bar{g}_{\sig} \big( \Ip, \mathcal{I}_{\sig} \big)}{\partial I^{2}_{\sig}} +  \frac{\partial \bar{g}_{\sig} \big( \Ip, \mathcal{I}_{\sig} \big)}{\partial I_{\sig}}  +  \frac{\partial^{2} \bar{g}_{\pmp} \big( \Ip, \mathcal{I}_{\sig} \big)}{\partial I^{2}_{\sig}} \right) \Big( \mathcal{I}_{\sig} -  I_{{\sig} 0} \Big) \Big( I_{\sig} - I_{\sig 0} \Big) \Big| \ \ ,
\end{align*}
which is bounded by
\begin{equation}\label{eq:HeatGenErrorBound}
    \text{error}_{Q} \le \left( \sum_{\ell \in \{\sig, \pmp \}} \frac{1}{2} \left| {\mathcal{I}_{\sig}^2}  \frac{\partial^{2} \bar{g}_{\ell} \big( \Ip, \mathcal{I}_{\sig} \big) }{\partial I^{2}_{\sig}} \right| \left| \frac{\Iso}{\mathcal{I}_{\sig}} \right|^2 \, \left| \frac{ \mathcal{I}_{{\sig}} - I_{{\sig}0}  }{ \Iso } \right|^{2}  \right) \ + \ \frac{1}{2} \Bigg| {\mathcal{I}_{\sig}} \frac{\partial \bar{g}_{\sig} \big( \Ip, \mathcal{I}_{\sig} \big)}{\partial I_{\sig}} \Bigg| \left| \frac{ \left( \mathcal{I}_{{\sig}} - I_{{\sig}0} \right)^{2}  }{ I_{\sig 0 } } \right| \ ,
\end{equation}
for some irradiance value $\mathcal{I}_{\sig}$. 
The first term in the sum is bounded by the same quantities used to bound~\eqref{eq:g_ell_err_bound}. 
The second term is bounded by~\eqref{eq:der-bound} and by the quantity $\max_{x, y, z} \left| \left(I_{\sig} - I_{\sig 0} \right)^{2} / I_{\sig 0} \right| \equiv \zeta \sim \mathcal{O}\left( 10^{-6} \right)$ (cf. claim in~\ref{eq:zeta}), which is supported by plot~(\subref{plot:Is0_ratio}) of Fig.~\ref{fig:SignalIrradianceComparedToItsComponents}. 
Thus, the error bound for the first-order Taylor expansion on the heat generation~\eqref{eq:HeatGenErrorBound} is also negligibly small. 

\FloatBarrier
%%%%%%%%%%%%%%%%%%%%%%%%%%%
% Appendix Subsection: Gain Kinetics of Thulium Oxide
%%%%%%%%%%%%%%%%%%%%%%%%%%%
\subsection{Gain Kinetics of Thulium Oxide}\label{subsec:TmDopant}

Tm-doped amplifiers are usually out-of-band-pumped at $\lp \simeq 790$ nm, exciting the outer-shell electron from the ground-state manifold $\prescript{3}{}{\mathbf{H}}_{6}$ (associated with $\calN_{0} \equiv \calN_{\text{ground}}$) to the excited-state manifold $\prescript{3}{}{\mathbf{H}}_{4}$ (associated with $\calN_{3}$)~\cite{power2009mccomb}. 
Natural decay processses bring the electrons down into the $\prescript{3}{}{\mathbf{H}}_{5}$ manifold (associated with $\calN_{2}$), and finally to the lowest excited-state manifold $\prescript{3}{}{\mathbf{F}}_{4}$ (associated with $\calN_{1} \equiv \calN_{\text{excited}}$), from which stimulated emission occurs.
The total dopant concentration is considered a known measured quantity such that 
\[
    \calN_{\text{total}}^{\text{Tm}} = \calN_{\text{ground}}^{\text{Tm}}(x, y, z, t) + \calN_{\text{excited}}^{\text{Tm}}(x, y, z, t) + 
    \calN_{2}^{\text{Tm}}(x, y, z, t) + \calN_{3}^{\text{Tm}}(x, y, z, t) \ \ \left[ \frac{\text{ions}}{\text{m}^{3}} \right] .
\]
Given specific pump and laser signal irradiance levels ($\{ \Ip, \Is \}$), the steady-state gain kinetics satisfy 
\begin{equation}\label{eq:TmFixedPoint}
	\begin{aligned}
		& \bs{\overline{\calN}}^{\text{Tm}} = 
		\bmat
			\overline{\calN}_{\text{ground}}^{\text{Tm}} \\
			\overline{\calN}_{\text{excited}}^{\text{Tm}} \\
			\overline{\calN}_{2}^{\text{Tm}} \\
			\overline{\calN}_{3}^{\text{Tm}}
		\ebmat = 
		\bmat
			\overline{\calN}_{\text{ground}}^{\text{Tm}} \\
			\frac{ \left( A_{\text{Tm}} + B_{\text{Tm}} \overline{\calN}_{\text{ground}}^{\text{Tm}} \right) \alpha_{\text{Tm}} \overline{\calN}_{\text{ground}}^{\text{Tm}} }{ 1 + \beta_{\text{Tm}} \overline{\calN}_{\text{ground}}^{\text{Tm}} } \\
			\frac{ \alpha_{\text{Tm}} \gamma_{\text{Tm}} \overline{\calN}_{\text{ground}}^{\text{Tm}} }{ 1 + \beta_{\text{Tm}} \overline{\calN}_{\text{ground}}^{\text{Tm}} } \\
			\frac{ \alpha_{\text{Tm}} \overline{\calN}_{\text{ground}}^{\text{Tm}} }{ 1 + \beta_{\text{Tm}} \overline{\calN}_{\text{ground}}^{\text{Tm}} } \\
		\ebmat \ \ \left[ \frac{\text{ions}}{\text{m}^{3}} \right] \text{  where,} \\
		& \overline{\calN}_{\text{ground}}^{\text{Tm}} = \frac{ \left( \beta_{\text{Tm}} \calN_{\text{total}}^{\text{Tm}} - \alpha_{\text{Tm}} \left[ A_{\text{Tm}} + \gamma_{\text{Tm}} + 1 \right] - 1 \right) }{ 2 \left( \alpha_{\text{Tm}} B_{\text{Tm}} + \beta_{\text{Tm}} \right) } \ + \\
			& \hspace*{12pt} \frac{ \sqrt{\left( \alpha_{\text{Tm}} \left[ A_{\text{Tm}} + \gamma_{\text{Tm}} + 1 \right] - \beta_{\text{Tm}} \calN_{\text{total}}^{\text{Tm}} + 1 \right)^{2} + 4 \left( \alpha_{\text{Tm}} B_{\text{Tm}} + \beta_{\text{Tm}} \right) \calN_{\text{total}}^{\text{Tm}}} }{ 2 \left( \alpha_{\text{Tm}} B_{\text{Tm}} + \beta_{\text{Tm}} \right) } \ \ ,
	\end{aligned}
\end{equation}
and where, assuming $\sigma_{\pmp}^{\text{ems}} = 0$, 
\begin{align*}
    \alpha_{\text{Tm}} & \defeq \frac{ \frac{\sigma_{\pmp}^{\text{abs}} \Ip}{\hbar \omega_{\pmp}} }{ \frac{\sigma_{\pmp}^{\text{ems}} \Ip}{\hbar \omega_{\pmp}} + \frac{1}{\tau_{32}^{\text{Tm}}} + \frac{1}{\tau_{31}^{\text{Tm}}} + \frac{1}{\tau_{30}^{\text{Tm}}} + \Gamma_{3}^{\text{Tm}} } = \frac{ \frac{\sigma_{\pmp}^{\text{abs}} \Ip}{\hbar \omega_{\pmp}} }{ \frac{1}{\tau_{32}^{\text{Tm}}} + \frac{1}{\tau_{31}^{\text{Tm}}} + \frac{1}{\tau_{30}^{\text{Tm}}} + \Gamma_{3}^{\text{Tm}} } \ , \\
	\beta_{\text{Tm}} & \defeq \frac{ \kappa_{\text{r}}^{\text{Tm}} }{ \frac{\sigma_{\pmp}^{\text{ems}} \Ip}{\hbar \omega_{\pmp}} + \frac{1}{\tau_{32}^{\text{Tm}}} + \frac{1}{\tau_{31}^{\text{Tm}}} + \frac{1}{\tau_{30}^{\text{Tm}}} + \Gamma_{3}^{\text{Tm}} } = \frac{ \kappa_{\text{r}}^{\text{Tm}} }{ \frac{1}{\tau_{32}^{\text{Tm}}} + \frac{1}{\tau_{31}^{\text{Tm}}} + \frac{1}{\tau_{30}^{\text{Tm}}} + \Gamma_{3}^{\text{Tm}} } \ \left[ \frac{\text{m}^{3}}{\text{ion}} \right] \ , \\
	\gamma_{\text{Tm}} & \defeq \frac{ \frac{1}{\tau_{32}^{\text{Tm}}} + \Gamma_{3}^{\text{Tm}} }{ \frac{1}{\tau_{21}^{\text{Tm}}} + \frac{1}{\tau_{20}^{\text{Tm}}} + \Gamma_{2}^{\text{Tm}} } \ ,
\end{align*} \vspace{-20pt}
\begin{align*}
	A_{\text{Tm}} & \defeq \frac{ \frac{1}{\tau_{31}^{\text{Tm}}} + \left( \frac{1}{\tau_{21}^{\text{Tm}}} + \Gamma_{2}^{\text{Tm}} \right) \gamma_{\text{Tm}} + \frac{\sigma_{\sig}^{\text{abs}} \Is}{\hbar \omega_{\sig} \alpha_{\text{Tm}}} }{ \frac{1}{\tau_{10}^{\text{Tm}}} + \Gamma_{1}^{\text{Tm}} + \frac{\sigma_{\sig}^{\text{ems}} \Is}{\hbar \omega_{\sig}} } \ , \text{ and} \\
	B_{\text{Tm}} & \defeq \frac{ 2 \kappa_{\text{r}}^{\text{Tm}} + \frac{\beta_{\text{Tm}}}{\alpha_{\text{Tm}}} \frac{\sigma_{\sig}^{\text{abs}} I_{\sig}}{\hbar \omega_{\sig}} }{ \frac{1}{\tau_{10}^{\text{Tm}}} + \Gamma_{1}^{\text{Tm}} + \frac{\sigma_{\sig}^{\text{ems}} \Is}{\hbar \omega_{\sig}} } \ \left[ \frac{\text{m}^{3}}{\text{ion}} \right] \ .
\end{align*}
Moreover, $\tau_{10}^{\text{Tm}}$, $\tau_{20}^{\text{Tm}}$, $\tau_{21}^{\text{Tm}}$, $\tau_{30}^{\text{Tm}}$, $\tau_{31}^{\text{Tm}}$, and $\tau_{32}^{\text{Tm}}$ are the mean radiative lifetimes [s] of the excited electrons, and $\Gamma_{1}^{\text{Tm}}$, $\Gamma_{2}^{\text{Tm}}$, and $\Gamma_{3}^{\text{Tm}}$ are the mean non-radiative rates [1/s] of the excited electrons, all of which are measured (known) quantities. 
Finally, $\kappa_{\text{r}}^{\text{Tm}} = (1.8 \cdot 10^{-47}) \calN_{\text{total}}^{\text{Tm}}$ is the cross-relaxation volumetric rate [m$^{3}$/(ion$\cdot$s)] of the Tm$_{2}$O$_{3}$ molecule. 

Noting that both $A_{\text{Tm}}$ and $B_{\text{Tm}}$ are dependent on $\Is$, their partial derivatives with respect to the signal irradiance are given by 
\begin{align}
    \pp{A_{\text{Tm}}}{\Is} & = \frac{ \frac{ \sigma_{\sig}^{\text{abs}} }{ \hbar \omega_{\sig} \alpha_{\text{Tm}} } \left( \frac{1}{\tau_{10}^{\text{Tm}}} + \Gamma_{1}^{\text{Tm}} + \frac{\sigma_{\sig}^{\text{ems}} \Is}{\hbar \omega_{\sig}} \right) - \left( \frac{1}{\tau_{31}^{\text{Tm}}} + \left( \frac{1}{\tau_{21}^{\text{Tm}}} + \Gamma_{2}^{\text{Tm}} \right) \gamma_{\text{Tm}} + \frac{\sigma_{\sig}^{\text{abs}} \Is}{\hbar \omega_{\sig} \alpha_{\text{Tm}}} \right) \frac{ \sigma_{\sig}^{\text{ems}} }{ \hbar \omega_{\sig} } }{ \left( \frac{1}{\tau_{10}^{\text{Tm}}} + \Gamma_{1}^{\text{Tm}} + \frac{\sigma_{\sig}^{\text{ems}} \Is}{\hbar \omega_{\sig}} \right)^{2} } \notag \\
    \pp{A_{\text{Tm}}}{\Is} & = \frac{ \frac{ \sigma_{\sig}^{\text{abs}} \Is }{ \hbar \omega_{\sig} \alpha_{\text{Tm}} } - A_{\text{Tm}} \frac{ \sigma_{\sig}^{\text{ems}} \Is }{ \hbar \omega_{\sig} } }{ \Is \left( \frac{1}{\tau_{10}^{\text{Tm}}} + \Gamma_{1}^{\text{Tm}} + \frac{\sigma_{\sig}^{\text{ems}} \Is}{\hbar \omega_{\sig}} \right) } \ \ \left[ \frac{\text{m}^{2}}{\text{W}} \right] \ \text{ and} \label{eq:dATm-dIs} \\
    \pp{B_{\text{Tm}}}{\Is} & = \frac{ \frac{ \beta_{\text{Tm}} \sigma_{\sig}^{\text{abs}} }{ \alpha_{\text{Tm}} \hbar \omega_{\sig} } \left( \frac{1}{\tau_{10}^{\text{Tm}}} + \Gamma_{1}^{\text{Tm}} + \frac{\sigma_{\sig}^{\text{ems}} \Is}{\hbar \omega_{\sig}} \right) - \left( 2 \kappa_{\text{r}}^{\text{Tm}} + \frac{\beta_{\text{Tm}}}{\alpha_{\text{Tm}}} \frac{\sigma_{\sig}^{\text{abs}} I_{\sig}}{\hbar \omega_{\sig}} \right) \frac{ \sigma_{\sig}^{\text{ems}} }{ \hbar \omega_{\sig} } }{ \left( \frac{1}{\tau_{10}^{\text{Tm}}} + \Gamma_{1}^{\text{Tm}} + \frac{\sigma_{\sig}^{\text{ems}} \Is}{\hbar \omega_{\sig}} \right)^{2} } \notag \\
    \pp{B_{\text{Tm}}}{\Is} & = \frac{ \frac{ \beta_{\text{Tm}} \sigma_{\sig}^{\text{abs}} \Is }{ \alpha_{\text{Tm}} \hbar \omega_{\sig} } - B_{\text{Tm}} \frac{ \sigma_{\sig}^{\text{ems}} \Is }{ \hbar \omega_{\sig} } }{ \Is \left( \frac{1}{\tau_{10}^{\text{Tm}}} + \Gamma_{1}^{\text{Tm}} + \frac{\sigma_{\sig}^{\text{ems}} \Is}{\hbar \omega_{\sig}} \right) } \ \ \left[ \frac{\text{m}^{5}}{\text{ion} \cdot \text{W}} \right] \ .  \label{eq:dBTm-dIs}
\end{align}
Using~\eqref{eq:TmFixedPoint}, the partial derivative of the ground-state Tm dopant concentration ($\bar{\calN}_{\text{ground}}^{\text{Tm}}$) with respect to the signal irradiance is given by 
\begin{align}
    \pp{ \bar{\calN}_{\text{ground}}^{\text{Tm}} }{ \Is } & = \frac{ \alpha_{\text{Tm}} }{ 2 \left( \alpha_{\text{Tm}} B_{\text{Tm}} + \beta_{\text{Tm}} \right)^{2} } \Bigg\{ -\pp{ A_{\text{Tm}} }{ \Is } \left( \alpha_{\text{Tm}} B_{\text{Tm}} + \beta_{\text{Tm}} \right) \ - \\
        & \hspace*{12pt} \left( \beta_{\text{Tm}} \calN_{\text{total}}^{\text{Tm}} - \alpha_{\text{Tm}} \left[ A_{\text{Tm}} + \gamma_{\text{Tm}} + 1 \right] - 1 \right) \pp{ B_{\text{Tm}} }{ \Is }  \ + \notag \\
		& \hspace*{12pt} \frac{ \left( \alpha_{\text{Tm}} B_{\text{Tm}} + \beta_{\text{Tm}} \right) \left( \pp{A_{\text{Tm}}}{\Is} \left( \alpha_{\text{Tm}} \left[ A_{\text{Tm}} + \gamma_{\text{Tm}} + 1 \right] - \beta_{\text{Tm}} \calN_{\text{total}}^{\text{Tm}} + 1 \right) + 2 \calN_{\text{total}}^{\text{Tm}} \pp{B_{\text{Tm}}}{\Is} \right) }{ \sqrt{ \left( \alpha_{\text{Tm}} \left[ A_{\text{Tm}} + \gamma_{\text{Tm}} + 1 \right] - \beta_{\text{Tm}} \calN_{\text{total}}^{\text{Tm}} + 1 \right)^{2} + 4 \left( \alpha_{\text{Tm}} B_{\text{Tm}} + \beta_{\text{Tm}} \right) \calN_{\text{total}}^{\text{Tm}} } } \ - \notag \\
        & \hspace*{12pt} \pp{ B_{\text{Tm}} }{ \Is } \sqrt{\left( \alpha_{\text{Tm}} \left[ A_{\text{Tm}} + \gamma_{\text{Tm}} + 1 \right] - \beta_{\text{Tm}} \calN_{\text{total}}^{\text{Tm}} + 1 \right)^{2} + 4 \left( \alpha_{\text{Tm}} B_{\text{Tm}} + \beta_{\text{Tm}} \right) \calN_{\text{total}}^{\text{Tm}}} \Bigg\} \notag \\
    \pp{ \bar{\calN}_{\text{ground}}^{\text{Tm}} }{ \Is } & = \frac{ \alpha_{\text{Tm}} }{ \alpha_{\text{Tm}} B_{\text{Tm}} + \beta_{\text{Tm}} } \Bigg\{ -\frac{1}{2} \pp{ A_{\text{Tm}} }{ \Is } - \bar{\calN}_{\text{ground}}^{\text{Tm}} \pp{ B_{\text{Tm}} }{ \Is }  \ + \label{eq:dNgroundTm-dIs} \\
		& \hspace*{12pt} \frac{ \pp{A_{\text{Tm}}}{\Is} \left( \alpha_{\text{Tm}} \left[ A_{\text{Tm}} + \gamma_{\text{Tm}} + 1 \right] - \beta_{\text{Tm}} \calN_{\text{total}}^{\text{Tm}} + 1 \right) + 2 \calN_{\text{total}}^{\text{Tm}} \pp{B_{\text{Tm}}}{\Is} }{ 2 \sqrt{ \left( \alpha_{\text{Tm}} \left[ A_{\text{Tm}} + \gamma_{\text{Tm}} + 1 \right] - \beta_{\text{Tm}} \calN_{\text{total}}^{\text{Tm}} + 1 \right)^{2} + 4 \left( \alpha_{\text{Tm}} B_{\text{Tm}} + \beta_{\text{Tm}} \right) \calN_{\text{total}}^{\text{Tm}} } } \Bigg\} \ \ . \notag
\end{align}
The partial derivative of the lowest excited-state Tm dopant concentration ($\bar{\calN}_{\text{excited}}^{\text{Tm}}$) with respect to the signal irradiance can be expressed as a function of ${\p}\bar{\calN}_{\text{ground}}^{\text{Tm}} / {\p}\Is$: 
\begin{align}
    \pp{ \bar{\calN}_{\text{excited}}^{\text{Tm}} }{ \Is } & = \alpha_{\text{Tm}} \bar{\calN}_{\text{ground}}^{\text{Tm}} \pp{}{\Is}\left[ \frac{ A_{\text{Tm}} + B_{\text{Tm}} \bar{\calN}_{\text{ground}}^{\text{Tm}} }{ 1 + \beta_{\text{Tm}} \bar{\calN}_{\text{ground}}^{\text{Tm}} } \right] + \alpha_{\text{Tm}} \pp{\bar{\calN}_{\text{ground}}^{\text{Tm}}}{\Is} \left( \frac{ A_{\text{Tm}} + B_{\text{Tm}} \bar{\calN}_{\text{ground}}^{\text{Tm}} }{ 1 + \beta_{\text{Tm}} \bar{\calN}_{\text{ground}}^{\text{Tm}} } \right) \notag \\
    \pp{ \bar{\calN}_{\text{excited}}^{\text{Tm}} }{ \Is } & = \frac{ \alpha_{\text{Tm}} \bar{\calN}_{\text{ground}}^{\text{Tm}} \left( \pp{A_{\text{Tm}}}{\Is} + B_{\text{Tm}} \pp{\bar{\calN}_{\text{ground}}^{\text{Tm}}}{\Is} + \pp{B_{\text{Tm}}}{\Is} \bar{\calN}_{\text{ground}}^{\text{Tm}} \right) - \bar{\calN}_{\text{excited}}^{\text{Tm}} \beta_{\text{Tm}} \pp{\bar{\calN}_{\text{ground}}^{\text{Tm}}}{\Is} }{ 1 + \beta_{\text{Tm}} \bar{\calN}_{\text{ground}}^{\text{Tm}} } \ + \label{eq:dNexcitedTm-dIs} \\
        & \hspace*{12pt} \frac{ \pp{\bar{\calN}_{\text{ground}}^{\text{Tm}}}{\Is} \bar{\calN}_{\text{excited}}^{\text{Tm}}}{ \bar{\calN}_{\text{ground}}^{\text{Tm}} } \ \ . \notag
\end{align}
With this information, the partial derivatives of the steady-state gain functions ($\{ \bar{g}_{\pmp}, \bar{g}_{\sig} \}$) with respect to the signal irradiance are determined to be 
\begin{align}
    \pp{\bar{g}_{\pmp}}{\Is} & = \sigma_{\pmp}^{\text{ems}} \pp{\overline{\calN}_{\text{excited}}^{\text{Tm}}}{\Is} - \sigma_{\pmp}^{\text{abs}} \pp{\overline{\calN}_{\text{ground}}^{\text{Tm}}}{\Is} \label{eq:Tm-dgp-dIs} \ \ \text{ and} \\
    \pp{\bar{g}_{\sig}}{\Is} & = \sigma_{\sig}^{\text{ems}} \pp{\overline{\calN}_{\text{excited}}^{\text{Tm}}}{\Is} - \sigma_{\sig}^{\text{ems}} \pp{\overline{\calN}_{\text{ground}}^{\text{Tm}}}{\Is} \ \ , \label{eq:Tm-dgs-dIs}
\end{align}
relying on relations~(\ref{eq:dATm-dIs}--\ref{eq:dNexcitedTm-dIs}). 

Tm-doped laser amplifiers can also be in-band-pumped at $\lp \simeq 1663$ nm, where the ground-state and first excited-state manifolds are the same as for the out-of-band-pumping configuration. 
Similar to the Yb-doped amplifier, in this case the total dopant concentration is considered a known measured quantity such that 
\[
    \calN_{\text{total}}^{\text{Tm}} = \calN_{\text{ground}}^{\text{Tm}}(x, y, z, t) + \calN_{\text{excited}}^{\text{Tm}}(x, y, z, t) \ \ \left[ \frac{\text{ions}}{\text{m}^{3}} \right] .
\]
The steady-state gain kinetics, driven by specific pump and laser signal irradiance levels ($\{ \Ip, \Is \}$), satisfy 
\begin{equation}\label{eq:InBandPumpedTmFixedPoint}
	\begin{aligned}
		\bs{\overline{\calN}}^{\text{Tm}} & = 
		\bmat
			\overline{\calN}_{\text{ground}}^{\text{Tm}} \\
			\overline{\calN}_{\text{excited}}^{\text{Tm}}
		\ebmat = 
		\bmat
			\frac{ 1 }{ 1 + C_{\text{Tm}} } \\
			\frac{ C_{\text{Tm}} }{ 1 + C_{\text{Tm}} }
		\ebmat \cdot \calN_{\text{total}}^{\text{Tm}} \ \ \left[ \frac{\text{ions}}{\text{m}^{3}} \right] , \text{ where} \\
            C_{\text{Tm}} & \defeq \frac{ \frac{\sigma_{\pmp}^{\text{abs}} \Ip}{\hbar \omega_{\pmp}} + \frac{\sigma_{\sig}^{\text{abs}} \Is}{\hbar \omega_{\sig}} }{ \frac{1}{\tau_{10}^{\text{Tm}}} + \Gamma_{1}^{\text{Tm}} + \frac{\sigma_{\pmp}^{\text{ems}} \Ip}{\hbar \omega_{\pmp}} + \frac{\sigma_{\sig}^{\text{ems}} \Is}{\hbar \omega_{\sig}} } \ \ .
	\end{aligned}
\end{equation}
Since $C_{\text{Tm}}$ is dependent on $\Is$, its partial derivative with respect to the signal irradiance is given by 
\begin{align}
    \pp{C_{\text{Tm}}}{\Is} & = \frac{ \frac{ \sigma_{\sig}^{\text{abs}} }{ \hbar \omega_{\sig} } \left( \frac{1}{\tau_{10}^{\text{Tm}}} + \Gamma_{1}^{\text{Tm}} + \frac{\sigma_{\pmp}^{\text{ems}} \Ip}{\hbar \omega_{\pmp}} + \frac{\sigma_{\sig}^{\text{ems}} \Is}{\hbar \omega_{\sig}} \right) - \left( \frac{\sigma_{\pmp}^{\text{abs}} \Ip}{\hbar \omega_{\pmp}} + \frac{\sigma_{\sig}^{\text{abs}} \Is}{\hbar \omega_{\sig}} \right) \frac{ \sigma_{\sig}^{\text{ems}} }{ \hbar \omega_{\sig} } }{ \left( \frac{1}{\tau_{10}^{\text{Tm}}} + \Gamma_{1}^{\text{Tm}} + \frac{\sigma_{\pmp}^{\text{ems}} \Ip}{\hbar \omega_{\pmp}} + \frac{\sigma_{\sig}^{\text{ems}} \Is}{\hbar \omega_{\sig}} \right)^{2} } \notag \\
    \pp{C_{\text{Tm}}}{\Is} & = \frac{ \frac{ \sigma_{\sig}^{\text{abs}} \Is }{ \hbar \omega_{\sig} } - C_{\text{Tm}} \frac{ \sigma_{\sig}^{\text{ems}} \Is }{ \hbar \omega_{\sig} } }{ \Is \left( \frac{1}{\tau_{10}^{\text{Tm}}} + \Gamma_{1}^{\text{Tm}} + \frac{ \sigma_{\pmp}^{\text{ems}} \Ip }{ \hbar \omega_{\pmp} } + \frac{ \sigma_{\sig}^{\text{ems}} \Is }{ \hbar \omega_{\sig} } \right) } \ \ \left[ \frac{\text{m}^{2}}{\text{W}} \right] \ . \label{eq:dCTm-dIs}
\end{align}
The partial derivative of the lowest excited-state concentration ($\overline{\calN}_{\text{excited}}^{\text{Tm}}$) with respect to the signal irradiance is given by 
\begin{equation}\label{eq:In-Band-dNexcitedTm-dIs}
    \pp{\bar{\calN}_{\text{excited}}^{\text{Tm}}}{\Is} = \calN_{\text{total}}^{\text{Tm}} \frac{ \pp{C_{\text{Tm}}}{\Is} \left( 1 + C_{\text{Tm}} \right) - C_{\text{Tm}} \pp{C_{\text{Tm}}}{\Is} }{ \left( 1 + C_{\text{Tm}} \right)^{2} } 
    = \pp{C_{\text{Tm}}}{\Is} \left( 1 - \frac{ \bar{\calN}_{\text{excited}}^{\text{Tm}} }{ \calN_{\text{total}}^{\text{Tm}} } \right) \bar{\calN}_{\text{ground}}^{\text{Tm}} \ \ .
\end{equation}
Just as with the in-band-pumped Yb molecule, the steady-state gain can be formulated without the ground-state concentration, as was done in~\eqref{eq:InBandPumpedSteadyStateGain}. 
Using relations~(\ref{eq:InBandPumpedTmFixedPoint}--\ref{eq:In-Band-dNexcitedTm-dIs}) with $\ell \in \{ \pmp, \sig \}$, the partial derivatives of the steady-state gain functions ($\{ \bar{g}_{\pmp}, \bar{g}_{\sig} \}$) with respect to the signal irradiance are given by 
\begin{align}
    \pp{\bar{g}_{\pmp}}{\Is} & = \big( \sigma_{\pmp}^{\text{abs}} + \sigma_{\pmp}^{\text{ems}} \big) \pp{\overline{\calN}_{\text{excited}}^{\text{Tm}}}{\Is} 
        = \big( \sigma_{\pmp}^{\text{abs}} + \sigma_{\pmp}^{\text{ems}} \big) \pp{C_{\text{Tm}}}{\Is} \left( 1 - \frac{ \bar{\calN}_{\text{excited}}^{\text{Tm}} }{ \calN_{\text{total}}^{\text{Tm}} } \right) \bar{\calN}_{\text{ground}}^{\text{Tm}} \label{eq:In-Band-Tm-dgp-dIs} \ \text{ and} \\
    \pp{\bar{g}_{\sig}}{\Is} & = \big( \sigma_{\sig}^{\text{abs}} + \sigma_{\sig}^{\text{ems}} \big) \pp{\overline{\calN}_{\text{excited}}^{\text{Tm}}}{\Is} 
        = \big( \sigma_{\sig}^{\text{abs}} + \sigma_{\sig}^{\text{ems}} \big) \pp{C_{\text{Tm}}}{\Is} \left( 1 - \frac{ \bar{\calN}_{\text{excited}}^{\text{Tm}} }{ \calN_{\text{total}}^{\text{Tm}} } \right) \bar{\calN}_{\text{ground}}^{\text{Tm}} \label{eq:In-Band-Tm-dgs-dIs} \ \ ,
\end{align}
where ${\p}C_{\text{Tm}} / {\p}\Is$ is stated in~\eqref{eq:dCTm-dIs}.

\FloatBarrier
%%%%%%%%%%%%%%%%%%%%%%%%%%%
% Appendix Subsection: Gain Kinetics of Holmium Oxide
%%%%%%%%%%%%%%%%%%%%%%%%%%%

\subsection{Gain Kinetics of Holmium Oxide}\label{subsec:HoDopant}

Ho-doped amplifiers are typically in-band-pumped at $\lp \simeq 1951$ nm, exciting the outer-shell electron from the ground-state manifold $\prescript{5}{}{\mathbf{I}}_{8}$ (associated with $\calN_{0} \equiv \calN_{\text{ground}}$) to the excited-state manifold $\prescript{5}{}{\mathbf{I}}_{7}$ (associated with $\calN_{1} \equiv \calN_{\text{excited}}$)~\cite{wang2018numerical, alharbi2022performance}. 
However, the Ho$_{2}$O$_{3}$ oxide molecule experiences a substantial up-conversion process, which can send some excited electrons up into the $\prescript{5}{}{\mathbf{I}}_{5}$ manifold (associated with $\calN_{3}$), which, afterwards, may decay into the $\prescript{5}{}{\mathbf{I}}_{6}$ (associated with $\calN_{2}$) excited manifold.
Therefore, the total dopant concentration is 
\[
    \calN_{\text{total}}^{\text{Ho}} = \calN_{\text{ground}}^{\text{Ho}}(x, y, z, t) + \calN_{\text{excited}}^{\text{Ho}}(x, y, z, t) + 
    \calN_{2}^{\text{Ho}}(x, y, z, t) + \calN_{3}^{\text{Ho}}(x, y, z, t) \ \ \left[ \frac{\text{ions}}{\text{m}^{3}} \right] .
\]
Again, at specific pump and laser signal irradiance levels, the steady-state gain kinetics satisfy 
\begin{equation}\label{eq:HoFixedPoint}
    \begin{aligned}
		& \bs{\overline{\calN}}^{\text{Ho}} = 
		\bmat
			\overline{\calN}_{\text{ground}}^{\text{Ho}} \\
			\overline{\calN}_{\text{excited}}^{\text{Ho}} \\
			\overline{\calN}_{2}^{\text{Ho}} \\
			\overline{\calN}_{3}^{\text{Ho}}
		\ebmat = 
		\bmat
			\calN_{\text{total}}^{\text{Ho}} - \overline{\calN}_{\text{excited}}^{\text{Ho}} - p_{4}^{\text{Ho}} \mathcal{U}_{\text{c}}^{\text{Ho}} \big( \overline{\calN}_{\text{excited}}^{\text{Ho}} \big)^{2} \\
			\overline{\calN}_{\text{excited}}^{\text{Ho}} \\
			\frac{ p_{2}^{\text{Ho}} p_{3}^{\text{Ho}} \mathcal{U}_{\text{c}}^{\text{Ho}} \big( \overline{\calN}_{\text{excited}}^{\text{Ho}} \big)^{2} }{ R_{21}^{\text{Ho}} } \\
			\frac{ p_{1}^{\text{Ho}} \mathcal{U}_{\text{c}}^{\text{Ho}} \big( \overline{\calN}_{\text{excited}}^{\text{Ho}} \big)^{2} }{ R_{31}^{\text{Ho}} }
		\ebmat \text{  where} \\
		& \overline{\calN}_{\text{excited}}^{\text{Ho}} = \frac{ -p_{b}^{\text{Ho}} + \sqrt{ \big( p_{b}^{\text{Ho}} \big)^2 - 4 p_{a}^{\text{Ho}} p_{c}^{\text{Ho}} } }{ 2 p_{a}^{\text{Ho}} } \ \ \left[ \frac{\text{ions}}{\text{m}^{3}} \right] ,
	\end{aligned}
\end{equation}
where 
\begin{align*}
    p_{1}^{\text{Ho}} & \defeq \frac{ R_{31}^{\text{Ho}} }{ R_{32}^{\text{Ho}} + R_{31}^{\text{Ho}} + R_{30}^{\text{Ho}} } \ , \\
    p_{2}^{\text{Ho}} & \defeq \frac{ R_{32}^{\text{Ho}} }{ R_{32}^{\text{Ho}} + R_{31}^{\text{Ho}} + R_{30}^{\text{Ho}} } \ , \\
    p_{3}^{\text{Ho}} & \defeq  \frac{ R_{21}^{\text{Ho}} }{ R_{21}^{\text{Ho}} + R_{20}^{\text{Ho}} } \ , \\
    p_{4}^{\text{Ho}} & \defeq \frac{ R_{21}^{\text{Ho}} + R_{20}^{\text{Ho}} + R_{32}^{\text{Ho}} }{ \left( R_{21}^{\text{Ho}} + R_{20}^{\text{Ho}} \right)\left( R_{32}^{\text{Ho}} + R_{31}^{\text{Ho}} + R_{30}^{\text{Ho}} \right) } \ \ \left[ \text{s} \right] , \\
    p_{a}^{\text{Ho}} & \defeq \left( 2 - p_{1}^{\text{Ho}} - p_{2}^{\text{Ho}} \cdot p_{3}^{\text{Ho}} + \left[ \frac{\sigma_{\pmp}^{\text{abs}} \Ip}{\hbar \omega_{\pmp}} + \frac{\sigma_{\sig}^{\text{abs}} \Is}{\hbar \omega_{\sig}} \right] p_{4}^{\text{Ho}} \right) \mathcal{U}_{\text{c}}^{\text{Ho}} \ \ \left[ \frac{ \text{m}^{3} }{ \text{ion} \cdot \text{s} } \right] , \\
    p_{b}^{\text{Ho}} & \defeq R_{10}^{\text{Ho}} + \frac{\left( \sigma_{\pmp}^{\text{abs}} + \sigma_{\pmp}^{\text{ems}} \right) \Ip}{\hbar \omega_{\pmp}} + \frac{\left( \sigma_{\sig}^{\text{abs}} + \sigma_{\sig}^{\text{ems}} \right) \Is}{\hbar \omega_{\sig}} \ \ \left[ \frac{1}{\text{s}} \right] , \text{ and} \\
    p_{c}^{\text{Ho}} & \defeq -\left( \frac{\sigma_{\pmp}^{\text{abs}} \Ip}{\hbar \omega_{\pmp}} + \frac{\sigma_{\sig}^{\text{abs}} \Is}{\hbar \omega_{\sig}} \right) \calN_{\text{total}}^{\text{Ho}} \ \ \left[ \frac{ \text{ions} }{ \text{m}^{3} \cdot \text{s} } \right] .
\end{align*}
Additionally, $R_{10}^{\text{Ho}}$, $R_{20}^{\text{Ho}}$, $R_{21}^{\text{Ho}}$, $R_{30}^{\text{Ho}}$, $R_{31}^{\text{Ho}}$, and $R_{32}^{\text{Ho}}$ are summations of the radiative and non-radiative manifold transition rates [1/s], and $\mathcal{U}_{\text{c}}^{\text{Ho}}$ is the measured up-conversion volumetric rate [m$^{3}$/(ion$\cdot$s)]. 

Noting that $p_{a}^{\text{Ho}}$, $p_{b}^{\text{Ho}}$, and $p_{c}^{\text{Ho}}$ are dependent on $\Is$, the partial derivatives of these parameters with respect to the signal irradiance are given by
\begin{subequations}\label{eq:HoParameterDerivatives}
    \begin{align}
        \pp{p_{a}^{\text{Ho}}}{\Is} & = \frac{ \sigma_{\sig}^{\text{abs}} p_{4}^{\text{Ho}} \mathcal{U}_{\text{c}}^{\text{Ho}} }{ \hbar \omega_{\sig} } \ \ \left[ \frac{ \text{m}^{5} }{ \text{ion} \cdot \text{s} \cdot \text{W} } \right] , \label{eq:HoAParameterDerivative} \\
        \pp{p_{b}^{\text{Ho}}}{\Is} & = \frac{ \sigma_{\sig}^{\text{abs}} + \sigma_{\sig}^{\text{ems}} }{ \hbar \omega_{\sig} } \ \ \left[ \frac{\text{m}^{2}}{\text{s} \cdot \text{W}} \right] , \text{ and} \label{eq:HoBParameterDerivative} \\
        \pp{p_{c}^{\text{Ho}}}{\Is} & = -\frac{ \sigma_{\sig}^{\text{abs}} \calN_{\text{total}}^{\text{Ho}} }{ \hbar \omega_{\sig} } \ \ \left[ \frac{ \text{ions} }{ \text{m} \cdot \text{s} \cdot \text{W} } \right] . \label{eq:HoCParameterDerivative}
    \end{align}
\end{subequations}
The partial derivative of the lowest excited-state concentration level with respect to the signal irradiance can be expressed as 
\begin{align}
    \pp{\overline{\calN}_{\text{excited}}^{\text{Ho}}}{\Is} & = \Bigg[ \Bigg( -\pp{p_{b}^{\text{Ho}}}{\Is} + \frac{ 2 p_{b}^{\text{Ho}} \pp{p_{b}^{\text{Ho}}}{\Is} - 4 \left[ \pp{p_{a}^{\text{Ho}}}{\Is} p_{c}^{\text{Ho}} + p_{a}^{\text{Ho}} \pp{p_{c}^{\text{Ho}}}{\Is} \right] }{ 2 \sqrt{ \left( p_{b}^{\text{Ho}} \right)^2 - 4 p_{a}^{\text{Ho}} p_{c}^{\text{Ho}} } } \Bigg) \left( 2 p_{a}^{\text{Ho}} \right) \ - \notag \\
        & \hspace*{12pt}  \left( -p_{b}^{\text{Ho}} + \sqrt{ \big( p_{b}^{\text{Ho}} \big)^2 - 4 p_{a}^{\text{Ho}} p_{c}^{\text{Ho}} } \right)  \left( 2 \pp{p_{a}^{\text{Ho}}}{\Is} \right) \Bigg] \left( 2 p_{a}^{\text{Ho}} \right)^{-2} \notag \\
    \pp{\overline{\calN}_{\text{excited}}^{\text{Ho}}}{\Is} & = \frac{1}{ p_{a}^{\text{Ho}} } \Bigg[ \frac{1}{2} \Bigg( -\pp{p_{b}^{\text{Ho}}}{\Is} + \frac{ p_{b}^{\text{Ho}} \pp{p_{b}^{\text{Ho}}}{\Is} - 2 \left[ \pp{p_{a}^{\text{Ho}}}{\Is} p_{c}^{\text{Ho}} + p_{a}^{\text{Ho}} \pp{p_{c}^{\text{Ho}}}{\Is} \right] }{ \sqrt{ \left( p_{b}^{\text{Ho}} \right)^2 - 4 p_{a}^{\text{Ho}} p_{c}^{\text{Ho}} } } \Bigg) \ - \label{eq:Ho_dNexc_dIs} \\
        & \hspace*{12pt}  \pp{p_{a}^{\text{Ho}}}{\Is} \overline{\calN}_{\text{excited}}^{\text{Ho}} \Bigg] \ \ \left[ \frac{\text{ions}}{\text{m} \cdot \text{W}} \right]  , \notag
\end{align}
which can be determined with~\eqref{eq:HoParameterDerivatives}. 
This can then be used to find the partial derivative of the ground-state concentration level with respect to the signal irradiance: 
\begin{align}
    \pp{\overline{\calN}_{\text{ground}}^{\text{Ho}}}{\Is} & = -\pp{\overline{\calN}_{\text{excited}}^{\text{Ho}}}{\Is} - 2 p_{4}^{\text{Ho}} \mathcal{U}_{\text{c}}^{\text{Ho}} \overline{\calN}_{\text{excited}}^{\text{Ho}} \pp{\overline{\calN}_{\text{excited}}^{\text{Ho}}}{\Is} \notag \\
    \pp{\overline{\calN}_{\text{ground}}^{\text{Ho}}}{\Is} & = -\left( 1 + 2 p_{4}^{\text{Ho}} \mathcal{U}_{\text{c}}^{\text{Ho}} \overline{\calN}_{\text{excited}}^{\text{Ho}} \right) \pp{\overline{\calN}_{\text{excited}}^{\text{Ho}}}{\Is} \label{eq:Ho_dNgnd_dIs} \ \ .
\end{align}
Using~\eqref{eq:HoFixedPoint} with $\ell \in \{ \pmp, \sig  \}$ and relations~\eqref{eq:HoParameterDerivatives}--\eqref{eq:Ho_dNgnd_dIs}, the partial derivatives of the steady-state gain functions ($\{ \bar{g}_{\pmp}, \bar{g}_{\sig} \}$) with respect to the signal irradiance are given by 
\begin{align}
    \pp{\bar{g}_{\pmp}}{\Is} & = \sigma_{\pmp}^{\text{ems}} \pp{\overline{\calN}_{\text{excited}}^{\text{Ho}}}{\Is} - \sigma_{\pmp}^{\text{abs}} \pp{\overline{\calN}_{\text{ground}}^{\text{Ho}}}{\Is} \label{eq:Ho-dgp-dIs} \ \text{ and} \\
    \pp{\bar{g}_{\sig}}{\Is} & = \sigma_{\sig}^{\text{ems}} \pp{\overline{\calN}_{\text{excited}}^{\text{Ho}}}{\Is} - \sigma_{\sig}^{\text{abs}} \pp{\overline{\calN}_{\text{ground}}^{\text{Ho}}}{\Is} \label{eq:Ho-dgs-dIs} \ \ .
\end{align}

\FloatBarrier
%\newpage
%%%%%%%%%%%%%%%%%%%%%%%%%%%%%%%%%%%%%%%%%%%%%%%%%%%%
% Section: Back Matter
%%%%%%%%%%%%%%%%%%%%%%%%%%%%%%%%%%%%%%%%%%%%%%%%%%%%
\section*{Back Matter}
   
    \subsection*{Funding}
    This research was supported in part by AFOSR grant FA9550-23-1-0103. This work also benefited from activities organized under the auspices of NSF RTG grant DMS-2136228.
    
    \subsection*{Acknowledgment}
    R.~Bryant gratefully acknowledges support from two internships with the AFRL Scholars Program and the DoD SMART Scholarship Program. 
    
    \subsection*{Disclosures}
    The authors declare no conflicts of interest.

    \subsection*{Data Availability}
    Data underlying the results presented in this article are not publicly available at this time, but may be obtained from the authors upon reasonable request.

    \subsection*{Disclaimers}
    This article has been approved for public release; distribution unlimited. 
    Public Affairs release approval {\#}AFRL-2025-5242. 
    The views expressed in this article are those of the authors and do not necessarily reflect the official policy or position of the Department of the Air Force, of the Department of Defense, nor of the U.S. government.

\end{document}